\documentstyle[12pt,bezier,epsf,epsfig]{article}
\newcommand{\bq}{\begin{equation}}
\newcommand{\eq}{\end{equation}}
\newcommand{\ba}{\begin{eqnarray}}
\newcommand{\ea}{\end{eqnarray}}
\def\theequation{\arabic{section}.\arabic{equation}}
\newcommand{\ezero}{\setcounter{equation}{0}}
\newcommand{\vph}{ \vphantom{\int\limits_t^t} }

\newcommand{\mr}{\rm}

\newcommand {\nll }{\nonumber \\              }
\newcommand {\nn  }{\noindent                 }
\newcommand {\lnl }{\nonumber \\ \\           }
\newcommand {\Sl  }{\mbox{$ S_{1}           $}}
\newcommand {\Qe  }{\mbox{$ Q^2_{e}         $}}
\newcommand {\Qel }{\mbox{$ {\tt Q}^2_{e}   $}}
\newcommand {\Qmul}{\mbox{$ {\tt Q}^2_{\mu} $}}
\newcommand {\Qmue}{\mbox{${\tt Q}_{\mu}{\tt Q}_e $}}

\newcommand {\ye  }{\mbox{$ y  _{e}         $}}
\newcommand {\me  }{\mbox{$ m^2_{e}         $}}
\newcommand {\Me  }{\mbox{$ m^2_{e}         $}}
\newcommand {\Mm  }{\mbox{$ m^2_{\mu}       $}}

\newcommand {\mmu }{\mbox{$ m^2_{\mu}       $}}
\newcommand {\Qmu }{\mbox{$ Q^2_{\mu}       $}}
\newcommand {\Qm  }{\mbox{$ Q^2_{\mu}       $}}

\newcommand {\ymu }{\mbox{$ y  _{\mu}       $}}
\newcommand {\xie }{\mbox{$ \xi_{e}         $}}
\newcommand {\ximu}{\mbox{$ \xi_{\mu}       $}}
\newcommand {\Pe  }{\mbox{$ P_{e}           $}}
\newcommand {\Pmu }{\mbox{$ P_{\mu}         $}}
\newcommand {\vkl }{\mbox{$ |\vec{k_1}|     $}}
\newcommand {\Ql} {\mbox{$Q^2_{\mu}$}}
\newcommand {\ql} {\mbox{$Q^2_{\mu}$}}
\newcommand {\yl} {\mbox{$y  _{\mu}$}}
\newcommand {\Qes}{\mbox{$Q^4_e    $}}
\newcommand {\Qms}{\mbox{$Q^4_{\mu}$}}
\newcommand {\ds }{\displaystyle                }
\newcommand {\SLS}{\mbox{$\sqrt{\lambda_S}  $}  }
\newcommand{\QMSS       }{\mbox{$  Q^4_\mu               $}}
\newcommand{\QllMMAXlll }{\mbox{$ {Q^6_{\mu}}^{\max}     $}}
\newcommand{\QllMMAX    }{\mbox{$ { Q^2_\mu}^{\max}      $}}
\newcommand{\QllM       }{\mbox{$ Q^2_{\mu}              $}}
\newcommand{\QllMMIN    }{\mbox{$ {Q^2_{\mu}}^{\min}     $}}
\newcommand{\QMMAXS     }{\mbox{$ {Q^4_{\mu}}^{\max}     $}}
\newcommand{\QMMAX      }{\mbox{$ {Q^2_{\mu}}^{\max}     $}}

\newcommand{\YMUlll     }{\mbox{$ y^3_{\mu}  $}}
\newcommand{\YMUS       }{\mbox{$ y^2_{\mu}  $}}

\newcommand{\YMU        }{\mbox{$ y_{\mu}    $}}
\newcommand{\AMUS       }{\mbox{$ m^4_{\mu}  $}}

\newcommand{\LLM        }{\mbox{${\rm L}_m         $}}
\newcommand{\LKl        }{\mbox{${\rm L}_{k1}      $}}
\newcommand{\LKll       }{\mbox{${\rm L}_{k2}      $}}
\newcommand{\LSS        }{\mbox{${\rm L}_{S}       $}}
\newcommand{\LSSl       }{\mbox{${\rm L}_{S_1}     $}}
\newcommand{\TMIN       }{\mbox{$t^{\min}    $}}
\newcommand{\TMAX       }{\mbox{$t^{\max}    $}}

\newcommand{\LOG        }{\mbox{$ \ln    $}}

\newcommand{\SPENCE     }{\mbox{$ {\rm Li}_2       $}}
\newcommand{\Litwo      }{\mbox{$ {\rm Li}_2       $}}
\newcommand{\litwo      }{\mbox{$ {\rm Li}_2       $}}
\newcommand{\DETl       }{\mbox{${ \rm D}_{k_1}       $}}
\newcommand{\DETll      }{\mbox{${ \rm D}_{k_2}       $}}
\newcommand{\LPl        }{\mbox{${\rm L}_{p1}         $}}
\newcommand{\VllMAX     }{\mbox{$  V^{\max}_2       $}}
\newcommand{\VllMAXS    }{\mbox{$ (V^{\max}_2)^2    $}}
\newcommand{\VllMIN     }{\mbox{$  V^{\min}_2       $}}
\newcommand{\VllMINS    }{\mbox{$ (V^{\min}_2)^2    $}}
\newcommand{\ALAMl      }{\mbox{$ {\lambda}_1        $}}
\newcommand{\ALAMll     }{\mbox{$ {\lambda}_2        $}}

\newcommand{\ALM        }{\mbox{$  \lambda_m          $}}
\newcommand{\SLAMM      }{\mbox{$\sqrt{\lambda^0_m}   $}}
\newcommand{\SLAMMS     }{\mbox{${\lambda^0_m  }      $}}
\newcommand{\SLAMMlll   }{\mbox{$(\lambda^0_m )^{3/2} $}}
\newcommand{\ALllMAX    }{\mbox{$ \lambda^{\max}_{k2} $}}
\newcommand{\ALlMAX     }{\mbox{$ \lambda^{\max}_{k1} $}}

\newcommand{\TlMAXS     }{\mbox{$(t^{\max}_1)^2 $}}
\newcommand{\TllMIN     }{\mbox{$t^{\min}_2     $}}
\newcommand{\TlMIN      }{\mbox{$t^{\min}_1     $}}

\newcommand{\TAMAX      }{\mbox{$t^{\max}_a     $}}
\newcommand{\TAMIN      }{\mbox{$t^{\min}_a     $}}
\newcommand{\TOl        }{\mbox{$t^0_1          $}}
\newcommand{\TAO        }{\mbox{$t^0_a          $}}
\newcommand{\TOlS       }{\mbox{$(t^{0}_{1})^2  $}}
\newcommand{\TOll       }{\mbox{$t^0_2          $}}
\newcommand{\TOllS      }{\mbox{$(t^{0}_{2})^2  $}}
\newcommand{\XTl        }{\mbox{$t_{11}         $}}
\newcommand{\XTll       }{\mbox{$t_{12}         $}}
\newcommand{\XMUl       }{\mbox{$x_{\mu1}       $}}

\newcommand{\TllMAX     }{\mbox{$t^{\max}_{2}   $}}
\newcommand{\XlTl       }{\mbox{$t_{11 }        $}}
\newcommand{\XlTll      }{\mbox{$t_{12 }        $}}
\newcommand{\XllTl      }{\mbox{$t_{21 }        $}}
\newcommand{\XllTll     }{\mbox{$t_{22 }        $}}
\newcommand{\XllDl      }{\mbox{$d_{ 1 }        $}}
\newcommand{\XllDll     }{\mbox{$d_{ 2 }        $}}
\newcommand{\XSDl       }{\mbox{$\sqrt{S_{d1}}  $}}

\newcommand{\RTl        }{\mbox{$t_{21}         $}}
\newcommand{\RTll       }{\mbox{$t_{22}         $}}
\newcommand{\ALME       }{\mbox{${\ds\ln\frac{S y_\mu}{\AME} }$}}
\newcommand{\ALVM       }{\mbox{${\ds\ln\frac{\VllMAX}{S\YMU}}$}}
\newcommand{\TA         }{\mbox{$t _{a}           $}}
\newcommand{\TO         }{\mbox{$t _{0}           $}}
\newcommand{\LQE        }{\mbox{${\rm L}_{Q_e}    $}}
\newcommand{\DKl        }{\mbox{${\rm D}_{k1}     $}}
\newcommand{\DKll       }{\mbox{${\rm D}_{k2}     $}}
\newcommand{\RMU        }{\mbox{$r_{\mu}          $}}
\newcommand{\LAMl       }{\mbox{$ {\lambda}_{1}   $}}
\newcommand{\LAMll      }{\mbox{$ {\lambda}_{2}   $}}
\newcommand{\RMUl       }{\mbox{$r_{\mu_1}        $}}
\newcommand{\SDETl      }{\mbox{$\sqrt{{\rm D}_{k_1}}   $}}
\newcommand{\SDETll     }{\mbox{$\sqrt{{\rm D}_{k_2}}   $}}
\newcommand{\BLAMl      }{\mbox{$  {\rm B}_{k_1}  $}}
\newcommand{\BLAMll     }{\mbox{$  {\rm B}_{k_2}  $}}
\newcommand{\TlMAX      }{\mbox{$t^{\max}_{1}     $}}
\newcommand{\Vl         }{\mbox{$V       _{1}     $}}
\newcommand{\Vls        }{\mbox{$V^2     _{1}     $}}
\newcommand{\Vll        }{\mbox{$V       _{2}     $}}
\newcommand{\Vlls       }{\mbox{$V^2     _{2}     $}}
\newcommand{\Zl         }{\mbox{$z       _{1}     $}}
\newcommand{\Zll        }{\mbox{$z       _{2}     $}}
\newcommand{\Zlls       }{\mbox{$z^2     _{2}     $}}
\newcommand{\Zls        }{\mbox{$z^2_1            $}}

\newcommand {  \Pm    }{\mbox{$  P_{\mu}     $}}

\newcommand {  \PIRMU }{\mbox{$ P^{IR}(\mu)  $}}
\newcommand {  \ym    }{\mbox{$  y_\mu       $}}
\newcommand {  \yms   }{\mbox{$  y^2_\mu     $}}
\newcommand {  \Rm    }{\mbox{$  R_{\mu}     $}}
\newcommand {  \Rms   }{\mbox{$  R^2_{\mu}   $}}
\newcommand {  \LSO   }{\mbox{$  l_{S}       $}}
\newcommand {  \LSl   }{\mbox{$  l_{S_1}     $}}
\newcommand {  \yml   }{\mbox{$  y_{\mu_1}   $}}

\newcommand {  \ymls  }{\mbox{$  y^2_{\mu_1} $}}
\newcommand {  \JKl   }{\mbox{$   J_K        $}}
\newcommand {  \KSO   }{\mbox{$  K_{S}       $}}
\newcommand {  \KSl   }{\mbox{$  K_{S_1}     $}}
\newcommand {  \JPl   }{\mbox{$   J_P        $}}

\newcommand{\QM   }{\mbox{$ Q^2_\mu           $}}
\newcommand{\AME  }{\mbox{$ m^2_e             $}}
\newcommand{\AMU  }{\mbox{$ m^2_\mu           $}}
\newcommand{\QlM  }{\mbox{$ Q^2_\mu           $}}

\newcommand{\QMS  }{\mbox{$ Q^4_\mu           $}}

\newcommand{\LAMS }{\mbox{$ \lambda_{_{S}}    $}}
\newcommand{\SLAMS}{\mbox{$ \sqrt{\lambda_S } $}}
\newcommand{\ALS  }{\mbox{$  \lambda_{_{S}}   $}}
\newcommand{\ALSS }{\mbox{$  \lambda^2_S      $}}

\newcommand{\SLAML}{\mbox{$ \sqrt{\lambda_l } $}}
\newcommand{\LAMU }{\mbox{$ \lambda_\mu       $}}
\newcommand{\SLAMU}{\mbox{$ \sqrt{\lambda_\mu}$}}
\newcommand{\PPl  }{\mbox{$ \lambda_\mu       $}}
\newcommand{\LAME }{\mbox{$ \lambda_e         $}}
\newcommand{\SLAME}{\mbox{$ \sqrt{\lambda_e } $}}
\newcommand{\SLX  }{\mbox{$ \sqrt{\lambda_l } $}}
\newcommand{\ALX  }{\mbox{$ \lambda_l         $}}
\newcommand{\ALXS }{\mbox{$ \lambda^2_l       $}}
\newcommand{\SMM  }{\mbox{$ \sqrt\lambda_m    $}}
\newcommand{\SLP  }{\mbox{$ S y_\mu +\sqrt\lambda_\mu $}}
\newcommand{\SLM  }{\mbox{$ S y_\mu -\sqrt\lambda_\mu $}}
\newcommand{\TAU  }{\mbox{$\tau          $}}
\newcommand{\Fl   }{\mbox{$F_1          $}}
\newcommand{\EKl  }{\mbox{$ S_{k1}       $}}

\newcommand{\EKll }{\mbox{$S_{k2}        $}}
\newcommand{\EPl  }{\mbox{$S_{p1}        $}}

\newcommand{\PKl  }{\mbox{$\sqrt{\lambda_{k1}}  $}}
\newcommand{\PKlS }{\mbox{$     {\lambda_{k1}}  $}}
\newcommand{\PKll }{\mbox{$\sqrt{\lambda_{k2}}  $}}
\newcommand{\PKllS}{\mbox{$     {\lambda_{k2}}  $}}
\newcommand{\VlS  }{\mbox{$V^2_{1}    $}}
\newcommand{\VllS }{\mbox{$V^2_{2}    $}}
\newcommand{\ZlS  }{\mbox{$z^2_{1}    $}}
\newcommand{\ZllS }{\mbox{$z^2_{2}    $}}
\newcommand{\born }{\mbox{$ \sigma^{^{\rm BORN}} $}}

\newcommand{\sskip   }{\vspace{.05cm} \\}
\newcommand{\smskip  }{\vspace{.1cm}  \\}

\newcommand{\REC } {\mbox{$ E^{^{RC}} $}}
\newcommand{\ERC } {\mbox{$ E^{^{RC}} $}}
\newcommand{\EBC } {\mbox{$ E^{^{BC}} $}}

\newcommand{\THMUBN}{\mbox{$ \theta^{^{\rm{BORN}}}_{\mu}$}}
\newcommand{\ANCM  }{\mbox{$ \bar \theta        $}}
\newcommand{\ANCE  }{\mbox{$ \bar \theta        $}}
\newenvironment{Feynman}[3]{\begin{center}
                            \setlength{\unitlength}{#3 mm}
                            \begin{picture}(#1)(#2)
                            \thicklines
                           }{\end{picture} \end{center}}
\begin{document}
%\include{m_title}
%-----------------------------------
\thispagestyle{empty}
\vspace{-1.4cm}
\begin{flushleft}
{DESY 97--230 \\}
{hep-ph/9712310\\}
December 1997
\end{flushleft}
\vspace{1.5cm}
\begin{Large}
\begin{center} 
%$\mu${\bf{e}}{\it{la}} \\
  QED Corrections for Polarized 
 Elastic 
 $\mu e$
 Scattering$^{a,b}$
\end{center}
\end{Large}
%-----------------------------------
\nn
{\large
\begin{center}
Dima Bardin$\;^{1}$ and
$\;$
Lida Kalinovskaya$\;^{1,2}$
\end{center}
}
\vspace*{1.5cm}
 
\begin{normalsize}
\begin{tabbing}
$^1$ \=
Laboratory for Nuclear Problems, JINR,
ul. Joliot-Curie 6, RU--141980 Dubna, Russia
\end{tabbing}
\begin{tabbing}
$^2$  \=
DESY-Zeuthen, Platanenallee 6, D-15735, Zeuthen, Germany
\end{tabbing}
\end{normalsize}
 
\vspace*{1.5cm}
\vfill
{\large
\centerline{\bf ABSTRACT}
}
\small
\nn
We present a new study of 
 polarized elastic muon-electron scattering.
The Born cross-section is calculated for arbitrary
polarization of muon and electron.
The complete photonic ${\cal O}(\alpha)$ radiative corrections
are determined for the case of longitudinally polarized
muons and electrons.
All calculations are done by two methods:
semianalytic, which
allows an implementation of the experimental
cuts used for the analysis of $\mu e$ scattering data from
the beam polarimeter of the SMC experiment at CERN and completely analytic,
which is used for cross checks.
 The {\tt FORTRAN} code
$\mu${\bf{e}}{\it{la}} realizes formulae of both approaches.
We prove that certain experimental cuts lead to negligible
radiative corrections in the muon beam polarization
experiment.
%------
\normalsize
\vfill
\vspace*{.5cm}
 
\bigskip
\vfill
\footnoterule
\nn
$^{a}$~Supported by the Heisenberg-Landau fund. \\
$^{b}$~Supported by the INTAS-93-744. \\
{\small
\begin{tabbing}
Emails: \= {\tt  bardin@nusun.jinr.ru}
\\      \> {\tt kalinov@nusun.jinr.ru}
\end{tabbing}
}
\newpage
%-----------------
\tableofcontents
%\include{m_introdsum}
%---------------------
\section{Introduction}
%-----------------------------------
 Polarized elastic $\mu e$ scattering is being  measured by the SMC
collaboration at CERN as a monitor of muon beam polarization~\cite{SMC}.
Since the measurement pretends to be very precise, the photonic corrections
have to be taken into account.

 The differential cross-section for this process in lowest order
may be cast into the simple form~\cite{fafei}
\bq
\frac{d\sigma^{^{\rm{BORN}}}}{dy\qquad } = \frac{2\pi\alpha^2}{m_eE_{\mu}}
     \left[\frac{(Y-y)}{y^2Y}\left(1-yP_{e}P_{\mu}\right)
     + \frac{1}{2}\left(1-P_{e}P_{\mu} \right)\right],
\label{bornol}
\eq
where the following notation is used:

$m_{\mu},\;m_e$ -- muon and electron masses,

$P_{\mu},\;P_{e}$ -- longitudinal polarizations of muon beam and electron
target,

$\ds{y=y_{\mu}=1-\frac{E^{'}_{\mu}}{E_{\mu}}}$
-- the measured energy loss of the muon,

$\ds{Y=\left(1+\frac{m_{\mu}}{2E_{\mu}}\right)^{-1}}$
-- its kinematical maximum,

$E_{\mu},\;E^{'}_{\mu},\;E^{'}_{e}$ 
-- muon (initial, final), electron final 
 energies in the laboratory frame.

  The polarization dependence of
$d\sigma$ is used to calculate
the measured electron spin-flip asymmetry $A^{\rm{exp}}_{\mu e}$
\bq
A^{\rm{exp}}_{\mu e} =\frac{\displaystyle
\frac{ d\sigma({\uparrow\downarrow})}{ dy \quad} -
\frac{ d\sigma({\uparrow  \uparrow})}{ dy \quad} }
                           {\displaystyle
\frac{ d\sigma({\uparrow\downarrow})}{ dy \quad} +
\frac{ d\sigma({\uparrow  \uparrow})}{ dy \quad} }\;.
\eq
The asymmetry is measured as a function of
the variable $y_{\mu}$.

 Previous calculations~\cite{blkwasborn}-\cite{kst},
presented results in terms of the variable
$y_e= {E^{'}_{e}}/{E_{\mu}}$,
and, only Ref.~\cite{kst} took into account the polarizations.

If elastic $\mu e$ scattering is treated in the Born approximation then
\bq
y_{\mu} = y_e = y ,
\label{borny}
\eq
and eq.~(\ref{bornol}) may be written in terms of either $y_{\mu}$
or $y_e$.

The situation changes drastically if one calculates  QED 
corrections. Due to the emission of non-observed photons
the identity~(\ref{borny}) does not hold anymore,
and one has to specify the variable to be used for the calculation of radiative
corrections. Their numerical values  may be very different in $y_{\mu}$ and $y_e$.

  Since the measurement and the analysis were performed in terms of
$y_{\mu}$~\cite{SMC},
the calculation of QED corrections must be done, of course, in
terms of the same variable. 
This is why a new calculation was neccessary.

 Our new calculation is the theoretical basis for the Fortran program
 $\mu${\bf{e}}{\it{la}},
~\cite{dblk}.
It is a complete, order ${\cal{O}}(\alpha^3)$, calculation.
It takes into account longitudinal polarizations of both $\mu$ and $e$,
finite muon mass effects (the electron mass is neglected wherever possible).
In the semianalytic approach
it is possible to apply all experimental cuts which were
used in the analysis of the experimental data:

-- a recoil electron energy cut, $E^{'}_e \geq~E^{^{RC}}$ 
($E^{^{RC}} = 35$ GeV);

-- an energy balance cut, $\left|E-E^{'}_\mu -E^{'}_e \right| \geq~E^{^{BC}}$
($E^{^{BC}} = 40$ GeV);

\newpage
-- angular cuts on both  $\mu$ and $e$,
 $\theta_\mu$ and $\theta_e$ in the laboratory system:
 $\left| \theta_e^{\rm meas}-\theta_e^{^{\rm BORN}}\right| \leq  \theta_{\min}$, \\
 $\left| \theta_\mu^{\rm meas}-\theta_\mu^{^{\rm BORN}}\right| \leq \theta_{\min}$
 ($\theta_{\min}=1$ mrad).
 In the above,
 $\theta_e^{\rm meas}$ and
 $\theta_\mu^{\rm meas}$ are the measured angles, while
 $\theta_\mu^{^{\rm BORN}}$ and $\theta_e^{^{\rm BORN}}$
 are angular values calculated using {\small{BORN}} kinematics.

\vspace*{3cm}

\hspace*{2.5cm}

\begin{figure}[h]
\hspace*{2.5cm}
\unitlength=0.80mm
\linethickness{0.6pt}
%\begin{picture}(142.00,155.00)(0,40)
\begin{picture}(142.0,145.0)(0,-30)
\put(9.00,135.00){\line(1,0){20.00}}
\bezier{16}(19.00,135.00)(17.00,135.00)(17.00,133.00)
\bezier{16}(17.00,133.00)(17.00,131.00)(19.00,131.00)
\bezier{16}(19.00,131.00)(21.00,131.00)(21.00,129.00)
\bezier{16}(21.00,129.00)(21.00,127.00)(19.00,127.00)
\bezier{16}(19.00,127.00)(17.00,127.00)(17.00,125.00)
\bezier{16}(17.00,125.00)(17.00,123.00)(19.00,123.00)
\bezier{16}(19.00,123.00)(21.00,123.00)(21.00,121.00)
\bezier{16}(21.00,121.00)(21.00,119.00)(19.00,119.00)
\bezier{16}(19.00,119.00)(17.00,119.00)(17.00,117.00)
\bezier{16}(17.00,117.00)(17.00,115.00)(19.00,115.00)
\put(9.00,115.00){\line(1,0){20.00}}
\put(45.00,135.00){\line(1,0){20.00}}
\bezier{16}(55.00,135.00)(53.00,135.00)(53.00,133.00)
\bezier{16}(53.00,133.00)(53.00,131.00)(55.00,131.00)
\bezier{16}(55.00,131.00)(57.00,131.00)(57.00,129.00)
\bezier{16}(57.00,129.00)(57.00,127.00)(55.00,127.00)
\bezier{16}(55.00,127.00)(53.00,127.00)(53.00,125.00)
\bezier{16}(53.00,125.00)(53.00,123.00)(55.00,123.00)
\bezier{16}(55.00,123.00)(57.00,123.00)(57.00,121.00)
\bezier{16}(57.00,121.00)(57.00,119.00)(55.00,119.00)
\bezier{16}(55.00,119.00)(53.00,119.00)(53.00,117.00)
\bezier{16}(53.00,117.00)(53.00,115.00)(55.00,115.00)
\put(45.00,115.00){\line(1,0){20.00}}
\bezier{16}(49.00,135.00)(49.00,137.00)(51.00,137.00)
\bezier{16}(51.00,137.00)(53.00,137.00)(53.00,139.00)
\bezier{16}(53.00,139.00)(53.00,141.00)(55.00,141.00)
\bezier{16}(55.00,141.00)(57.00,141.00)(57.00,139.00)
\bezier{16}(57.00,139.00)(57.00,137.00)(59.00,137.00)
\bezier{16}(59.00,137.00)(61.00,137.00)(61.00,135.00)
\put(85.00,135.00){\line(1,0){20.00}}
\bezier{16}(95.00,135.00)(93.00,135.00)(93.00,133.00)
\bezier{16}(93.00,133.00)(93.00,131.00)(95.00,131.00)
\bezier{16}(95.00,131.00)(97.00,131.00)(97.00,129.00)
\bezier{16}(97.00,129.00)(97.00,127.00)(95.00,127.00)
\bezier{16}(95.00,127.00)(93.00,127.00)(93.00,125.00)
\bezier{16}(93.00,125.00)(93.00,123.00)(95.00,123.00)
\bezier{16}(95.00,123.00)(97.00,123.00)(97.00,121.00)
\bezier{16}(97.00,121.00)(97.00,119.00)(95.00,119.00)
\bezier{16}(95.00,119.00)(93.00,119.00)(93.00,117.00)
\bezier{16}(93.00,117.00)(93.00,115.00)(95.00,115.00)
\put(85.00,115.00){\line(1,0){20.00}}
\put(115.00,135.00){\line(1,0){20.00}}
\bezier{16}(125.00,135.00)(123.00,135.00)(123.00,133.00)
\bezier{16}(123.00,133.00)(123.00,131.00)(125.00,131.00)
\bezier{16}(125.00,131.00)(127.00,131.00)(127.00,129.00)
\bezier{16}(127.00,129.00)(127.00,127.00)(125.00,127.00)
\bezier{16}(125.00,127.00)(123.00,127.00)(123.00,125.00)
\bezier{16}(123.00,125.00)(123.00,123.00)(125.00,123.00)
\bezier{16}(125.00,123.00)(127.00,123.00)(127.00,121.00)
\bezier{16}(127.00,121.00)(127.00,119.00)(125.00,119.00)
\bezier{16}(125.00,119.00)(123.00,119.00)(123.00,117.00)
\bezier{16}(123.00,117.00)(123.00,115.00)(125.00,115.00)
\put(115.00,115.00){\line(1,0){20.00}}
\bezier{16}(89.00,147.00)(87.00,147.00)(87.00,145.00)
\bezier{16}(87.00,145.00)(87.00,143.00)(89.00,143.00)
\bezier{16}(89.00,143.00)(91.00,143.00)(91.00,141.00)
\bezier{16}(91.00,141.00)(91.00,139.00)(89.00,139.00)
\bezier{16}(89.00,139.00)(87.00,139.00)(87.00,137.00)
\bezier{16}(87.00,137.00)(87.00,135.00)(89.00,135.00)
\bezier{16}(130.00,147.00)(128.00,147.00)(128.00,145.00)
\bezier{16}(128.00,145.00)(128.00,143.00)(130.00,143.00)
\bezier{16}(130.00,143.00)(132.00,143.00)(132.00,141.00)
\bezier{16}(132.00,141.00)(132.00,139.00)(130.00,139.00)
\bezier{16}(130.00,139.00)(128.00,139.00)(128.00,137.00)
\bezier{16}(128.00,137.00)(128.00,135.00)(130.00,135.00)
\put(80.00,110.00){\line(0,1){40.00}}
\put(140.00,150.00){\line(0,-1){40.00}}
\put(70.00,112.00){\line(0,1){31.00}}
\put(4.00,143.00){\line(0,-1){31.00}}
\bezier{16}(4.00,143.00)(4.00,145.00)(6.00,145.00)
\bezier{16}(4.00,112.00)(4.00,110.00)(6.00,110.00)
\bezier{16}(70.00,112.00)(70.00,110.00)(68.00,110.00)
\bezier{16}(70.00,143.00)(70.00,145.00)(68.00,145.00)
\put(34.00,112.00){\line(0,1){31.00}}
\bezier{16}(34.00,112.00)(34.00,110.00)(32.00,110.00)
\bezier{16}(34.00,143.00)(34.00,145.00)(32.00,145.00)
\put(40.00,143.00){\line(0,-1){31.00}}
\bezier{16}(40.00,143.00)(40.00,145.00)(42.00,145.00)
\bezier{16}(40.00,112.00)(40.00,110.00)(42.00,110.00)
\put(37.00,125.00){\circle*{2.00}}
\put(73.00,125.00){\line(1,0){4.00}}
\put(75.00,123.00){\line(0,1){4.00}}
\put(108.00,125.00){\line(1,0){4.00}}
\put(110.00,127.00){\line(0,-1){4.00}}
\put(70.00,147.00){\line(1,0){4.00}}
\put(72.00,145.00){\line(0,1){4.00}}
\put(142.00,149.00){\makebox(0,0)[cc]{$2$}}
\put(45.00,85.00){\line(1,0){20.00}}
\bezier{16}(55.00,85.00)(53.00,85.00)(53.00,83.00)
\bezier{16}(53.00,83.00)(53.00,81.00)(55.00,81.00)
\bezier{16}(55.00,81.00)(57.00,81.00)(57.00,79.00)
\bezier{16}(57.00,79.00)(57.00,77.00)(55.00,77.00)
\bezier{16}(55.00,77.00)(53.00,77.00)(53.00,75.00)
\bezier{16}(53.00,75.00)(53.00,73.00)(55.00,73.00)
\bezier{16}(55.00,73.00)(57.00,73.00)(57.00,71.00)
\bezier{16}(57.00,71.00)(57.00,69.00)(55.00,69.00)
\bezier{16}(55.00,69.00)(53.00,69.00)(53.00,67.00)
\bezier{16}(53.00,67.00)(53.00,65.00)(55.00,65.00)
\put(45.00,65.00){\line(1,0){20.00}}
\put(85.00,85.00){\line(1,0){20.00}}
\bezier{16}(95.00,85.00)(93.00,85.00)(93.00,83.00)
\bezier{16}(93.00,83.00)(93.00,81.00)(95.00,81.00)
\bezier{16}(95.00,81.00)(97.00,81.00)(97.00,79.00)
\bezier{16}(97.00,79.00)(97.00,77.00)(95.00,77.00)
\bezier{16}(95.00,77.00)(93.00,77.00)(93.00,75.00)
\bezier{16}(93.00,75.00)(93.00,73.00)(95.00,73.00)
\bezier{16}(95.00,73.00)(97.00,73.00)(97.00,71.00)
\bezier{16}(97.00,71.00)(97.00,69.00)(95.00,69.00)
\bezier{16}(95.00,69.00)(93.00,69.00)(93.00,67.00)
\bezier{16}(93.00,67.00)(93.00,65.00)(95.00,65.00)
\put(85.00,65.00){\line(1,0){20.00}}
\put(115.00,85.00){\line(1,0){20.00}}
\bezier{16}(125.00,85.00)(123.00,85.00)(123.00,83.00)
\bezier{16}(123.00,83.00)(123.00,81.00)(125.00,81.00)
\bezier{16}(125.00,81.00)(127.00,81.00)(127.00,79.00)
\bezier{16}(127.00,79.00)(127.00,77.00)(125.00,77.00)
\bezier{16}(125.00,77.00)(123.00,77.00)(123.00,75.00)
\bezier{16}(123.00,75.00)(123.00,73.00)(125.00,73.00)
\bezier{16}(125.00,73.00)(127.00,73.00)(127.00,71.00)
\bezier{16}(127.00,71.00)(127.00,69.00)(125.00,69.00)
\bezier{16}(125.00,69.00)(123.00,69.00)(123.00,67.00)
\bezier{16}(123.00,67.00)(123.00,65.00)(125.00,65.00)
\put(115.00,65.00){\line(1,0){20.00}}
\bezier{16}(89.00,65.00)(87.00,65.00)(87.00,63.00)
\bezier{16}(87.00,63.00)(87.00,61.00)(89.00,61.00)
\bezier{16}(89.00,61.00)(91.00,61.00)(91.00,59.00)
\bezier{16}(91.00,59.00)(91.00,57.00)(89.00,57.00)
\bezier{16}(89.00,57.00)(87.00,57.00)(87.00,55.00)
\bezier{16}(87.00,55.00)(87.00,53.00)(89.00,53.00)
\bezier{16}(130.00,65.00)(128.00,65.00)(128.00,63.00)
\bezier{16}(128.00,63.00)(128.00,61.00)(130.00,61.00)
\bezier{16}(130.00,61.00)(132.00,61.00)(132.00,59.00)
\bezier{16}(132.00,59.00)(132.00,57.00)(130.00,57.00)
\bezier{16}(130.00,57.00)(128.00,57.00)(128.00,55.00)
\bezier{16}(128.00,55.00)(128.00,53.00)(130.00,53.00)
\put(80.00,51.00){\line(0,1){40.00}}
\put(140.00,91.00){\line(0,-1){40.00}}
\put(70.00,57.00){\line(0,1){31.00}}
\bezier{16}(70.00,57.00)(70.00,55.00)(68.00,55.00)
\bezier{16}(70.00,88.00)(70.00,90.00)(68.00,90.00)
\put(40.00,88.00){\line(0,-1){31.00}}
\bezier{16}(40.00,88.00)(40.00,90.00)(42.00,90.00)
\bezier{16}(40.00,57.00)(40.00,55.00)(42.00,55.00)
\put(37.00,76.00){\circle*{2.00}}
\put(73.00,76.00){\line(1,0){4.00}}
\put(75.00,74.00){\line(0,1){4.00}}
\put(108.00,76.00){\line(1,0){4.00}}
\put(110.00,74.00){\line(0,1){4.00}}
\put(70.00,92.00){\line(1,0){4.00}}
\put(72.00,90.00){\line(0,1){4.00}}
\put(142.00,90.00){\makebox(0,0)[cc]{$2$}}
\bezier{16}(49.00,65.00)(49.00,63.00)(51.00,63.00)
\bezier{16}(51.00,63.00)(53.00,63.00)(53.00,61.00)
\bezier{16}(53.00,61.00)(53.00,59.00)(55.00,59.00)
\bezier{16}(55.00,59.00)(57.00,59.00)(57.00,61.00)
\bezier{16}(57.00,61.00)(57.00,63.00)(59.00,63.00)
\bezier{16}(59.00,63.00)(61.00,63.00)(61.00,65.00)
\put(9.00,85.00){\line(1,0){20.00}}
\put(9.00,65.00){\line(1,0){20.00}}
\bezier{16}(19.00,85.00)(17.00,85.00)(17.00,83.00)
\bezier{16}(17.00,83.00)(17.00,81.00)(19.00,81.00)
\bezier{16}(19.00,81.00)(21.00,81.00)(21.00,79.00)
\bezier{16}(21.00,79.00)(21.00,77.00)(19.00,77.00)
\bezier{16}(19.00,77.00)(17.00,77.00)(17.00,75.00)
\bezier{16}(17.00,75.00)(17.00,73.00)(19.00,73.00)
\bezier{16}(19.00,73.00)(21.00,73.00)(21.00,71.00)
\bezier{16}(21.00,71.00)(21.00,69.00)(19.00,69.00)
\bezier{16}(19.00,69.00)(17.00,69.00)(17.00,67.00)
\bezier{16}(17.00,67.00)(17.00,65.00)(19.00,65.00)
\put(4.00,88.00){\line(0,-1){31.00}}
\bezier{16}(4.00,88.00)(4.00,90.00)(6.00,90.00)
\bezier{16}(4.00,57.00)(4.00,55.00)(6.00,55.00)
\put(34.00,57.00){\line(0,1){31.00}}
\bezier{16}(34.00,57.00)(34.00,55.00)(32.00,55.00)
\bezier{16}(34.00,88.00)(34.00,90.00)(32.00,90.00)
\put(40.00,31.00){\line(0,-1){31.00}}
\bezier{16}(40.00,31.00)(40.00,33.00)(42.00,33.00)
\bezier{16}(40.00,0.00)(40.00,-2.00)(42.00,-2.00)
\put(37.00,15.00){\circle*{2.00}}
\put(9.00,25.00){\line(1,0){20.00}}
\bezier{16}(19.00,25.00)(17.00,25.00)(17.00,23.00)
\bezier{16}(17.00,23.00)(17.00,21.00)(19.00,21.00)
\bezier{16}(19.00,21.00)(21.00,21.00)(21.00,19.00)
\bezier{16}(21.00,19.00)(21.00,17.00)(19.00,17.00)
\bezier{16}(19.00,17.00)(17.00,17.00)(17.00,15.00)
\bezier{16}(17.00,15.00)(17.00,13.00)(19.00,13.00)
\bezier{16}(19.00,13.00)(21.00,13.00)(21.00,11.00)
\bezier{16}(21.00,11.00)(21.00,9.00)(19.00,9.00)
\bezier{16}(19.00,9.00)(17.00,9.00)(17.00,7.00)
\bezier{16}(17.00,7.00)(17.00,5.00)(19.00,5.00)
\put(9.00,5.00){\line(1,0){20.00}}
\put(4.00,31.00){\line(0,-1){31.00}}
\bezier{16}(4.00,31.00)(4.00,33.00)(6.00,33.00)
\bezier{16}(4.00,0.00)(4.00,-2.00)(6.00,-2.00)
\put(34.00,0.00){\line(0,1){31.00}}
\bezier{16}(34.00,0.00)(34.00,-2.00)(32.00,-2.00)
\bezier{16}(34.00,31.00)(34.00,33.00)(32.00,33.00)
\put(45.00,25.00){\line(1,0){30.00}}
\put(75.00,5.00){\line(-1,0){30.00}}
\bezier{16}(55.00,25.00)(53.00,25.00)(53.00,23.00)
\bezier{16}(53.00,23.00)(53.00,21.00)(55.00,21.00)
\bezier{16}(55.00,21.00)(57.00,21.00)(57.00,19.00)
\bezier{16}(57.00,19.00)(57.00,17.00)(55.00,17.00)
\bezier{16}(55.00,17.00)(53.00,17.00)(53.00,15.00)
\bezier{16}(53.00,15.00)(53.00,13.00)(55.00,13.00)
\bezier{16}(55.00,13.00)(57.00,13.00)(57.00,11.00)
\bezier{16}(57.00,11.00)(57.00,9.00)(55.00,9.00)
\bezier{16}(55.00,9.00)(53.00,9.00)(53.00,7.00)
\bezier{16}(53.00,7.00)(53.00,5.00)(55.00,5.00)
\bezier{16}(65.00,25.00)(63.00,25.00)(63.00,23.00)
\bezier{16}(63.00,23.00)(63.00,21.00)(65.00,21.00)
\bezier{16}(65.00,21.00)(67.00,21.00)(67.00,19.00)
\bezier{16}(67.00,19.00)(67.00,17.00)(65.00,17.00)
\bezier{16}(65.00,17.00)(63.00,17.00)(63.00,15.00)
\bezier{16}(63.00,15.00)(63.00,13.00)(65.00,13.00)
\bezier{16}(65.00,13.00)(67.00,13.00)(67.00,11.00)
\bezier{16}(67.00,11.00)(67.00,9.00)(65.00,9.00)
\bezier{16}(65.00,9.00)(63.00,9.00)(63.00,7.00)
\bezier{16}(63.00,7.00)(63.00,5.00)(65.00,5.00)
\bezier{16}(94.00,21.00)(94.00,19.00)(96.00,19.00)
\bezier{16}(96.00,19.00)(98.00,19.00)(98.00,17.00)
\bezier{16}(98.00,17.00)(98.00,15.00)(100.00,15.00)
\bezier{16}(100.00,15.00)(102.00,15.00)(102.00,17.00)
\bezier{16}(102.00,17.00)(102.00,19.00)(104.00,19.00)
\bezier{16}(104.00,19.00)(106.00,19.00)(106.00,21.00)
\bezier{16}(94.00,9.00)(94.00,11.00)(96.00,11.00)
\bezier{16}(96.00,11.00)(98.00,11.00)(98.00,13.00)
\bezier{16}(98.00,13.00)(98.00,15.00)(100.00,15.00)
\bezier{16}(100.00,15.00)(102.00,15.00)(102.00,13.00)
\bezier{16}(102.00,13.00)(102.00,11.00)(104.00,11.00)
\bezier{16}(104.00,11.00)(106.00,11.00)(106.00,9.00)
\bezier{16}(94.00,21.00)(94.00,23.00)(92.00,23.00)
\bezier{16}(92.00,23.00)(90.00,23.00)(90.00,25.00)
\bezier{16}(106.00,21.00)(106.00,23.00)(108.00,23.00)
\bezier{16}(108.00,23.00)(110.00,23.00)(110.00,25.00)
\bezier{16}(106.00,9.00)(106.00,7.00)(108.00,7.00)
\bezier{16}(108.00,7.00)(110.00,7.00)(110.00,5.00)
\bezier{16}(94.00,9.00)(94.00,7.00)(92.00,7.00)
\bezier{16}(92.00,7.00)(90.00,7.00)(90.00,5.00)
\put(85.00,25.00){\line(1,0){30.00}}
\put(115.00,5.00){\line(-1,0){30.00}}
\put(80.00,17.00){\line(0,-1){4.00}}
\put(78.00,15.00){\line(1,0){4.00}}
\put(120.00,0.00){\line(0,1){31.00}}
\bezier{16}(120.00,0.00)(120.00,-2.00)(118.00,-2.00)
\bezier{16}(120.00,31.00)(120.00,33.00)(118.00,33.00)
\put(120.00,35.00){\line(1,0){4.00}}
\put(122.00,33.00){\line(0,1){4.00}}
\put(80.00,-16.00){\line(1,0){20.00}}
\bezier{16}(90.00,-16.00)(88.00,-16.00)(88.00,-18.00)
\bezier{16}(88.00,-18.00)(88.00,-20.00)(90.00,-20.00)
\bezier{16}(90.00,-20.00)(92.00,-20.00)(92.00,-22.00)
\bezier{16}(92.00,-22.00)(92.00,-24.00)(90.00,-24.00)
\bezier{16}(90.00,-24.00)(88.00,-24.00)(88.00,-26.00)
\bezier{16}(88.00,-26.00)(88.00,-28.00)(90.00,-28.00)
\bezier{16}(90.00,-28.00)(92.00,-28.00)(92.00,-30.00)
\bezier{16}(92.00,-30.00)(92.00,-32.00)(90.00,-32.00)
\bezier{16}(90.00,-32.00)(88.00,-32.00)(88.00,-34.00)
\bezier{16}(88.00,-34.00)(88.00,-36.00)(90.00,-36.00)
\put(80.00,-36.00){\line(1,0){20.00}}
\put(110.00,-16.00){\line(1,0){20.00}}
\bezier{16}(120.00,-16.00)(118.00,-16.00)(118.00,-18.00)
\bezier{16}(118.00,-18.00)(118.00,-20.00)(120.00,-20.00)
\bezier{16}(120.00,-20.00)(122.00,-20.00)(122.00,-22.00)
\bezier{16}(122.00,-22.00)(122.00,-24.00)(120.00,-24.00)
\bezier{16}(120.00,-24.00)(118.00,-24.00)(118.00,-26.00)
\bezier{16}(118.00,-26.00)(118.00,-28.00)(120.00,-28.00)
\bezier{16}(120.00,-28.00)(122.00,-28.00)(122.00,-30.00)
\bezier{16}(122.00,-30.00)(122.00,-32.00)(120.00,-32.00)
\bezier{16}(120.00,-32.00)(118.00,-32.00)(118.00,-34.00)
\bezier{16}(118.00,-34.00)(118.00,-36.00)(120.00,-36.00)
\put(110.00,-36.00){\line(1,0){20.00}}
\bezier{16}(84.00,-36.00)(82.00,-36.00)(82.00,-38.00)
\bezier{16}(82.00,-38.00)(82.00,-40.00)(84.00,-40.00)
\bezier{16}(84.00,-40.00)(86.00,-40.00)(86.00,-42.00)
\bezier{16}(86.00,-42.00)(86.00,-44.00)(84.00,-44.00)
\bezier{16}(84.00,-44.00)(82.00,-44.00)(82.00,-46.00)
\bezier{16}(82.00,-46.00)(82.00,-48.00)(84.00,-48.00)
\bezier{16}(125.00,-36.00)(123.00,-36.00)(123.00,-38.00)
\bezier{16}(123.00,-38.00)(123.00,-40.00)(125.00,-40.00)
\bezier{16}(125.00,-40.00)(127.00,-40.00)(127.00,-42.00)
\bezier{16}(127.00,-42.00)(127.00,-44.00)(125.00,-44.00)
\bezier{16}(125.00,-44.00)(123.00,-44.00)(123.00,-46.00)
\bezier{16}(123.00,-46.00)(123.00,-48.00)(125.00,-48.00)
\put(75.00,-50.00){\line(0,1){40.00}}
\put(135.00,-10.00){\line(0,-1){40.00}}
\put(68.00,-29.00){\line(1,0){4.00}}
\put(70.00,-31.00){\line(0,1){4.00}}
\put(103.00,-29.00){\line(1,0){4.00}}
\put(105.00,-31.00){\line(0,1){4.00}}
\put(32.00,-29.00){\line(1,0){4.00}}
\put(34.00,-31.00){\line(0,1){4.00}}
\put(9.00,-24.00){\line(1,0){20.00}}
\bezier{16}(19.00,-24.00)(17.00,-24.00)(17.00,-26.00)
\bezier{16}(17.00,-26.00)(17.00,-28.00)(19.00,-28.00)
\bezier{16}(19.00,-28.00)(21.00,-28.00)(21.00,-30.00)
\bezier{16}(21.00,-30.00)(21.00,-32.00)(19.00,-32.00)
\bezier{16}(19.00,-32.00)(17.00,-32.00)(17.00,-34.00)
\bezier{16}(17.00,-34.00)(17.00,-36.00)(19.00,-36.00)
\bezier{16}(19.00,-36.00)(21.00,-36.00)(21.00,-38.00)
\bezier{16}(21.00,-38.00)(21.00,-40.00)(19.00,-40.00)
\bezier{16}(19.00,-40.00)(17.00,-40.00)(17.00,-42.00)
\bezier{16}(17.00,-42.00)(17.00,-44.00)(19.00,-44.00)
\put(9.00,-44.00){\line(1,0){20.00}}
\put(39.00,-24.00){\line(1,0){20.00}}
\bezier{16}(49.00,-24.00)(47.00,-24.00)(47.00,-26.00)
\bezier{16}(47.00,-26.00)(47.00,-28.00)(49.00,-28.00)
\bezier{16}(49.00,-28.00)(51.00,-28.00)(51.00,-30.00)
\bezier{16}(51.00,-30.00)(51.00,-32.00)(49.00,-32.00)
\bezier{16}(49.00,-32.00)(47.00,-32.00)(47.00,-34.00)
\bezier{16}(47.00,-34.00)(47.00,-36.00)(49.00,-36.00)
\bezier{16}(49.00,-36.00)(51.00,-36.00)(51.00,-38.00)
\bezier{16}(51.00,-38.00)(51.00,-40.00)(49.00,-40.00)
\bezier{16}(49.00,-40.00)(47.00,-40.00)(47.00,-42.00)
\bezier{16}(47.00,-42.00)(47.00,-44.00)(49.00,-44.00)
\put(39.00,-44.00){\line(1,0){20.00}}
\bezier{16}(13.00,-12.00)(11.00,-12.00)(11.00,-14.00)
\bezier{16}(11.00,-14.00)(11.00,-16.00)(13.00,-16.00)
\bezier{16}(13.00,-16.00)(15.00,-16.00)(15.00,-18.00)
\bezier{16}(15.00,-18.00)(15.00,-20.00)(13.00,-20.00)
\bezier{16}(13.00,-20.00)(11.00,-20.00)(11.00,-22.00)
\bezier{16}(11.00,-22.00)(11.00,-24.00)(13.00,-24.00)
\bezier{16}(54.00,-12.00)(52.00,-12.00)(52.00,-14.00)
\bezier{16}(52.00,-14.00)(52.00,-16.00)(54.00,-16.00)
\bezier{16}(54.00,-16.00)(56.00,-16.00)(56.00,-18.00)
\bezier{16}(56.00,-18.00)(56.00,-20.00)(54.00,-20.00)
\bezier{16}(54.00,-20.00)(52.00,-20.00)(52.00,-22.00)
\bezier{16}(52.00,-22.00)(52.00,-24.00)(54.00,-24.00)
\put(4.00,-50.00){\line(0,1){40.00}}
\put(64.00,-10.00){\line(0,-1){40.00}}
\bezier{16}(4.00,-10.00)(4.00,-8.00)(6.00,-8.00)
\bezier{16}(4.00,-50.00)(4.00,-52.00)(6.00,-52.00)
\bezier{16}(64.00,-50.00)(64.00,-52.00)(62.00,-52.00)
\bezier{16}(64.00,-10.00)(64.00,-8.00)(62.00,-8.00)
\bezier{16}(75.00,-10.00)(75.00,-8.00)(77.00,-8.00)
\bezier{16}(75.00,-50.00)(75.00,-52.00)(77.00,-52.00)
\bezier{16}(135.00,-50.00)(135.00,-52.00)(133.00,-52.00)
\bezier{16}(135.00,-10.00)(135.00,-8.00)(133.00,-8.00)
\put(137.00,-9.00){\line(1,0){4.00}}
\put(139.00,-7.00){\line(0,-1){4.00}}
\put(40.00,-69.00){\line(0,-1){31.00}}
\bezier{16}(40.00,-69.00)(40.00,-67.00)(42.00,-67.00)
\bezier{16}(40.00,-100.00)(40.00,-102.00)(42.00,-102.00)
\put(37.00,-84.00){\circle*{2.00}}
\put(9.00,-75.00){\line(1,0){20.00}}
\bezier{16}(19.00,-75.00)(17.00,-75.00)(17.00,-77.00)
\bezier{16}(17.00,-77.00)(17.00,-79.00)(19.00,-79.00)
\bezier{16}(19.00,-79.00)(21.00,-79.00)(21.00,-81.00)
\bezier{16}(21.00,-81.00)(21.00,-83.00)(19.00,-83.00)
\bezier{16}(19.00,-83.00)(17.00,-83.00)(17.00,-85.00)
\bezier{16}(17.00,-85.00)(17.00,-87.00)(19.00,-87.00)
\bezier{16}(19.00,-87.00)(21.00,-87.00)(21.00,-89.00)
\bezier{16}(21.00,-89.00)(21.00,-91.00)(19.00,-91.00)
\bezier{16}(19.00,-91.00)(17.00,-91.00)(17.00,-93.00)
\bezier{16}(17.00,-93.00)(17.00,-95.00)(19.00,-95.00)
\put(9.00,-95.00){\line(1,0){20.00}}
\put(4.00,-69.00){\line(0,-1){31.00}}
\bezier{16}(4.00,-69.00)(4.00,-67.00)(6.00,-67.00)
\bezier{16}(4.00,-100.00)(4.00,-102.00)(6.00,-102.00)
\put(34.00,-100.00){\line(0,1){31.00}}
\bezier{16}(34.00,-100.00)(34.00,-102.00)(32.00,-102.00)
\bezier{16}(34.00,-69.00)(34.00,-67.00)(32.00,-67.00)
\put(74.00,-100.00){\line(0,1){31.00}}
\bezier{16}(74.00,-100.00)(74.00,-102.00)(72.00,-102.00)
\bezier{16}(74.00,-69.00)(74.00,-67.00)(72.00,-67.00)
\put(74.00,-65.00){\line(1,0){4.00}}
\put(76.00,-67.00){\line(0,1){4.00}}
\put(4.00,-60.00){\makebox(0,0)[lc]{\mbox{Vacuum Polarization}}}
\put(4.00,40.00){\makebox(0,0)[lc]{$\mu e$ - \mbox{interference}}}
\put(4.00,97.00){\makebox(0,0)[lc]{\mbox{Electronic RC}}}
\put(4.00,155.00){\makebox(0,0)[lc]{\mbox{Muonic RC}}}
\bezier{16}(57.00,-86.00)(55.00,-86.00)(55.00,-88.00)
\bezier{16}(55.00,-88.00)(55.00,-90.00)(57.00,-90.00)
\bezier{16}(57.00,-90.00)(59.00,-90.00)(59.00,-92.00)
\bezier{16}(59.00,-92.00)(59.00,-94.00)(57.00,-94.00)
\bezier{16}(57.00,-94.00)(55.00,-94.00)(55.00,-96.00)
\bezier{16}(55.00,-96.00)(55.00,-98.00)(57.00,-98.00)
\bezier{16}(57.00,-98.00)(59.00,-98.00)(59.00,-100.00)
\put(57.00,-84.00){\circle*{5.20}}
\bezier{16}(55.00,-70.00)(55.00,-72.00)(57.00,-72.00)
\bezier{16}(57.00,-72.00)(59.00,-72.00)(59.00,-74.00)
\bezier{16}(59.00,-74.00)(59.00,-76.00)(57.00,-76.00)
\bezier{16}(57.00,-76.00)(55.00,-76.00)(55.00,-78.00)
\bezier{16}(55.00,-78.00)(55.00,-80.00)(57.00,-80.00)
\bezier{16}(57.00,-80.00)(59.00,-80.00)(59.00,-82.00)
\put(47.00,-70.00){\line(1,0){20.00}}
\put(67.00,-100.00){\line(-1,0){20.00}}
\end{picture}
\vspace{6cm}
\caption[]{Feynman diagrams for the elastic $\mu  e$ scattering in order
${\cal {O}}(\alpha^3)$.\label{feynmand}}
\vspace{-2cm}
\end{figure}

\clearpage
It  order ${\cal{O}}(\alpha^3)$, the
14 Feynman graphs, shown in Fig.~\ref{feynmand}, contribute to the 
cross-section.
The latter may be subdivided into 
{\bf 12=2$\times$6} separately gauge invariant contributions: 
\bq
\frac{d\sigma^{^{\rm{QED}}}}{dy_{\mu}\quad}=\sum_{l=1}^2\sum_{k=1}^6
\frac{d\sigma_k^l}{dy_{\mu}},
\label{cont12}
\eq 
where the indices $k$ and $l$ have the following meaning
\[
\begin{array}{rccl}
l=&1& - & {\mbox{unpolarized contribution,}}\;l={\rm{unpol}};\\
  &2& - & {\mbox{polarized contribution (the terms proportional to} }\;P_eP_{\mu}\;
          {\mbox{in~(\ref{bornol})) }},\;l={\rm{pol}}. \\
k=&1&-& {\mbox{Born cross-section,}}\;k=b;\\
  &2&-& {\mbox{Radiative corrections (RC) for the muonic current:
           vertex + bremsstrahlung,}}\;
        k={\mu\mu};\\
  &3&-& {\mbox{contribution of the anomalous magnetic moment
               of the muon,}}\;k={\rm{amm}};\\
  &4&-& {\mbox{RC for the electronic current: vertex + bremsstrahlung,}}\;
        k={ee};\\
 %  & & & ({\mbox{no amm here due to neglecting}}\;m_e);\\
  &5&-& \mu e\;{\mbox{interference: two-photon exchange +}}\\
  & & & {\mbox{muon-electron bremsstrahlung interference,}}\;
        k={\mu e};\\
  &6&-& {\mbox{Vacuum polarization correction, running}}\;\alpha,\;k={\rm{vp}}.
\end{array}
\]

\noindent
The resulting QED corrected cross-section is given by the sum

\begin{eqnarray}
\frac{d\sigma^{^{\rm{QED }}}}{dy_{\mu}\quad}=\sum_k\left(
\frac{d\sigma_k^{\rm{unpol}}}{dy_{\mu}\quad\;}+P_eP_{\mu}
\frac{d\sigma_k^{\rm{pol  }}}{dy_{\mu}\quad}\right).
\label{dsym}
\end{eqnarray}

\noindent
The cross-sections with $k={\mu\mu},{ee},{\mu e}$ 
have similar generic structure

\begin{eqnarray}
\frac{d\sigma_k}{dy_{\mu}}=
\frac{\alpha}{\pi}\delta_k^{^{\rm{VR}}}
\frac{d\sigma^{^{\rm{BORN}}}}{dy_{\mu}\quad\;}
+\frac{d\sigma_k^{^{\rm{BREM}}}}{dy_{\mu}\quad\;},
\end{eqnarray}
with a factorized part $\delta_k^{_{\rm{VR}}}$ originating from infrared 
divergent virtual (V) and real soft photon (R)
contri\-bu\-ti\-ons\footnote{
In the following we will always present the formulae in the form
~(\ref{dsym}), i.e. summed over $l$.}.

\begin{figure}[htbp]
\begin{center}
\mbox{
\epsfysize=12.cm
\epsffile[0 0 730 730]{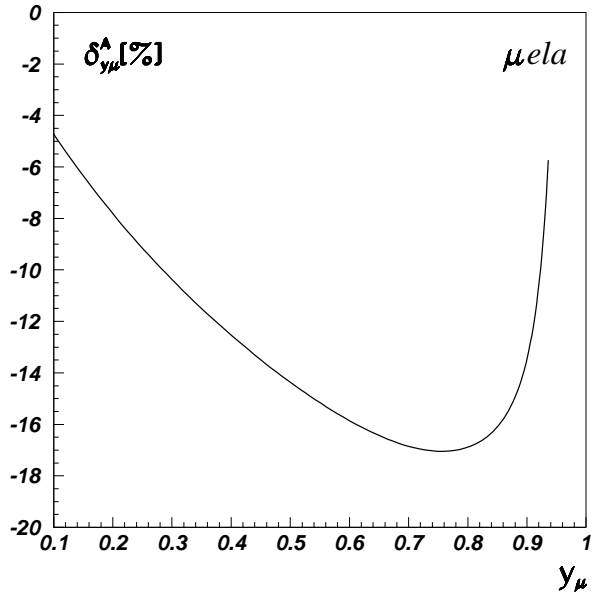}
}
\end{center}
\vspace{-3.0cm}
\caption{\it QED corrections to the polarization asymmetry without
experimental cuts.
\label{fiia1}
}
\end{figure}

\begin{figure}[htbp]
\begin{center}
\mbox{
\epsfysize=12.cm
\epsffile[0 0 730 730]{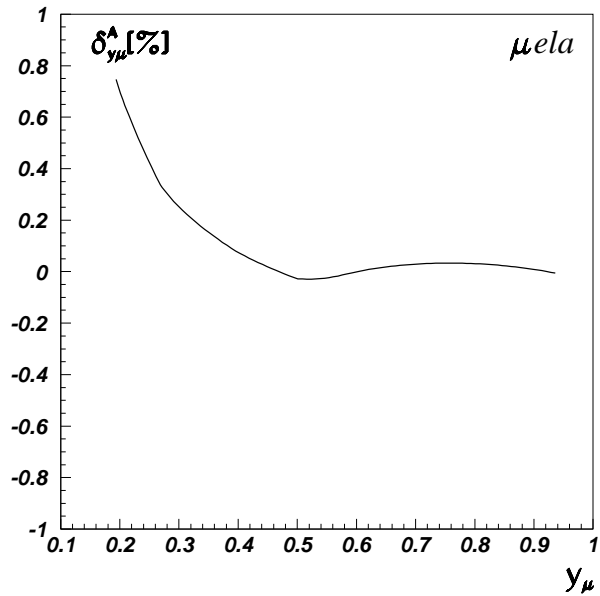}
}
\end{center}
\vspace{-3.0cm}
\caption{\it ~QED ~corrections ~to ~the ~polarization ~asymmetry ~with
~experimental ~cuts: 
$E^{^{RC}} = 35$ GeV, ~$E^{^{BC}} = 40$,
~$\theta_{e,\min}=\theta_{\mu.\min}=1$ mrad.
\label{fiia2}
}
\end{figure}

The main conclusion of this study is illustrated in 
Figs.~\ref{fiia1} and~\ref{fiia2}, which
present radiative corrections to the asymmetry as a function of the variable
$y_{\mu}$ for two cases: without any cuts and with experimental cuts
described above.

The asymmetry  $A_{\mu e}^{a}$, and the
radiative correction to it, $\delta^A_{y_{\mu}}$ are defined as follows:
\bq
A_{\mu e}^{a} =\frac{\displaystyle
\frac{ d\sigma^{a}(\uparrow\downarrow)}{ dy_{\mu}\qquad } -
\frac{ d\sigma^{a}(\uparrow  \uparrow)}{ dy_{\mu}\qquad } }
                        {\displaystyle
\frac{ d\sigma^{a}(\uparrow\downarrow)}{ dy_{\mu}\qquad} +
\frac{ d\sigma^{a}(\uparrow  \uparrow)}{ dy_{\mu}\qquad } },
\qquad 
a = {\mbox{\small{BORN,~QED}}},
\qquad 
\delta^A_{y_{\mu}} = \frac{A^{^{\rm{QED }}}_{\mu e}}
                          {A^{^{\rm{BORN}}}_{\mu e}}-1.
\eq

As is seen from the figures, the corrections without cuts 
are very large and reach up to -20\%.
When the
four above mentioned cuts are taken into account they reduce $\delta$ 
to values below 1\%.
Actually, for a wide range of 
$y_{\mu}$ they are even well below 1\%.

The main conclusion of our investigation is that one may safely neglect
radiative corrections in the
determination of the muon beam polarization
with the SMC set-up.

%-------------
%-----------

%-------------
%-----------
%\input{m_born}
%--------------
\section{The Born cross-section}
%-------------------------------
\subsection{Kinematics and phase space}
%--------------------------------------
We consider the elastic scattering process

\ba
\mu(k_1) + e(p_1) \to \mu(k_2) + e(p_2) ,
\label{b1}
\ea
in a fixed target experiment, i.e. with the initial state electron at rest,
$k_1=(\vec{0},im_e)$.

Since the typical incident 
muon energy,$\;E_{\mu}$, in present-day fixed target experiments
is ${\cal O}(10^2-10^3\;{\mbox {GeV}})$, the maximal c.m.s. energy,
\ba
s &=& -(k_1+p_1)^2=m^2_{\mu}+m^2_e+2m_eE_{\mu},
\label{s}
\ea
is very small, $\sqrt{s} \leq 1\;{\mbox {GeV}}$.
Therefore, we may completely neglect  $Z$-boson exchange.

In fact, for the energy used by the SMC collaboration, $E_{\mu}=190\;{\mbox {GeV}}$,
the invariant $s$ is only 20 times bigger than the muon mass squared. Therefore,
we {\it can  not} neglect effects of the finite muon mass. Of course,
the  electron mass may be completely neglected.

While calculating the Born cross-section,
we will perform all derivations exactly even in $m_e$, since the
resulting expressions are very compact even if $m_e$ is kept,
but at the end of calculations we will neglect the electron mass\footnote{
We refer to this approximation as to 
the ``Ultra-Relativistic Approximation (URA) in $m_e$''.}.
The Born process is characterized by one  kinematical
variable, besides $s$.
 We will use the dimensionless variable $y$:
\bq
    y = \frac{p_1(k_1-k_2)}{p_1k_1} = \frac{k_1^0-k_2^0}{k_1^0}
      \equiv 1-\frac{E^{'}_{\mu}}{E_{\mu}}.
\label{yvar}
\eq

We will introduce also the transferred momentum squared
\bq
    Q^2 = (k_1-k_2)^2 = -t.
\label{q2var}
\eq

\noindent
It is easy to derive the identity:
\bq
    Q^2 = Sy,
\label{qyid}
\eq
where
\bq
    S = s - m^2_e - m^2_{\mu}.
\label{bigs}
\eq

 For the Born cross-section, we have
\ba
d\sigma^{^{\rm{BORN}}} = \frac{1}{2\sqrt{\lambda_{_S}}} 
|M^{^{\rm{BORN}}}|^2 d\Gamma_{2},
\label{born1}
\ea
with 
\bq
\lambda_{_S} = S^2 - 4 m^2_e m^2_\mu,
\label{lambda_s}
\eq
\ba
    d\Gamma_{2} =(2\pi)^4
    \frac{d^3 \vec {k_2}}{(2\pi)^3 2k_2^0}
    \frac{d^3 \vec {p_2}}{(2\pi)^3 2p_2^0}
    \delta(k_1+p_1-k_2-p_2).
\label{ps1}
\ea
In terms of $y$ the differential phase space reads
%---
\ba
d\Gamma_{2} = \frac{1}{16\pi^2}\frac{dQ^2}{\sqrt{\lambda_{_S}}} d\varphi
=\frac{S}{16 \pi^2\sqrt{\lambda_{_S}}}dy d\varphi.
\label{ps2}
\ea
%---

\subsection{Spin degrees of freedom}
%-----------------------------------

Since we are going to deal with the scattering of polarized particles, there
will be additional essential variables, besides $s$ and $y$, which are supposed to describe the
{\it spin degrees of freedom} of the problem. Their
description uses  the language of spin density matrix (
for details we refer e.g. to Appendix C
of ~\cite{bilbook}).

For non-polarized particles, we use projection operators in trace calculations,
i.e. summing and averaging over spin indices looks as

\bq
{\overline {\sum_s}} u^s(p) {\bar u}^s(p) = \frac{1}{2}\Lambda(p),
\label{projop1}
\eq
with
\bq
\Lambda(p) = -i{\hat{p}} + m.
\label{projop2}
\eq

For polarized particles, we use the spin density matrix instead

\bq
\sum_s u^s(p) {\bar u}^s(p) = \frac{1}{2}(1+i\gamma_5 \hat{\xi}) \Lambda(p),
\label{spdema}
\eq
where $\xi$ is the {\it polarization} four-vector
\footnote{
 A naive use of
longitudinal polarizations from the early beginning of calculations, i.e.
use of the spin density matrices in the form
\bq
\sum_s u^s(p) {\bar u}^s(p) =
                 \frac{1}{2}(1+\lambda \gamma_5) \Lambda(p),
\eq
does not 
properly reproduce the finite muon mass terms in the $P_{e}P_{\mu}$ part
of the cross-section.}.

 In the particle rest frame,
$\vec{p}=0$, it is:

\bq
\xi=(P\vec{n},0),
\label{xirest}
\eq
where $\vec{n}$ is a unit vector in the direction of spin quantization,
and $P$ is the {\it polarization}, defining the degree of spin
orientation along the direction $\vec{n}$. For instance, $P=1$ means that
the probability
of a particle to have its spin projection {\it along} the direction
$\vec{n}$ is equal to 1 (right handed longitudinal polarization, if vector
$\vec{n}$ is chosen along particle momentum $\vec{p}\,$).
From (\ref{xirest}) in the particle rest frame, we have
\ba
\xi p &=& 0, \nll
\xi^2 &=& P^2.
\label{xipr}
\ea
Due to Lorenz invariance, the properties (\ref{xipr}) are fulfilled
in {\it any} Lorenz frame.

 The initial electron with the four-momentum $p_1$ is at rest in the
laboratory frame. Using then the direction of incoming muon as the 
direction of spin quantization, i.e.
$\vec{n}=\vec{k_1}/|\vec{k_1}|$, we get
the four-vector of the electron polarization from (\ref{xirest})
\bq
\xi_{e}= P_{e}\left(\frac{\vec{k_1}}{|\vec{k_1}|},0\right).
\label{xie}
\eq

 The four-vector $\xi_{\mu}$
may be obtained from the expression similar to (\ref{xie})
in the muon rest frame
\bq
\xi_{\mu}= P_{\mu}\left(\frac{\vec{k_1}}{|\vec{k_1}|},0\right)
\label{ximur}
\eq
by Lorenz boost to the electron rest frame along the beam axis:
\bq
\xi_{\mu}=\Pmu\frac{k_1^0}{m_{\mu}}
\left(\frac{\vec{k}_1}{\vkl},\frac{\vkl}{k_1^0}\right).
\label{ximu}
\eq

We will consider the Born cross-section with arbitrary orientations of
electron and muon spins. 
We choose the laboratory frame with $z$-axis oriented
along the incoming muon 3-momentum $\vec{k}_1$ and with the plane ($x$,$z$) 
coinciding with the reaction plane. Another plane is spanned by the vectors
$\vec{k_1}$ and the projection $(\vec{\xi_{e}})_{xy}$ of vector
$\vec{\xi_e}$ to a plane perpendicular to $z$-axis.
In this frame, the relevant 4-vectors are written as follows:
\ba
k_1 &=& \left( 0,~0,~|\vec{k}_1|, k^0_1 \right), \nll                      
k_2 &=& \left( |\vec{k}_2|\sin\theta_\mu,~0,~|\vec{k}_2|\cos\theta_\mu, k^0_2 \right), \nll
p_1 &=& \left( 0,~0,~0,~m_e \right), \nll
p_2 &=& \left( |\vec{p}_2|\sin\theta_e,~0,~|\vec{p}_2|\cos\theta_e, p^0_2 \right).
\label{transv4p}
\ea

For the spin vector $\xi_e$
arbitrarily oriented in 3D-space, we have in the choosen laboratory frame
instead of (\ref{xie}) the following generalization
\bq
\xie = \Pe \left(\sin\vartheta_e\cos\varphi_e,~\sin\vartheta_e\sin\varphi_e,
~\cos\vartheta_e,~0 \right).
\label{arb4srfe}
\eq

We may identify the angle $\varphi_e$ in~(\ref{transv4p}) with
$\varphi$ of the phase space parametrization in~(\ref{ps2})
and therefore $\varphi$ becomes an essential degree of freedom
in presence of a transverse polarization.

For arbitrarily oriented $\xi_{\mu}$ , we have instead of (\ref{ximur})
in the corresponding rest frame
\bq
\ximu = \Pmu\left(\sin\vartheta_{\mu}\cos\varphi_{\mu},~\sin\vartheta_{\mu}
\sin\varphi_{\mu},~\cos\vartheta_{\mu},~0 \right).
\label{arb4srf}
\eq
Now we boost $\xi_{\mu}$ from the muon rest frame to the
laboratory frame 
\ba
\ximu &=& \Pmu \left(\sin\vartheta_{\mu}\cos\varphi_{\mu},~\sin\vartheta_{\mu}
\sin\varphi_{\mu},
~\frac{k^0_1}{m_{\mu}}\cos\vartheta_{\mu},~\frac{|\vec{k_1}|}{m_{\mu}}
\cos\vartheta_{\mu} \right).
\label{arb4slfmu}
\ea 

Using the explicit representations (\ref{transv4p}), (\ref{arb4srfe})
and (\ref{arb4slfmu}), we can easily
derive all scalar products involving polarization vectors.

The doubly differential (in $y$ and $\varphi$ 
\footnote{
For $\varphi$ two choices are possible: 
1) $\varphi= \varphi_e$, then
$\varphi_\mu=\varphi_e-~\delta_{_(\varphi_e-\varphi_{\mu}) }$ or
2) $\varphi= \varphi_\mu$, then
$\varphi_e=~\varphi_\mu+~\delta_{_(\varphi_e-\varphi_{\mu})}$.})
cross-section exact in both masses reads:
\ba 
\frac{d\sigma^{^{\rm{BORN}}}}{dyd\varphi} &=& \frac{2\alpha^2S}{\lambda_{_S}}
     \Biggl\{\frac{1}{y^2}-\frac{s}{yS}+\frac{1}{2}
+P_{e}P_{\mu}
\Biggr[
\Biggl(-\frac{1}{y}+\frac{1}{Y}-\frac{1}{2}\Biggr)\cos\vartheta_e
\cos\vartheta_{\mu}
\nll
&& + \frac{m_{\mu}|\vec{p_2}|}{y\sqrt{\lambda_{_S}}}\left(1+\frac{2m^2_e}{S}
\right)
            \sin\theta_e\cos\vartheta_e\sin\vartheta_{\mu}\cos\varphi_{\mu}
\nll
&& - \frac{m_{e}|\vec{k_2}|}{y\sqrt{\lambda_{_S}}}
\left(1+\frac{2m^2_{\mu}}{S}\right)
          \sin\theta_{\mu}\cos\vartheta_{\mu}\sin\vartheta_e\cos\varphi_e
\nll
&& -2\frac{m_e m_{\mu}}{y^2S}\sin\vartheta_e \sin\vartheta_{\mu}
\Biggl( \frac{|\vec{k_2}||\vec{p_2}|}{S}\sin\theta_e\sin\theta_{\mu}
\cos\varphi_e\cos\varphi_{\mu}
\nll       
%&&\hspace{4cm} + y \cos\left( \varphi_e - \varphi_{\mu} \right)\Biggl)
&&\hspace{4cm} + y \cos\delta_{_(\varphi_e-\varphi_{\mu})} \Biggl)
\Biggr]
\Biggl\}.
\label{born_arb}
\ea

The expressions  
for $\sin\theta_e$ and $\sin\theta_{\mu}$ exact in $m_e$
are
\ba
\sin\theta_e    &=& \frac{2m_e \sqrt{S\hat y} }{\sqrt{\lambda^0_e}},
\label{sinte}
\\
\sin\theta_{\mu}&=& \frac{2m_e \sqrt{S\hat y} }{\sqrt{\lambda_l  }},
\label{sintmu}
\ea
where  
\ba
\lambda_{l} &=& S^2(1-y)^2 - 4 m^2_e m^2_\mu, \nll
\lambda^0_e &=& S^2y^2+4 m^2_e Sy,
\label{twolambdas}
\\
{\hat y} &=& y \left(1-\frac{y}{Y}\right)
\ea
and 
\ba
Y &=& \frac{\lambda_{_S}}{sS}\;\approx\;\left(1+\frac{m^2_{\mu}}{S}\right)^{-1}
\label{ymax}
\ea
is the kinematical maximum of $y$-variation.

The substitution of these variables into~(\ref{born_arb}) exhibits an 
interesting property of the general Born 
cross-section which becomes
\ba 
\frac{d\sigma^{^{\rm{BORN}}}}{dyd\varphi} &=& \frac{2\alpha^2S}{\lambda_{_S}}
     \Biggl\{\frac{1}{y^2}-\frac{s}{yS}+\frac{1}{2}
+P_{e}P_{\mu}
\Biggr[
\Biggl(-\frac{1}{y}+\frac{1}{Y}-\frac{1}{2}\Biggr)
\cos\vartheta_e\cos\vartheta_{\mu}
\nll
&& + \frac{m_{\mu}\sqrt{S\hat y}}{y\sqrt{\lambda_{_S}}}
\left(1+\frac{2m^2_e}{S}\right)
                  \cos\vartheta_e\sin\vartheta_{\mu}\cos\varphi_{\mu}
\nll
&& - \frac{m_{e}  \sqrt{S\hat y}}{y\sqrt{\lambda_{_S}}}
\left(1+\frac{2m^2_{\mu}}{S}\right)
                  \cos\vartheta_{\mu}\sin\vartheta_e\cos\varphi_e
\nll
&& -2\frac{m_e m_{\mu}}{yS}\sin\vartheta_e \sin\vartheta_{\mu}
\Biggl( \left(1-\frac{y}{Y}\right) \cos\varphi_e\cos\varphi_{\mu} 
 + \cos\delta_{_(\varphi_e-\varphi_{\mu})} \Biggl)
\Biggr]
\Biggl\}.
\label{born_arbint}
\ea
From the last presentation it is clearly seen that while terms related 
to the transverse electron polarization are small
since they are suppressed by the electron mass (third and fourth lines), 
the term induced by the transverse muon beam
polarization (second line) is {\bf not small}, 
since it appears to be proportional to the {\bf muon mass}.

By trivial algebra one may show that the expression~(\ref{born_arb})
reduces to a very short form in two particular cases.
In case of tranverse electron and longitudinal muon
polarizations in the URA in the electron mass we obtain
\ba 
\frac{d\sigma^{^{\rm{BORN}}}}{dyd\varphi} = \frac{2\alpha^2}{S}
     \left[\frac{1}{y^2}-\frac{1}{yY}+\frac{1}{2}
      +P_{e}P_{\mu}\cos\varphi\sin\theta_{\mu}\frac{1-y}{y}
\left(\frac{1}{2}-\frac{1}{Y}   
\right)\right].
\label{born_trans}
\ea
The corresponding 
expression for the case of longitudinal polarization of both particles,
but exact in both masses reads:
\ba
\frac{d\sigma^{^{\rm{BORN}}}}{dy\qquad} = \frac{4\pi\alpha^2 S}{\lambda_{_S}}
     \left[\frac{1}{y^2}-\frac{s}{yS}+\frac{1}{2}
    +P_{e}P_{\mu}\left(-\frac{1}{y} + \frac{1}{Y} - \frac{1}{2} \right)\right].
\label{born_fin}
\ea
Having in mind that $r_e=\alpha/m_e$ and $S=2m_eE_{\mu}$, we immediately
identify (\ref{born_fin}) with the corresponding expressions
from Section 2.2 of ref.~\cite{fafei} if one neglects
terms of ${\cal O}(m^2_e)$ here.
In the URA in $m_e$, equation (\ref{born_fin}) may be rewritten
in a form, which is explicitly positive definite:
\ba
\frac{d\sigma^{^{\rm{BORN}}}}{dy\qquad} = \frac{4\pi\alpha^2}{S}
     \left[\frac{(Y-y)}{y^2Y}\left(1-yP_{e}P_{\mu}\right)
     + \frac{1}{2}\left(1-P_{e}P_{\mu} \right)\right].
\label{born_fins}
\ea
%-----------------
%\input{m_brem_kin}
%-----------------
\section{Complete ${\cal {O}}(\alpha)$ Radiative Corrections}
\subsection{Kinematics of $\mu e \to \mu e \gamma$\label{skin} }
\ezero
The reaction
%-----------
\bq
\mu(k_1) + e(p_1) \rightarrow \mu(k_2) + e(p_2)
\label{born}
\eq
is accompanied by the bremsstrahlung of non-observed photon(s)
\bq
\mu(k_1) + e(p_1) \rightarrow \mu(k_2) + e(p_2) + (n)\gamma(p).
\label{brem}
\eq
First of all we will study the kinematics of one-photon bremsstrahlung.
We want $\ymu$ to be the last integration variable
out of a set of four variables (besides $S$). There is some freedom 
in doing this.

We will use the definitions:
%-------------
\ba
\Qmu = (k_1 - k_2)^2, \hspace{1.cm}
\ymu = \frac{p_1 (k_1 - k_2)}{ p_1 k_1},
\label{muonic}
\ea
%-------------
and
%-------------
\ba
Q_e^2 = (p_2 - p_1)^2, \hspace{1.cm}
y_e = \frac{p_1 (p_2 - p_1)}{ p_1 k_1}.
\label{electronic}
\ea

The other invariants are:
\ba
z_1 &=& -2pk_1,
\qquad
z_2 \;=\; -2pk_2,
\nll
V_1 &=& -2pp_1,
\qquad
V_2 \;=\; -2pp_2.
\label{zv12}
\ea
Of course, only four of the invariants are independent.

Using 4-momentum conservation, it is easy to derive the following relations among
them:
\ba
z_1 + V_1 &=& z_2 + V_2,
\nll
V_1 &=& S \ymu - \Qe,
\nll
V_2 &=& S \ymu - \Qmu,
\nll
\Qe &=& S \ye.
\label{relat}
\ea

   We will use the following set of independent variables
\bq
S,\quad \Qmu,\quad \ymu,\quad \Qe,\quad z_{2(1)}.
\label{fiveind}
\eq

The last of eqs.~(\ref{relat}) deserves a comment. 
Our choice of the 4-momentum $p_1$ in
definitions~(\ref{muonic}) and~(\ref{electronic}) introduces the asymmetry 
between $\ymu$ and $\ye$.
This is reason why $\Qmu$ and $\ymu$ may be
chosen as independent variables, while there exists a relation between
$\Qe$ and $\ye$.
 
%--------------------
For the moduli of particle momenta and the energies,
in the electron rest system $\vec{p}_1=0$, we obtain
%--------------------
\bq
\displaystyle
%==========================================================
\begin{array}{rclcrcl}
%==========================================================
\vphantom{\int\limits_t^t}
|\vec{p}|
&=& \ds {
\frac{\displaystyle\sqrt{\lambda_p} }
{\displaystyle 2m_{e}}},
&\hspace{1cm}&
p^0
&=& \ds{
\frac{\displaystyle S(\yl-\ye)}{\displaystyle 2m_{e}}},
%============
\nll  %\nll
%============
\vphantom{\int\limits_t^t}
{|\vec{k}_1|}
&=&
{\ds
\frac{\displaystyle \sqrt{\lambda_{_{S}}} }{\displaystyle 2m_{e}}},
&\hspace{1cm}&
k_1^0
&=& \ds{
\frac{\displaystyle S}{\displaystyle 2m_{e}}},
%============
\nll  %\nll
%============
\vphantom{\int\limits_t^t}
|\vec{k}_2|
&=& \ds{
\frac{\displaystyle \sqrt{\lambda_{l}} }{\displaystyle 2m_{e}}},
&\hspace{1cm}&
k_2^0
&=& \ds{
\frac{\displaystyle S_1 }{\displaystyle 2m_{e}}},
%============
\nll  %\nll
%============
\vphantom{\int\limits_t^t}
|\vec{Q}_l|
&=&  \ds{
\frac{\displaystyle \sqrt{\lambda_{\mu}} }{\displaystyle 2m_{e}}},
&\hspace{1cm}&
Q_{\mu}^0
&=&  \ds{
\frac{\displaystyle S \yl}{\displaystyle 2m_{e}}},
%============
\nll   %\nll
%============
\vphantom{\int\limits_t^t}
|\vec p_2|
&=&
|\vec Q_e|
 = \ds{
\frac{\displaystyle\sqrt{\lambda_e} }{\displaystyle 2m_{e}}},
&\hspace{1cm}&
p_2^0
&=& m_{e}+\ds {\frac{\displaystyle S y_e }{\displaystyle 2m_{e}}}.
%============
\nll   %\nll
%============
\vphantom{\int\limits_t^t}
\vec{k}_1 \cdot\vec{k}_2
      &=& \ds{\frac{\ds S S_1}{\ds 4m^2_{e}}}
         -\ds{\frac{\ds \Qmu}{\ds 2}-m^2_{\mu}},   
&\hspace{1cm}&
\vec{k}_1 \cdot\vec{p}_2
      &=& \ds{\frac{\ds S  (\ds {\Qe + 2m^2_e})}{\ds 4m^2_e}}
         -\ds{\frac{\ds {S-\Qe-z_2}}{\ds 2}}, 
\end{array}
\label{eq06}
\eq
with
\bq
 S_1=S(1-\ymu).
\eq

The corresponding equations for the Born kinematics 
($|\vec{p}|=0$) can be easily derived from (\ref{eq06}) setting
\bq
V_1=V_2=z_1=z_2=0.
\eq
In this limit
\bq
\Qe=\Qmu=Q^2 \qquad \mbox{and}\qquad \ye-\yl=y.
\eq

 Many relations and useful notations can now be taken from~\cite{mimi}.
As usual, we introduce the relevant kinematical $\lambda$-functions:
%--------------------
\bq
\begin{array}{rclcl}
\vph
\lambda_p &\equiv& \lambda[ -(p_1+p)^2, -p_1^2, -p^2 ]
        &=& S^2 (\yl-\ye)^2,
\nll
\vph
\lambda_{_{S}}
&\equiv&
\lambda[ -(p_1+k_1)^2, -p_1^2, -k_1^2 ]
        &=& S^2 - 4 \mmu \me,
\nll
\vph
\lambda_l
&\equiv& \lambda[ -(p_1+k_2)^2, -p_1^2, -k_2^2 ]
        &=& S_1^2  - 4 \mmu \me,
\nll
\vph
\lambda_{\mu}
&\equiv& \lambda[ -(p_1+Q_{\mu})^2, -p_1^2, -\Qmu ]
        &=& S^2 \yl^2 + 4 \me \Qmu,
\nll
\vph
\lambda_e &\equiv& \lambda[ -(p_1+p_2)^2, -p_1^2, -p_2^2 ]
        &=& S^2 \ye^2 + 4 \me \Qe,
\end{array}
\label{lamb1}
\eq
%----------------------------
where
%--------------------
\ba
\lambda (x,y,z)
   &=& x^2 + y^2 + z^2 - 2xy - 2xz - 2yz.
\ea

\subsection{Kinematic boundaries}

%--------------------
The boundary conditions may be taken from~\cite{mimi}.
The first one, (B.3) of~\cite{mimi}, remains unchanged
%--------------------
\bq
\me \Qms +  S^2 \yl \Ql + \mmu S^2 \yl^2 - \lambda_S \Ql = 0,
\label{eq21}
\eq
%--------------------
while the second condition, (B.4) of~\cite{mimi}, changes,
since, contrary to~\cite{mimi}, $y_e$ is not an independent variable here.
This is due to the fact that we are dealing now with elastic scattering
rather than with the deep inelastic scattering in~\cite{mimi}.
 Using the last of eqs.~(\ref{relat}),
eq.~(B.4) of~\cite{mimi} takes the form:
%--------------------
\bq
 S^2 \yl^2 \Qe + \Qes \ql
 - \me(\Ql-\Qe)^2 - S \yl \Qe (\Ql+\Qe) = 0.
\label{eq39}
\eq
 
 The physical region ${\cal E}_{\mu} = (\Ql,\yl)$
is given by two inequalities
(see~\cite{mimi}, subsection {B.2.1}),
which are derived from ~(\ref{eq21})
%---------------------------
\bq
0 \leq \Ql \leq \ds{\frac {\lambda_{_S}} {S+\mmu+\me}} \equiv
{\bar{\ql}}.
\label{eq23}
\eq
%--------------------
where
%--------------------
\bq
y_{\mu}^{\min}(\ql) = \frac{\ql}{S} \leq \yl \leq y_{\mu}^{\max}(\ql),
\label{eq24}
\eq
%--------------------
\bq
y_{\mu}^{\max}(\ql)
= \frac{1}{2\mmu} \left[ \frac{1}{S}\sqrt{\lambda_{_S}\lambda_m} -
\Ql \right],
\label{eq25}
\eq
%--------------------
and
%---------------
\ba
\lambda_m &=& \Qmu(\Qmu+4\mmu).
\label{lambda_m}
\ea
%---------------------------------------

 The solution of eq.(\ref{eq39}) is
\bq
(\Qe)^{\max,\min} =
      \frac{S\yl ( S\yl-\Ql)+2\me\Ql\pm(S\yl-\Ql)\sqrt{\lambda_{\mu}}}
                        {2(S\yl-\Ql+\me)}.
\label{Qelimits}
\eq

%---------------------------------------
\subsection{Another set of independent variables}
%---------------------------------------
 Besides of (\ref{fiveind}) we  use
\bq
S,\quad  \ymu,\quad V_2,\quad V_1,\quad z_{2(1)},
\label{fiveindpr}
\eq
 To write down the limits in these
invariants, we reorder first
the physical region ${\cal E}_{\mu} = (\Ql,\yl) \to (\yl,\Ql)$.
Trivial manipulations with (\ref{eq23})--(\ref{eq25}) lead to
\ba
0 \leq \ymu \leq y^{\max}_{\mu},
\label{ymu_limits}
\ea
\ba
(\Qmu)^{\min} \leq \Qmu \leq \min\{(\Qmu)^{\max},S\ymu\}.
\label{q2mu_limits}
\ea
 Here
\ba
(\Qmu)^{\max,\min} = \frac{\lambda_{_S}-S^2\ymu
\pm\sqrt{\lambda_{_S} \lambda_l}}{2\me}.
\label{q2mu1_limits}
\ea
 Solving the equation
\ba
(\Qmu)^{\max}=S \ymu,
\ea
we find a maximal value ${y}^{\max}_{\mu}$
\ba
y^{\max}_{\mu} \equiv \frac{\lambda_{_S}}{S(S+\mmu+\me)},
\ea
where the two upper limit branches of (\ref{q2mu_limits}) meet each other.

 From (\ref{q2mu_limits}) and the definitions of $V_i$, we easily derive the limits
of $V_2$ as function of $\ymu$
\ba
0 \leq V_2 \leq
\frac{S \ymu (S+2\me) - \lambda_{_S} + \sqrt{\lambda_{_S}\lambda_l}}{2\me}.
\label{v2_limits}
\ea
The second solution
\ba
V^{\min}_2 = \frac{S \ymu (S+2\me) - \lambda_{_S}
- \sqrt{\lambda_{_S}\lambda_l}}{2\me},
\label{v2_min}
\ea
is unphysical. It is negative in the physical region 
(\ref{ymu_limits}) of $\ymu$.
Finally, from (\ref{Qelimits}) we derive the limits of $V_1$ as functions
of $\ymu$ and $V_2$
\ba
V^{\min}_1 \leq V_1 \leq V^{\max}_1,
\label{v1_limitss}
\ea
where
\ba
V^{\max,\min}_1 = V_2\frac{S \ymu +
            2\me\pm\sqrt{\lambda_{\mu}}}{2(V_2+\me)}\;.
\label{v1_limits}
\ea

Examining (\ref{v2_limits}), one may see that the invariant $V_2$ is positive
only in the interval
\ba
0 \leq \ymu \leq y^{\max}_{\mu}.
\label{ymu_limits2}
\ea

 To complete the study of kinematics of the reaction (\ref{brem}), we have to
give  the limits of variation of the variable $z_{1(2)}$. We may take all the
relevant formulae from~\cite{mimi} and simply list them for completenes:
\bq
z_{1(2)}^{\max,\min}(\yl,\ql,\ye)=\frac{B_{1(2)} \pm \sqrt{D_z}}{A_{1(2)}}
= \frac{C_{1(2)}}{B_{1(2)}\mp \sqrt{D_z}},
\label{ax}
\eq
and the Gram  determinant
%--------------------
\ba
R_z &=& -
A_1 z_1^2 + 2 B_1 z_1 - C_1 \equiv -
A_2 z_2^2 + 2 B_2 z_2 - C_2,
\label{rz}
\\
A_2 &=&  \lambda_{\mu} \equiv A_1,
\label{eq09}
 \\
B_2 &=&\Bigl\{ 2 \me \Ql ( \Ql - \Qe )
      +   S^2 (1-\yl) (\yl \Qe - \ye \Ql)
%\nonumber \\    & &
 +S^2 (1) \Ql (\yl - \ye) \Bigr\}
\nll
&\equiv&~
-~B_1  \Bigl\{ (1) \leftrightarrow - (1-\yl) \Bigr\},
\label{eq10}
\\
C_2 &=& \Bigl\{S(1-\yl) \Qe - S \Ql \left[ (1)- \ye \right] \Bigr]^2
     + 4\mmu \Bigl[S^2 (\yl-\ye)(\yl \Qe-\ye \Ql) 
     -  \me(\Qe - \Ql )^2  \Bigr\}
\nll
   &\equiv&~
C_1 \Bigl\{  (1) \leftrightarrow - (1-\yl) \Bigr\},
\label{eq11}
\\
D_z &=& B^2_{1(2)}-A_{1(2)}C_{1(2)}.
\ea
Here we now understand that $\Qe\equiv S\ye$. This makes no simplifications,
therefore we did not substitute it in order not to destroy a nice symmetry
of these equations.

The inequalities (\ref{ymu_limits2}),~(\ref{v2_limits}), 
(\ref{v1_limits}) and~(\ref{ax})
are the kinematical limits in the sequence: $\ymu,\;V_2,\;V_1,\;z_2$.

%-----------------------
\subsection{Phase spaces}
%------------------------

 In order to have a convincing proof that all
the boundaries are correctly derived, we always write a supporting 
{\tt FORTRAN}
program to numerically check that the phase space volume in any sequence of variables
is exactly the same.
 We  calculated as many
phase space integrals analytically as is possible.

  For example, we realized the cross check of our sequence
 (\ref{fiveindpr}). Three integrations may be easily performed
 analytically, yielding:
\ba
\Gamma
&=& \frac{\pi^2S}{4\sqrt{\lambda_{_S}}}
\int_{0}^{y^{\max}_{\mu}}d\ymu
\left(V_2^{\max}-{\me}\ln\frac{V_2^{\max}+\me}{\me}\right).
\label{ps_3}
\ea
The numerical check returned for this integral exactly
the same  value  as for the basic
sequence, (C.1) of~\cite{mimi}.

 We will always integrate over $z_2$ first. Then,
only three variables $ \Vl ,\Vll, \ym $ remain.
 The physical regions
${\cal{E}}=(\ymu,V_2)$ and ${\cal{I}}=(V_2,V_1)$, 
as derived from (\ref{v2_limits}) and (\ref{v1_limits}), 
are shown in Figs. 4 and 5.
\begin{figure}[ht]
\centering
\begin{minipage}[c]{6.8cm}
\raggedleft
\epsfig{file=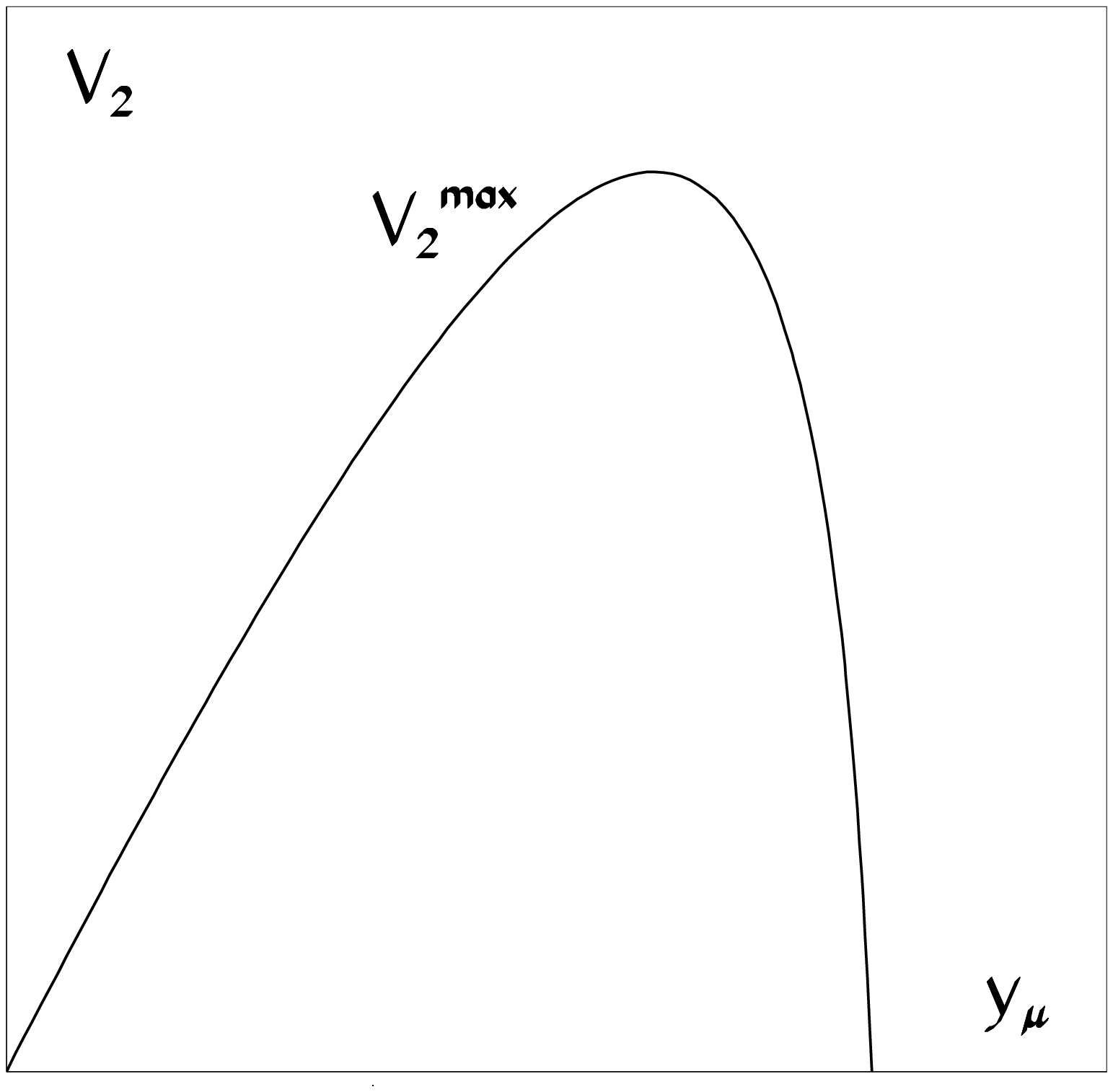, width=7  cm}
\caption{
 \it $(V_2,\ym)$-plot\label{fik1} }
\end{minipage}
\hspace*{1cm}
\begin{minipage}[c]{6.8cm}
\raggedright
\epsfig{file=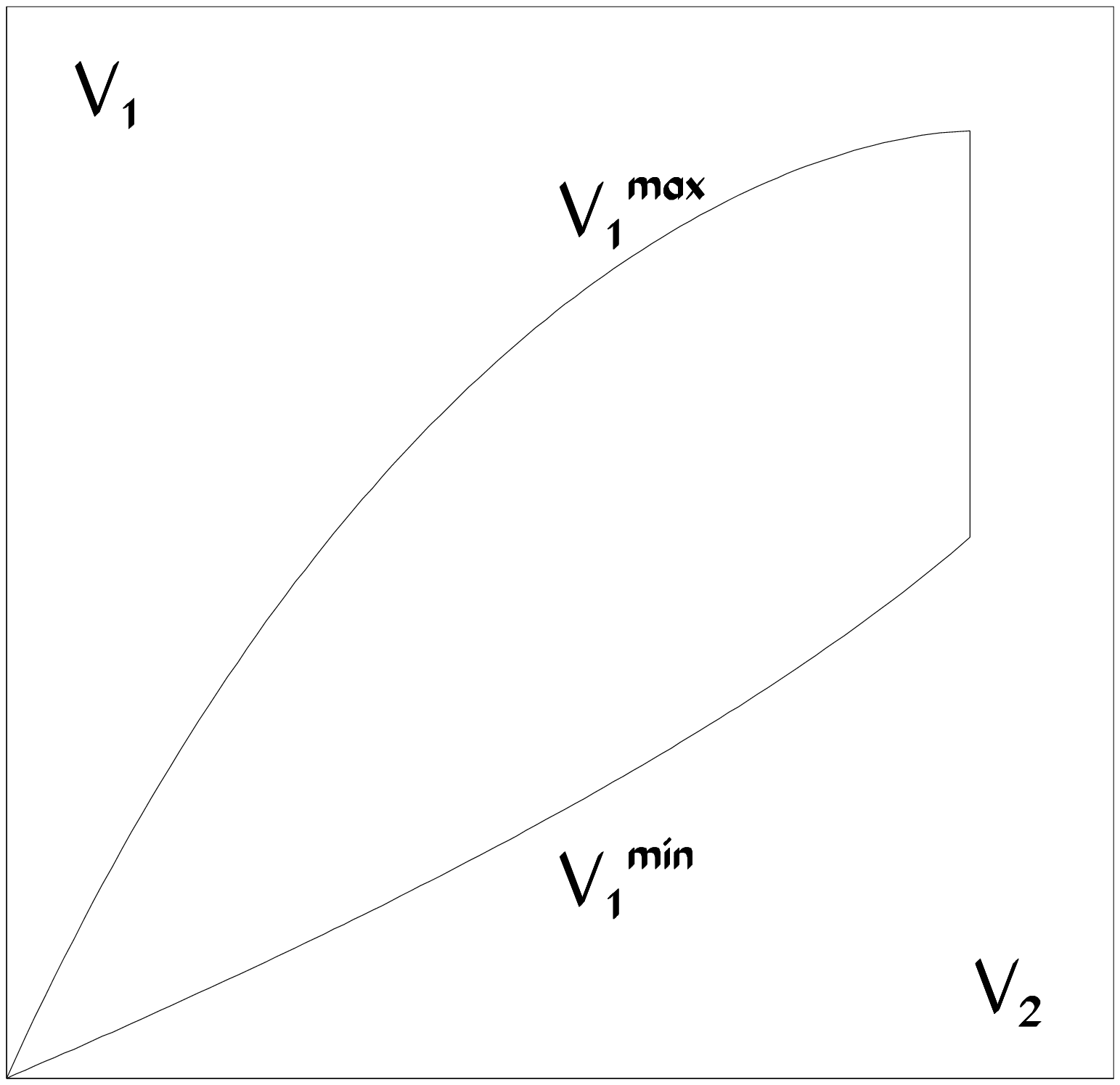,width=7  cm}
\caption{\it $(V_1,V_2)$-plot\label{fik2}
}
\end{minipage}
\end{figure}
%-----------
%\input{m_brem_cr_s}
%-------------------
%m_brem_cr_s.tex
%-------------------------------------------------------------------
\subsection{Bremsstrahlung  cross section \label{m_brem_cr_s}}
%-------------------------------------------------------------
For the normalized bremsstrahlung cross-section, we have
\ba
d\sigma^{^{\rm BREM}} = \frac{1}{2\sqrt{\lambda_{_S}}} 
\left|M^{^{\rm BREM}}\right|^2 d\Gamma_{3},
\label{brem1}
\ea
with the $2\to 3$ phase space
\ba
    d\Gamma_{3} =(2\pi)^4 \frac{d^3 \vec {k_2}}{(2\pi)^3 2k_2^0}
                          \frac{d^3 \vec {p_2}}{(2\pi)^3 2p_2^0}
                          \frac{d^3 \vec {p  }}{(2\pi)^3 2p  ^0}
                          \delta(k_1+p_1-k_2-p_2-p).
\label{phvbr}
\ea
In terms of the variables~(\ref{fiveindpr}) the differential phase space looks
as follows
\ba
  d\Gamma_{3}=\frac{S}{2^7\pi^4 \sqrt{\lambda_{_S}}} d\ym d \Vll \ d \Vl
              \frac{d\Zl}{\sqrt{R_z}}.
\label{br_ps3}
\ea

 We developped two branches to present the final result:
\begin{itemize}
\item calculation within the {\it Numerical Approach},
with the possibility to impose arbitrary experimental cuts;
\item calculation within an {\it Analytic Approach} with
limited possibility to apply cuts.
\end{itemize}

 We note that the
integrals over $z_1$ were calculated analytically with arbitrary 
limits ${\hat{z}}^{\min,\max}_{1}$
(see, Appendix~(\ref{radiators})),
while integrals over $V_{2(1)}$ were computed
purely numerically. This is the essence of our
{\it numerical approach}, the {\tt muenum} 
branch of $\mu${\bf e}{\it la}.

 We also performed the complete ${\cal{O}(\alpha)}$ RC calculations
within an {\it Analytic Approach}, the {\tt mueana}
branch of $\mu${\bf e}{\it la}.
In this description, we will present only a sketch of the calculations.

 Within the analytic approach we have integrated the
bremsstrahlung contribution
over the variables $z_1$ and $V_1$ over the full 
photonic phase space without imposing
experimental cuts.
The aim of this is twofold: first, we performed an
independent calculation of the bremsstrahlung cross section and its 
integration which was used to cross check the formulae 
of the numerical approach
and their coding in the {\tt FORTRAN} program
${\mu}{\bf{e}}{\it{la}}$. 
%Needless to say
%that a complete numerical agreement was achieved;
 Second, these analytic
formulae are rather elegant, although lengthy, in the SIngle Ultra 
Relativistic Approximation (SIURA) in the electron mass.
Due to analytic integration, they run at the computer incredibly fast, several orders of 
magnitude faster than the numerical integration within the numerical approach.
We note that this relative simplicity may be reached only if we integrate
over the full phase space of photons. From formulae in the SIURA it is easy 
to derive the formulae in the DOubly Ultra Relativistic Approximation, DOURA.
% They represent an independent value and
% of course might be used in a potential comparison with another independent
% calculation.
 It is important to stress, that in the latter case
satisfactory precision of the calculations is not ensured.
          
The derivation of analytic results is quite
straightforward. We 
consequently integrate in the sequence $dV_2dV_1dz_1$. In reality,
it is convenient to calculate and substitute the twofold integrals:
$dV_1 dz_1$. These are presented in the Table in Appendix~(\ref{f_s_a}).
After substituting the $dV_1dz_1$ integrals, 
we arrive at another Table of integrals
over $V_2$ which are listed in Appendix~(\ref{third_a}).
They are much more complicated than
the first, two fold integrals. The latter integrals were calculated 
in limits $[0, {\hat{V}}^{\max}_2]$ with an arbitrary upper limit. 
This allows us to
impose $V_2$-dependent experimental cuts
even within the analytic approach. 
An example is the muon angular cut.
                 
 A comment on DOURA is in due here. Since  here we neglect $\Mm$,
a problem might arise from  the fact that $\Qmu^{\min}$ is proportional
to the muon mass. Apparently, $\Mm$ can't be neglected in
$\Qmu^{\min}$, since it enters in the denominator of the photon propagator.
However, we  assume that a cut is imposed on $V_2$, which
prevents $\Qmu^{\min}$ from reaching its kinematical limit. In other words
we assume that
\ba
\Qmu \geq (\Qmu)^{cut} \equiv( \Qmu)^{c} \gg \Mm.
\ea

\subsubsection{Final expression for the normalized bremsstrahlung cross section}
%-------------------------------------------------------------------------------
  Defining the {\it z-integrated} contributions,
\bq
\Biggl[\left|{M^{^{\rm BREM}}} \right|^2\Biggr]_z  =
\int^{{\hat{z}}^{\max}_1}_{{\hat{z}}^{\min}_1} \frac{d\Zl}{\pi \sqrt{R_z}}
\left| {M^{^{\rm BREM}}}       \right|^2
\label{zintegration}
\eq
and neglecting $m_e$ wherever possible, we finally obtain
\ba
\frac{d\sigma^{^{\rm BREM}}}{d\ym\qquad} = \frac{\alpha^3}{S}
\int^{{\hat{V}}^{\max}_2}_{0} d\Vll
\int^{{\hat{V}}^{\max}_1}_{V^{\min}_1} d\Vl
  \left[ \left| M^{^{\rm BREM}}\right| \right]^2_z.
\label{bremm1}
\ea

 In~(\ref{bremm1}), the limits 
${\hat{V}}^{\max}_{2(1)},\;{\hat{z}}^{\min,\max}_{1}$
are functions 
of experimental cuts. In this way, the cuts are implemented
within our numerical approach.

 There are three gauge-invariant contributions to the bremsstrahlung
differential cross section:
{\it muonic}, {\it electronic} and 
$\mu e$ {\it interference},
each of them is represented as the sum of an
InfraRed divergent contribution `IR' and a finite (Regular) part `R': 
\ba
\label{brem1_frm}
\Biggl[\left|{M^{^{\rm{BREM}}}}   \right|^2_{\mu \mu}\Biggr]_z &=&
        \Qmul \left({\cal{B}}F^{\rm{IR}}_{\mu \mu}+S^{\rm{R}}_{\mu\mu}\right),
  \\
\Biggl[\left|{M^{^{\rm{BREM}}}}   \right|^2_{e e}\Biggr]_z &=&
        \Qel  \left({\cal{B}}F^{\rm{IR}}_{e e} + S^{\rm{R}}_{e e}\right),
  \\
\Biggl[\left|{M^{^{\rm{BREM}}}}   \right|^2_{\mu  e }\Biggr]_z &=&
        \Qmue \left({\cal{B}}F^{\rm{IR}}_{\mu  e } + S^{\rm{R}}_{\mu e}\right),
\label{brem3_frm}
\ea
where $\cal{B}$ is the Born factor
\bq
{\cal{B}} = 4 \frac{Y - y}{y^2 Y} \left(1-y\Pe\Pm \right)
                       +2 \left(1- \Pe\Pm \right)
\equiv\frac{1}{16\pi^2\alpha^2}\left|{M^{^{\rm BORN}}}\right|^2.
\label{bornfac}         
\eq

The $F^{^{\rm IR}}_i$ are infrared factors:
\ba
{F}_{\mu\mu}^{\rm{IR}}&=&
   \left(  S\ym +2\mmu \right)
        \Biggl[\frac{1}{\Zl \Zll} \Biggr]_z
 - \mmu \Biggl[\frac{1}{\Zls    } \Biggr]_z
 - \mmu \Biggl[\frac{1}{\Zlls   } \Biggr]_z,
\label{brem20_frm}
 \\
{F}_{ee}^{\rm{IR}}&=&
   \left(\frac{S\ym}{\Vl \Vll}-\frac{\me}{\Vls}-\frac{\me}{\Vlls}\right)
   \Biggl[1 \Biggr]_z,
\label{brem21_frm}    
 \\
{F}_{\mu e }^{\rm{IR}}&=&
 - \frac{S  }{\Vl }  \Biggl[\frac{1}{\Zl  } \Biggr]_z
 - \frac{S  }{\Vll}  \Biggl[\frac{1}{\Zll } \Biggr]_z
 + \frac{\Sl}{\Vl }  \Biggl[\frac{1}{\Zll } \Biggr]_z
 + \frac{\Sl}{\Vll}  \Biggl[\frac{1}{\Zl  } \Biggr]_z.
\label{brem22_frm}
\ea

In~(\ref{brem20_frm})-(\ref{brem22_frm}) 
and in the
cumbersome functions $S^{^{\rm R}}_{i}$ given 
in Appendix~(\ref{radiators}),    
the $z$-integration~(\ref{zintegration}) with arbitrary limits
is assumed to be done.

Tables of $z$-integrals with cuts and equations for 
limits ${\hat{V}}^{\max}_{2(1)},\;{\hat{z}}^{\min,\max}_{1}$
in terms of experimental cuts are presented below 
in Appendix B:~(\ref{V1maxcut}),~(\ref{V2maxcut}),~(\ref{zmaxcut})-(\ref{zmincut}).

\subsubsection{Treatment of the infrared divergent part}
%-------------------------------------------------------
The three terms with $F^{\mr{IR}}$ 
in~(\ref{brem1_frm})-(\ref{brem3_frm}) cannot be
simply integrated in (\ref{bremm1}) because of
the infrared divergency at $V_2=0$. 
It is treated  by  dimensional regularization.

Substituting all terms with $F^{\mr{IR}}$ into (\ref{br_ps3}),
we define the {\bf IR Part} of the bremsstrahlung cross-section.   

\ba
d\sigma^{\rm{IR}} &\equiv& \frac{2^7\pi^3\alpha^3}{\sqrt{\lambda_S}}{\cal{B}}
\left(\Qmul F^{\mr{IR}}_{\mu\mu}
+ \Qmue F^{\mr{IR}}_{\mu e}
+ \Qel F^{\mr{IR}}_{e e}
\right)
d\Gamma_3
\nll
 & = &\frac{2^7\pi^3\alpha^3}{\sqrt{\lambda_S}} {\cal{B}}
\left(\Qmul F^{\mr{IR}}_{\mu\mu}
+ \Qmue F^{\mr{IR}}_{\mu e}
+ \Qel F^{\mr{IR}}_{e e}
\right)
\left[\theta(\varepsilon-p^0)+\theta(p^0-\varepsilon)\right]
d\Gamma_3
\nll
 &\equiv &d\sigma^{\rm{IR,soft}}+d\sigma^{\rm{IR,hard}}.
\label{br_ird}
\ea

We will treat $d\sigma^{\rm{IR}}$ in the R-frame which
is defined in Appendix \ref{Rframe} by the condition

\bq
\vec{p}_2+\vec{p} = 0.
\eq
\noindent
In this frame we find
\ba
V_2=-2p.p_2=-(p_2+p)^2-\me = (p^0_2+p^0)^2-\me.
\ea
For sufficiently small $\varepsilon$ we have at the point of separation of
{\em soft} and {\it hard} photons, $p^0=\varepsilon$,
\ba
V_2=\bar{V}_2=2m_e\varepsilon,  
\label{vbar}
\ea
which can be choosen to be much smaller then any typical invariant of the 
process.

From a study of the R-system it may be understood that there is an unique
 correspondence
between angular integrations in the R-system and invariant integrations, i.e.
we may use invariant limits of $z$ in (\ref{zintegration}) and of $V_1$ 
in~(\ref{bremm1})
instead of angular limits in the R-frame in order to compute the {\em hard} part of
$d\sigma^{\rm{IR}}$
\ba
d\sigma^{\rm{IR,hard}} &=&\frac{2^7\pi^3\alpha^3}{\sqrt{\lambda_S}} {\cal{B}}
\left(\Qmul F^{\mr{IR}}_{\mu\mu}
+ \Qmue F^{\mr{IR}}_{\mu e}
+ \Qel F^{\mr{IR}}_{e e}
\right)  d\Gamma_3
\theta(p^0-\varepsilon).
\label{ir_hard1}
\ea
This is very convenient, since {\em hard} photons are in general affected by 
experimental 
cuts and the cuts are implemented in our approach by means of {\em limits} of  
a numerical integration over $z_1$ and $V_1$. 
Finally,  we get from ~(\ref{br_ird})
\ba
\frac{d\sigma^{\rm{IR,hard}}}{d\ym\qquad} &=&
\frac{\alpha^3}{S}{\cal{B}}
\int^{{\hat{V}}^{\max}_2}_{\bar{V}_2} d\Vll
\int^{{\hat{V}}^{\max}_1}_{V^{\min}_1} d\Vl
\left(\Qmul \Biggl[ F^{\mr{IR}}_{\mu\mu} \Biggr]_z
+ \Qmue     \Biggl[ F^{\mr{IR}}_{\mu  e} \Biggr]_z
+ \Qel      \Biggl[ F^{\mr{IR}}_{e    e} \Biggr]_z
\right)            \nll
&=& \frac{d\sigma^{^{\rm{BORN}}}}{d\ym\qquad}\frac{\alpha}{\pi}\delta^{\rm{IR,hard}}
\label{ir_hard2}
\ea
with
\ba
\delta^{\rm{IR,hard}}=
\int^{{\hat{V}}^{\max}_2}_{\bar{V}_2} d\Vll
\int^{{\hat{V}}^{\max}_1}_{V^{\min}_1} d\Vl
\left(\Qmul \Biggl[ F^{\mr{IR}}_{\mu\mu}\Biggr]_z
+     \Qmue \Biggl[ F^{\mr{IR}}_{\mu e} \Biggr]_z
+     \Qel  \Biggl[ F^{\mr{IR}}_{e e}   \Biggr]_z
\right).
\label{del_hard}
\ea
Again, in (\ref{ir_hard2})-(\ref{del_hard}) the $z$-integration 
(\ref{zintegration}) with experimental cuts is assumed to have been done.

%=============================================================================
\subsubsection{A short form of the completely differential 
bremsstrahlung cross-section}
The IR finite $z$-integrated contributions~(\ref{srmm}-\ref{srem})
are rather cumbersome indeed. For the unpolarized squared matrix elements
of~(\ref{brem1_frm}-\ref{brem3_frm}) with not yet separated infrared 
parts~(\ref{brem20_frm}-\ref{brem22_frm}) a compact representation
is known in the literature~\cite{Kuraevization}.
 We present it here, in notations
of our paper, even in a more compact form:
% Then, we discuss, why it can't be used for our purposes.
% So, analogs of~(\ref{brem1_frm}-\ref{brem3_frm}) are:
%---------------
\ba
\Biggl[\left|{M_{\rm{form}}^{^{\rm{BREM}}}}\right|^2_{\mu\mu}\Biggr]^{\rm{unpol}} &=&
\Qmul \Biggl[
\frac{{\cal{F}}^{^R}}{\Qe z_1 z_2}
      +\frac{2}{\Qes} \left(\Mm \Delta^{\mu}_{\mu\mu}+\Me \Delta^{e}_{\mu\mu}\right)
      \Biggr],
\nll
\Biggl[\left|{M_{\rm{form}}^{^{\rm{BREM}}}}\right|^2_{e e}\Biggr]^{\rm{unpol}} &=&
\Qel  \Biggl[
\frac{{\cal{F}}^{^R}}{\Qm V_1 V_2}
      +\frac{2}{\Qms} \left(\Me \Delta^{e}_{ee}+\Mm \Delta^{\mu}_{ee}\right)
      \Biggr],
\nll
\Biggl[\left|{M_{\rm{form}}^{^{\rm{BREM}}}}\right|^2_{\mu  e}\Biggr]^{\rm{unpol}} &=&
\Qmue \Biggl[
-\frac{{\cal{F}}^{^R}}{\Qe \Qm}
          \left(\frac{S}{V_1 z_1}+\frac{S'}{V_2 z_2}
          +\frac{U}{V_1 z_2}+\frac{U'}{z_1 V_2}\right)
         -\frac{2 \Mm}{\Qe\Qm} \Delta_{\mu e}
      \Biggr],
\ea
%--
where
\ba
{\cal{F}}^{^R}&=&S^2+S'^2+U^2+U'^2
\nll 
\Delta^{\mu}_{\mu\mu}&=&-\frac{S'^2+U^2}{z_1^2}-\frac{S^2+U'^2}{z_2^2}
        -2 \frac{S U+S' U'}{z_1z_2}
\nll
&&
        -2 \Qe \left(\frac{1}{z_1}-\frac{1}{z_2}\right)
        +2 \Mm \Qe \left(\frac{1}{z_1}-\frac{1}{z_2}\right)^2,
\nll
\Delta^{e}_{\mu\mu}&=&-\frac{S+U'}{z_1}+\frac{S'+U}{z_2}
                     +4\frac{\Mm\Qe}{z_1z_2},
\nll
\Delta^{e}_{ee}&=&-\frac{S'^2+U'^2}{V_1^2}-\frac{S^2+U^2}{V_2^2}
           +2\Mm\Qm\left(\frac{1}{V_1^2}+\frac{1}{V_2^2}\right),
\nll
\Delta^{\mu}_{ee}&=&-\frac{(S+U)^2+(S'+U')^2}{V_1 V_2},
\nll 
\Delta_{\mu e}&=&-\left(\Qe+\Qm\right) 
 \left(\frac{S}{V_1z_1}+\frac{S'}{V_2z_2}
      +\frac{U}{V_1z_2}+\frac{U'}{z_1V_2}\right)
\nll
&&
           +2\frac{S'-U}{z_1}+2\frac{U'-S}{z_2},
\label{arbkur1}
\ea            
and
\ba
S'&=&  S-z_1-V_1, 
\nll
U &=&-S+\Qe+V_1,
\nll
U'&=&-S+\Qe+z_2. 
\label{arbkur2}
\ea

The representation~(\ref{arbkur1})-(\ref{arbkur2}) is really very
elegant, and it may be used in a Monte Carlo code.
It can't be used, however, within the numerical approach of this paper 
for the following reasons:
\begin{enumerate}
\item It does not take into account polarizations (main reason).
\item It includes the IRD part which
      has to be separated out.
\item Our bremsstrahlung cross-section, as we emphazed already, is integrated
once over the invariant $z_{1,2}$ within arbitrary limits 
(by making use of a table of indefinite integrals). To do so, we need 
the canonical representation
\bq
S^{^R}_{i}=\sum_{l,j} f_l(z_j)K_{il}(S,\ym,V_2,V_1).
\quad i=ee,e\mu,\mu\mu, \quad j=1,2,\quad l=1-24,
\label{canon}
\eq
see Appendix~(\ref{ddd}).
 We gain at least one order 
of magnitude in CPU-time, since instead of a three-fold numerical integration
over $V_2,V_1$ and $z_{1,2}$, we need only two-fold: $(V_2,V_1)$.
This gain of CPU-time makes our code very fast and user friendly.
In order to arrive at this level, however, one has to substitute $S',\;U,\;U'$,
and the compact expressions, actually, become much more cumbersome
(see Appendix A).
\item Finally, the canonical representation~(\ref{canon}) seems
 to be the only road
towards an analytically integrated differential cross-section,
see the discussion at the beginning of this section.
\end{enumerate}
%--------------
%\input{m_ammvp_net}
\subsection{The muon anomalous magnetic moment contribution}
%-------------------------------------
\newcommand{\BETA}{\mbox{$\beta                     $}}
\newcommand{\SAMM}{\mbox{$\ds{\frac{d\sigma^{\rm{amm}\quad}}{dy} }$}}

 The contribution of the anomalous magnetic moment of the muon to the cross-section
can be expressed as (here we follow decomposition of ref.~\cite{ABL}):
\bq
%----
\ds{\frac{d\sigma^{\rm{amm}}}{dy\qquad} }
 = \frac{\alpha^3}{S}
        \;   F^{\rm amm}_\mu   \;
 \Biggl( -\frac{4 \AMU}{ S y^2} \Biggr)
            \left( \frac{1}{y} -\frac{1}{Y}  \right)
          \left(  2 - y \Pe \Pm  \right),
\eq
where
%--------------
\bq
 F^{\rm amm}_\mu = - \frac{1}{\BETA} \log\frac{\BETA+1}{\BETA-1},
\eq
%----
and
\bq
      \BETA   = \sqrt{1+\frac{4\AMU}{Q^2}}.
\eq
%--
\subsection{The running electromagnetic coupling}
%------------------------------------------------
The correction due to the running QED coupling (vacuum polarization)
can be implemented as
\bq
\frac{d\sigma_{\rm{vp}}}{dy\;\;}=
\left[\left(\frac{\alpha(Q^2)}{\alpha}\right)^2-1
\right]
\frac{d\sigma^{^{\rm{BORN}}}}{dy\qquad}=
\delta_{\rm vp}
\frac{d\sigma^{^{\rm{BORN}}}}{dy\qquad},
\eq
%--
where
\ba
\alpha(Q^2) = \frac{\alpha}{1-\Delta\alpha(Q^2)}.
\label{al_run}
\ea
The correction $\Delta\alpha$ contains two parts:
\ba
\Delta \alpha(Q^2) = \Delta \alpha_{l} + \Delta \alpha_{udcsb},
\label{sum_d}
\ea
the first contribution is due to the charged leptons,
the second one is due to the light quarks ($u,d,c,s,b$).
All details about the calculation of $\delta \alpha_{l}$
are described in~\cite{HECTOR}.
%The only thing left to discuss is what we should take 
For the calculation of the 
$ \Delta \alpha_{udcsb}$
we use the parametrization of ref.~\cite{FrEg}.
\subsection{The net ${\cal{O}}(\alpha^3)$ cross-section} 
%-------------------------------------------------------
Now we have collected all the ingredients to construct the net QED 
cross-section
up to order ${\cal{O}}(\alpha^3)$. It has the form
\ba
\frac{d\sigma^{^{\rm{QED }}}}{dy_{\mu}\quad}&=&
 \frac{d\sigma^{^{\rm{BORN}}}}{dy_{\mu}\quad \;\;} \left( 1+\delta_{\rm vp} \right)
+\frac{d\sigma^{\rm{amm}}}{dy_{\mu}\quad \; }
\nll
&&+\sum_{k=\mu\mu,ee,\mu e}\left(
\frac{\alpha}{\pi}\delta_k^{_{\rm{VR}}}
\frac{d\sigma^{^{\rm{BORN}}}}{dy_{\mu}\quad \;\; }
+\frac{d\sigma_k^{^{\rm{BREM}}}}{dy_{\mu}\;\; \quad}
\right),
\label{net3}
\ea
where the corrections $\delta_k^{_{\rm{VR}}}$ 
are given by~(\ref{dvrmm}),~(\ref{dvree}) 
and(\ref{dvrme}) and the last sum in~(\ref{net3}) reads
\bq
\sum_k\frac{d\sigma^{^{\rm{BREM}}}_k}{dy_{\mu}\; \;\quad}=
\frac{\alpha^3}{S}
\int^{{\hat{V}}^{max}_2}_{\bar{V}_2} d\Vll 
\int^{{\hat{V}}^{max}_1}_{V^{min}_1} d\Vl
\left(\Qmul S^{\rm{R}}_{\mu\mu}
    + \Qel  S^{\rm{R}}_{ee}  
    + \Qmue S^{\rm{R}}_{\mu e}
\right),
\label{netb}
\eq
with $S^{\rm{R}}_{k}$ given by~(\ref{srmm}),~(\ref{sree}) and~(\ref{srem}).
The limits of integration in~(\ref{netb}) are discussed in the Appendix B.
%--------------------------
%\input{m_acknowledgements}
\section{Acknowledgments}
%------------------------
 We thank our colleagues from the SMC collaboration, especially B.~Badelek for
 triggering this study, E.~Burtin, F.~Feinstein, V.~Hughes
 and A.~Magnon for useful discussions and for inviting us to the SMC 
 collaboration meetings at CERN, and F.~Simeoni and J.~Cranshaw
 for their interest in this study.

 We are very grateful to F.~Jegerlehner and T.~Riemann for reading 
the manuscript and helpful discussions.
 Our gratitude goes to T.~Riemann and the DESY Directorate, especially 
to P.~S\"oding,  
 for inviting us to DESY-Zeuthen and for the kind hospitally during our stays.
 DB is indebted to A.~Arbuzov and E.~Kuraev for a discussion
 of a short form of the unpolarized bremsstahlung cross-section.

 This work was supported in part by DESY-Zeuthen, by the
 Heisenberg-Landau Program and by the grant INTAS-93-744.

% APPENDIX
%---------
%\input{m_brem}
%--------------

\appendix

\def\theequation{\Alph{section}.\arabic{equation}}

\section{Finite contributions to the bremsstrahlung cross-\-sec\-ti\-on
\label{radiators}}
%----------------------------------------------------------------
\vspace*{-5mm}
Here we list three finite contributions,
$S^{^{\rm{R}}}_{i}$,
to the hard part of $d\sigma$
which enter the 
formulae~(\ref{brem1_frm}-\ref{brem3_frm}):
\ba
\label{srmm}
  S^{\rm{R}}_{\mu \mu}&=&
      \Pe \Pm \Biggl\{
%-------------------------------------
      \Biggl[\frac{1}{\Zls} \Biggr]_z
%---------
 2\Rm
\Biggl[ \frac{-2\Me S}{\Qes}
    \left(\Vl \ym-2 \Vll\ym+\frac{\Vlls}{S}\right)
  \nll  &&\hspace{2cm}
  + \frac{S}{\Qe}
 \left(\Vl \left(2\ym-3+\frac{2}{\ym}\right)+\Vll\right)-\Vl\Biggr]
  \nll  &&
%--------------------------------------
      +\Biggl[\frac{1}{\Zlls} \Biggr]_z
%---------
 \frac{4\Rm S \Vl}{\Qe\ym}
  \nll  &&
%--------------------------------------
      +\Biggl[\frac{ 1}{\Zl \Zll }  \Biggr]_z
%---------
 \Biggl[ \frac{2 \Me}{\Qes}
 \Biggl( - S \Vl  \yms (1+2\Rm)^2
 + S \Vll \yms (3+8\Rm+4\Rms)
  \nll  &&\hspace{2cm}
-\Vlls\ym (3+4\Rm)
+\frac{V^3_2}{S}
\Biggr)
    \nll  &&
 + \frac{S^2 \yml}{\Qes}\left[    \Vl \left( \ym + 2(2+\ym)\Rm \right)
                           - 2\Vll\left(\ym+(2+\ym)\Rm\right)
                                 +\frac{\Vlls}{S} \right]
    \nll  &&
 -\frac{S}{\Qe} \Biggl[ \Vl \left(1-2\ym-2
 \left(2\ym+3-\frac{4}{\ym}\right)\Rm\right)
 -\Vll (-3\ym+2-2\Rm)\Biggr]
    \nll  &&
                -\Vl (1+2 \Rm)+\Vll (1-4\Rm)
 \Biggr]
  \nll  &&
%-------------------------------------
      +\Biggl[\frac{1}{\Zl} \Biggr]_z
%---------
\Biggl[
    \frac{4 \Me}{\Qes}
\left(\Vl \ym \left(1+\Rm+2\Rms\right)
       -\Vll \ym (2+\Rm)
+\frac{\Vlls}{S}
\right)
 \nll &&
 -\frac{S \yml}{\Qes}
\left[ \Vl \left( 1+\frac{2(2+\ym)}{\ym}\Rm \right)-\Vll\right]
 \nll &&
 -\frac{1}{\Qe}
\left[\frac{2\Vl}{\ym}
\left(1-\left(1- \frac{2}{\ym}\right)\Rm \right)+\Vll\right]
 \nll &&
 +2 \left(1-\frac{1}{\ym}
+\frac{1}{\ym}\left(1-\frac{2}{\ym}\right) \Rm  \right)
 \Biggr]
 \nll  &&
%-------------------------------------
      +\Biggl[\frac{1}{\Zll}  \Biggr]_z
%---------
 \Biggl[
 \frac{2 \Me}{\Qes}
\left(-\Vl \ym (1+2\Rm)^2
       +2\Vll \ym (1+2\Rm)
-\frac{\Vlls}{S}
\right)
 \nll &&
                +\frac{S \yml}{\Qes}
 \left(\Vl
 \left(1+\frac{2 (2+\ym)}{\ym}\Rm \right)-\Vll\right)
                +\frac{1}{\Qe}
 \left(2 S \left(1-\left(1-\frac{2}{\ym}\right)\Rm\right)-\Vll\right)\Biggr]
 \nll  &&
%--------------------------------------
      +\Biggl[ 1  \Biggr]_z
%---------
      \frac{4 \Me}{S\Qes}(\Vl-\Vll)
% \nll  &&
             \Biggr\}
 \nll  &&
%--------------------------------------
      -\Biggl[\frac{1}{\Zls} \Biggr]_z
%---------
 \frac{4 \Rm S \Vl}{\Qe\ym}
 \left(\frac{S \ymls}{\Qe}+\frac{\yml}{\ym}-\Rm\right)
 \nll  &&
%--------------------------------------
      -\Biggl[ \frac{1}{\Zlls} \Biggr]_z
%---------
 \frac{4\Rm S \Vl}{\Qe\ym}
 \left(\frac{S}{\Qe}+\frac{\yml}{\ym}-\Rm\right)
 \nll  &&
%--------------------------------------
      +\Biggl[\frac{1}{\Zl \Zll}    \Biggr]_z
%---------
\Biggl[
       -\frac{4 \Me}{\Qes}
\left( S \Vl (\ym+2\Rm)-2 S\Vll (\ym+\Rm)+\Vlls \right)
  \nll  &&
   + \frac{S^2}{\Qes}
\Biggl[
\Vl \left(4\yml+\yms+2\left(\ym+\frac{4\yml}{\ym} \right) \Rm\right)
-\Vll \left(4\yml+2\ym\Rm\right)-\frac{\Vlls \ym}{S}\Biggr]
  \nll &&
 +\frac{S}{\Qe}
\left[ 4 \Vl \left( \frac{\yml}{\ym}-\Rm\right)
             \left( 1+\frac{2\Rm}{\ym}  \right)
              +\Vll (\ym+ 8\Rm)         \right] -2\Vll
 \Biggr]
  \nll  &&
%--------------------------------------
      +\Biggl[\frac{1}{\Zl} \Biggr]_z
%---------
 \Biggl[ \frac{2\Me}{\Qes}\left(\Vl \left(1+\frac{4}{\ym}\right)-\Vll\right)
  -\frac{S}{\Qes}
        \left(\Vl \left(\frac{4 \yml}{\ym}+\ym+2\Rm\right)+\Vll\ym\right)
      \nll &&
  -\frac{1}{\Qe }
        \left(\frac{2\Vl}{\ym}
        \left( \frac{\yml(2+\ym)}{\ym}-\Rm\right)+\Vll\right)
 % \nll  &&
   +1+\frac{2}{\ym}-\frac{4}{\yms}+\frac{2\Rm}{\ym} \Biggr]
      \nll  &&
%--------------------------------------
      +\Biggl[\frac{1}{\Zll} \Biggr]_z
%---------
 \Biggl[-\frac{2 \Me}{\Qes}
 \left(\Vl \left(1+\frac{4 \Rm}{\ym} \right)-\Vll   \right)
  +\frac{S    }{\Qes} \left(\Vl
 \left(\frac{4\yml}{\ym}+\ym+2\Rm\right)+\Vll\ym\right)
      \nll &&
  -\frac{1}{\Qe }\left(\frac{2 \Vl}{\ym}
\left(3-\frac{2}{\ym}+\Rm\right)-\Vll\right)
  +1-\frac{6}{\ym}+\frac{4}{\yms}-\frac{2 \Rm}{\ym} \Biggr]
      \nll  &&
%--------------------------------------
      +\Biggl[ 1  \Biggr]_z
%---------
 2\left(-\frac{2 \Me \Vl}{\Qes S \ym}+\frac{1}{\Qe} \right),
\ea
%=============================================================================
\ba
\label{sree}
S^{\rm{R}}_{ee}&=&
  \Pe \Pm \Biggl\{
        \Biggl[ z^2_1 \Biggr]_z
 \frac{4 \Me}{S \Vls \Qm}
 \nll &&
      +         \Biggl[ \Zl \Biggr]_z
           \left[ \frac{ 2 \Me}{S \Vls }
                  \left( 1-\frac{2 S}{\ds \Qm}(1-\ym \Rm) \right)
                      - \frac{2}{\Vll\Qm }
                      + \frac{2}{\Vl \Vll} \right]
\nll &&
      +         \Biggl[  1  \Biggr]_z
   \Biggl[-\frac{2 \Me}{\Vls \Vll \Qm}
               \left[
                1-\frac{2}{\ym}+2\left(\frac{S \ym}{\Qm} +1 \right)
                    \Rm \right]
               -\frac{\Vl}{\Qm \Vll}(1-2\Rm)
               +\frac{2}{\Qm\ym}   
\nll &&
               +2\left( \frac{ S \ym \Vll }{ \Vl} - \Vl \right)
                \frac{ \Rm}{\Qms   }
               +\frac{\Vll}{\Vl\Qm }
              -2\left(
                \frac{S\ym}{\Vl \Qm} -\frac{1}{\Vll}
                \right)
                \left( \frac{1}{\ym} - 2 \Rm \right)
                \Biggr]
     \Biggr\}
   \nll &&
      +  \Biggl[ \Zls  \Biggr]_z
       2\left( -\frac{2 \Me}{\Vls\Qm}
               + \frac{1}{\Vl\Vll} \right) \frac{1}{\Qm}
   \\    &&
      +   \Biggl[  \Zl  \Biggr]_z
   2\left[
         \left( \frac{2\Me}{\Vls \Qm}- \frac{1}{\Vl\Vll}\right)
                \left( \frac{2 S}{\Qm}-1 \right)
               +\frac{1}{\Vll \Qm}
   \right]
\nll  &&
      +  \Biggl[   1  \Biggr]_z
  \Biggl[ \frac{4 \Me \Vll}{\Vls\Qm\ym}
               \left( 1-\frac{1}{\ym} -\frac{S}{\Qm}+\Rm \right)
\nll  &&
          +\frac{\Vl}{\Vll \Qm}\left( 1- \frac{2\Rm}{\ym} \right)
          -\frac{2}{\ym\Qm} (1-2\Rm)
          -2\left(
           \frac{ S \Vll}{\Vl}
          +\frac{\Vl    }{\ym}\right) \frac{\Rm}{\Qms}
\nll  &&
              +\frac{\Vll}{\Vl\Qm}
                \left(1-\frac{2}{\ym}
                   +\frac{4}{\yms}-\frac{4\Rm}{\ym} \right)
               -\frac{2}{\Vl\ym} \left(1 -\frac{2}{\ym}+2\Rm \right)
%\nll  &&
               -\frac{2}{\Vll\ym}(1-2\Rm)\Biggr] \nonumber,
\ea
%=============================================================================
\ba
\label{srem}
S^{\rm{R}}_{\mu e} &=&
  \Pe \Pm \Biggl\{
            \Biggl[ 1 \Biggr]_z
              2 \left[
                   \frac{1     }{S \ym} \left( 2
                  +\frac{  \Vl }{  \Qe}
                  +\frac{  \Vll}{  \Qm} \right)
                  -\frac{1}{\Vl}-\frac{1}{\Vll} \right]
\nll  &&
      +\Biggl[ \frac{1}{\Zl} \Biggr]_z
   \Biggl[S
       \left(  \frac{  \Vl \yml }{\Qe\Vll  }
              -\frac{  \Vll     }{\Vl\Qm   } \right)
       \left( 1-\frac{2}{\ym}+2\Rm           \right)
\nll  && 
   +\frac{1}{\Qe}
                       \frac{\yml}{\ym}
    \left[2\Vl\left(\frac{\yml}{\ym} +\Rm\right) +\Vll \right]
\nll  &&
      +\frac{\Vl }{\Qm}
       \left[\frac{1}{\ym}-2
             \left(2+\frac{1}{\ym}\right)\Rm\right]
      -\frac{2 \Vll}{\Qm \ym}
             \left(\frac{1}{\ym}-2\yml\Rm   \right)
\nll  &&
    + \frac{\Vll}{\Vl \ym}
               +\frac{\Vl\yml}{\Vll\ym}(1-2\Rm)
    +2\left(1-\frac{2}{\ym}-\left(1-\frac{3}{\ym}\right) \Rm\right)
          \Biggr]     
\nll   &&
         +\Biggl[ \frac{1}{\Zll}  \Biggr]_z
          \Biggl[
          -\frac{\Vl}{\Qe}
          \left[
          \frac{S}{\Vll} \left(1-\frac{2}{\ym}+2\Rm\right)
            +\frac{2}{\ym}\left(\frac{1}{\ym}+\Rm\right)\right]
\nll  &&
         +\frac{\Vll }{\Qe \ym}
         +\frac{\Vl  }{\ym    }
               \left(\frac{\yml}{\Qm}+\frac{1}{\Vll}\right)
         (1+2\Rm)
%\nll   &&
         +\frac{\Vll\yml}{\Vl}
          \left[\frac{S}{\Qm}\left(1-\frac{2}{\ym}+2\Rm \right)
         +\frac{1}{\ym} \right]
\nll   &&
         +\frac{2 \Vll}{\Qm\ym}
          \left(\frac{\ymls}{\ym}-2\Rm\right)
%\nll   &&
         +2\left(1-\frac{2}{\ym}+\left(2-\frac{3}{\ym}\right)\Rm\right)
          \Biggr]
         \Biggr\}
\nll  &&
      + \Biggl[ \Zl  \Biggr]_z
        \left[-2\left(\frac{1}{\Qe \Vll }
                    + \frac{1}{\Vl \Qm  } \right) \right]
\nll   &&
     +      \Biggl[ 1 \Biggr]_z
      \Biggl[- \frac{2       }{S\yms}
       \left(  \frac{\Vl \yml}{\Qe  }
             + \frac{\Vll}{\Qm} \right)
\nll  &&
          +2\left(\frac{1}{S\ym}-\frac{1}{\Vl}-\frac{1}{\Vll}\right)
                           \left(1 - \frac{2}{\ym} \right)
          -\frac{2 \Vl}{\Qe \Vll}
                           \left( 2 - \frac{3}{\ym} \right)
            -\frac{2 \Vll }{\Vl \Qm}
                            \left( 1 - \frac{3}{\ym} \right) \Biggr]
\nll  &&
      + \Biggl[ \frac{1}{\Zl}  \Biggr]_z
             \Biggl[
             -\frac{8 S \Vll\yml\Rm}{\Qe \Qm \ym}
       +\frac{S \Vl}{\Qe \Vll}
              \left(-3\ym + 9
                      - \frac{10      }{\ym }
                      + \frac{4       }{\yms}
                      - \frac{2\yml\Rm}{\ym }        \right)
\nll  &&
       +\frac{2 \Vl\yml}{\Qe \yms}(\yml-3\Rm)
       +\frac{\Vll \yml}{\Qe\ym}-\frac{\Vl}{\ym}
                     \left(\frac{1}{\Qm}+\frac{\yml}{\Vll}\right)
\nll  &&
             +\frac{\Vll}{\Vl}
               \Biggl[ \frac{S}{\Qm}
               \left(-1+\frac{2}{\ym}
             -\frac{4}{\yms}
             +\frac{2\Rm}{\ym} \right)
             +\frac{1}{\ym}\Biggr]
\nll  &&
             +\frac{2 \Vll}{\Qm \yms} [1 -(1+ 2 \ym) \Rm]
             +2\left[1-\frac{2}{\ym}
             +\frac{2}{\yms}+ \frac{1}{\ym}
              \left(1 - \frac{4}{\ym}\right)\Rm\right]\Biggr]
\nll  &&
      +\Biggl[\frac{1}{\Zll} \Biggr]_z
             \Biggl[
             \frac{2 \Vl}{\Qe \yms} (1+3 \Rm)
             +\frac{S \Vl}{\Qe \Vll}
             \left( - 1 + \frac{2}{\ym} - \frac{4}{\yms}
                       + \frac{2\Rm}{\ym} \right)
\nll  &&
             +\frac{\Vll    }{\Qe \ym}
              \left( -1+\frac{8S\Rm}{\Qm} \right)
%\nll  &&
            +\frac{\Vll}{\Vl}
             \left[\frac{S}{\Qm}\left(
            - 3 \ym +9 - \frac{10}{\ym}
            +\frac{4 }{\yms}
            - \frac{2\yml\Rm}{\ym}\right)-\frac{\yml}{\ym}\right]
\nll  &&
         +\frac{\Vl }{\ym}
          \left(\frac{\yml}{\Qm }
         +      \frac{1   }{\Vll} \right)
             +\frac{2\Vll}{\Qm \yms}
                \left(\ymls+(1-3\ym)\Rm \right)
\nll  &&
             + 2\left(1-\frac{2}{\ym}
                       +\frac{2}{\yms }- \frac{1}{\ym}
              \left(3-\frac{4}{\ym}\right)\Rm \right)
             \Biggr].
\ea
%---------------------------------------------------
The indefinite integrals, in terms of which the
quantities $S^{^{\rm{R}}}_k$, are presented here,
is the first series of integrals over $z_1$ or $z_2$, which is given 
in this paper. 
It correspontds to the first, innermost and the only 
semianalytical integration,
within our {\em numerical approach}. In the R-frame, it corresponds
to the integration over the angle $\varphi_R$ of the photon, see eq.~(\ref{379}).

The three first integrals of this series are
presented in Appendix D.1 of~\cite{mimi}, eqns.(D.6)-(D.8).
Here we recall the definition of integration, and present two
additional integrals.
%==========================
\ba
\begin{array}{lclclcl}
\vph
  && {\ds \Biggl[ {\cal A}  \Biggr]_z }
 &= &{\ds \frac{1}{\pi}        }
     {\ds \int_{{\hat{z}}^{\min}_{1(2)}}
              ^{{\hat{z}}^{\max}_{1(2)}}
        \frac{dz_{1(2)}}{\sqrt{R_z}} {\cal A} },
\lnl
%===========================
 \vph
    4)&& {\ds \Biggl[ z_{1(2)}\Biggr]_z } &=&
%%%      {\ds \frac{\BZ}{\lambda^{3/2}_\mu}         }
-\ds{\frac{1}{\pi\lambda_q}} {\ds \sqrt{R_z(z_{1(2)})}}
\Big|_{z}
+ \ds{\frac{B_{1(2)}}{\lambda_q}}
    {\ds \Biggl[  1 \Biggr]_z },
              \lnl
%-----------------
 \vph
    5)&& {\ds \Biggl[  z^2_{1(2)} \Biggr]_z } &=&
 %%%       \frac    {\ds   3 \BZ- {\lambda_\mu}  \CZ }
 %%%                {\ds  2{\lambda^{5/2}_ \mu}      }
-\frac{\ds 1}{\ds 2\pi \lambda_q}\left( z_{1(2)}+3 \frac{\ds {B_{1(2)}}}{\ds \lambda_q}
                                \right)       \sqrt{R_z(z_{1(2)})} \Big|_{z}
+ \ds{\frac{1}{ 2\lambda^2_q}}
 \left(  3 {\ds B^2_{1(2)}} - {\ds \lambda_q} C_{1(2)}\right)
    {\ds \Biggl[  1 \Biggr]_z }.
%-----------------
\end{array}
\ea
The functions $ A, B, C $ are defined in ~(\ref{eq09})-(~\ref{eq11}).

In the table, we introduced the abbreviation
%--------------
\ba
\Big|_{z}
\equiv \Big|_{ {\hat z}_{1(2)}^{\min}} ^{ {\hat z}_{1(2)}^{\max}}
\label{hatz}.
\ea
%----------------
The limits of the integration ${\hat z}_{1(2)}^{\min,\max}$
may be arbitrary. The particular realization of these limits,
depending on realistic experimental cuts, which we implement within
our numerical approach, has been derived in Appendix~\ref{z1depcut}.
%----------------
%\input{m_cut}
%------------
\section{Numerical approach.
 Implementation of experimental cuts}
%-----------------------------------------------
Within our numerical approach, the cuts are applied via integration limits
in~(\ref{netb}). There are cuts in variables $\Vll,\;\Vl$ and $z_1$.

\subsection{$\Vll$-dependent cut}
\begin{itemize}
\item{Angular cut of muon}
%--------------------------------------------------------

 Here we consider the cut on the angle of the scattered muon.
This angle may be easily calculated from the Born and
bremsstrahlung kinematics.
In both cases
\bq
 -2k_1 k_2 = 2 \left( k^0_1 k^0_2
  - |\vec k_1| |\vec k_2| \cos \theta_\mu  \right). 
\label{eq_theta_mu}
\eq
 For the case of Born kinematics one can easily derive
\bq
\cos \theta^{^{\rm{BORN}}}_\mu =
 \frac{S^2 (1-\ym)-2\Me(S\ym+2\Mm)}{\SLAMS  \SLAML}.
 \label{cosborn}
\eq

 From eq.~(\ref{cosborn}), keeping only terms ${\cal{O}(\Me)}$, we get
\bq 
\sin^2 \THMUBN =1- \cos^2\theta^{^{\rm BORN}}_\mu  =
       \frac{ 4 \AME V^{\rm \max}_2}{S^2(1-\YMU)},
\label{sintborn}
\eq   
so, one may derive the $\THMUBN$. Here $V^{\rm \max}_2$
\bq
 V_2^{\rm {\max}}
   = S\YMU-\frac{\Mm \YMU^2}{1- \YMU}
   = \frac{S\YMU}{1-\YMU}\left(1-\frac{\YMU}{Y}\right),
\nll
\eq
is the upper limit of $V_2$ neglecting $m_e$. The exact expression
was given in~(\ref{v2_limits}). 

   Using (\ref{eq06}),
one may receive from (\ref{eq_theta_mu}) the following expression for
$\cos \theta^{^{\rm{BREM}}}_\mu$:
\bq
 \cos \theta^{^{\rm{BREM}}}_\mu  =
\frac{S^2 (1-\ym)  - 2 \me \left(S \ym-\Vll+2 \Mm \right)}
                 {\sqrt{ \lambda_{_S}}  \sqrt{\lambda_l}  }.
\eq

  Keeping again only terms linear in $m^2_e$, we may derive the value
of sine of the muon scattering angle in the bremsstrahlung process:
\bq
 \sin^2 \theta_\mu^{^{\rm{BREM}}}
     =  \frac{4 \Me}{S^2 (1-\ym)}
        \left(V^{\rm \max}_2 - V_2  \right).
\label{sintbr}  
\eq

 It is obvious that:
\bq
 \theta_\mu^{^{\rm{BREM}}} \leq \theta_\mu^{^{\rm{BORN}}}.
\eq
Therefore, we may introduce the upper limit of the difference
\bq
 \theta_\mu^{^{\rm{BORN}}} - \theta_\mu^{^{\rm{BREM}}} \leq
 \ANCM,
\eq
or
\bq
   (\theta_\mu^{^{\rm{BREM}}})^2 \geq
   (\theta_\mu^{^{\rm{BORN}}})^2
-2  \theta_\mu^{^{\rm{BORN}}}   \ANCM + \ANCM^2.
\eq
 
 By substitution of expressions (\ref{sintborn}) and (\ref{sintbr}), 
we receive
\bq
 V_2 \leq V^c_2  =  V^{\rm \max}_2
         \frac{ \ANCM (2 \THMUBN- \ANCM)}{(\THMUBN)^2}.
\eq
%--

\item{Kinematical limitation}

 The kinematical limit on $V_2^{\rm{\max}}$ was written in
(\ref{v2_limits}).

\end{itemize}

 So, the absolute $\max $ of $\Vll$
should be a minimum of
the two possible maximal values of $\Vll$
\bq
\hat{V}^{\rm \max}_2
          = \min\left[ V^{\rm \max}_2,  V^c_2  \right].
\label{V2maxcut}
\eq

\subsection{$V_1$-dependent cuts}
\begin{itemize}
\item {Energy recoil cut $\ERC$.}

%---------------------------------
The electron energy in the final state is limited from below by:
$E^{'}_{el} = p_2^0 \geq \ERC $,
where $\ERC$ is some cut on recoil electron energy.
In numerical calculation
 we use the SMC value
$\ERC = 35$ GeV.
So,
\bq
 2 m_e p_2^0 \geq 2 m_e \ERC.
\label{ERC2}
\eq
From the other hand:
%---------------------------------
\bq
-2p_1 p_2 = 2m_e p_2^0 = Q^2_e +2 \Me.  \nll
\label{nn_ERC}
\eq
This corresponds to
\bq
S \YMU - V_1 \geq 2 m_e (\ERC - m_e),
\label{n_ERC}
\eq
%--------------------------------

and we receive one of the possible cuts on $V_1$:
\bq
    V_1 \leq  V_1^{^{ERC}} = S \YMU - 2 m_e (\REC-m_e).
\label{cut_ERC}
\eq

\item{Energy balance cut $\EBC$.}

 Photon energy $p^0$ also could be implicitely 
limited by an experimental condition, emerging from checking of overall
energy balance:
\bq
   E_\mu+m_e-E^{'}_\mu-E^{'}_e=p_0 \leq \EBC.
\eq
The SMC value is $\EBC = 40$ GeV.
 It is another energy cut related to the
invariant $V_1$:
\bq
 2 m_e p^0 \leq 2 m_e \EBC,
\eq
\bq
   V_1 \equiv -2p_1 p = 2 m_e p^0 \leq  2 m_e \EBC \equiv V_1^{^{EBC}}.
\eq

\item{Kinematical limitation}

 Still we must
take into account the kinematical limitation
on $V_1^{\rm{\max}}$.
The boundaries of the allowed phase-space region are defined by
(\ref{v1_limits}).

\end{itemize}

 These three cuts allow us to make the right choice for
an absolute maximum of $V_1$
\bq
\hat{V}_1^{\max}   =  \min
        \left[ V_1^{\rm \max}, V_1^{^{ERC}}, V_1^{^{EBC}} \right].
\label{V1maxcut}
\eq

\subsection{$z_1$-dependent cut}

\begin{itemize}
\item{Cut on the electron angle \label{z1depcut}}
 The electron angle can be determined from
\ba
-2k_1 p_2 = 2 k^0_1 p^0_2 - 2|\vec k_1| |\vec p_2| \cos \theta_e
\ea
It differs for the Born and bremsstrahlung kinematics.
 We consider first the radiative case, where 
it may be rewritten as:
\ba
\cos \theta^{^{\rm BREM}}_e
  &=& \frac{
     S\left[(S \ym -\Vl) +2\Me\right]  -\left[S(1-\ym)+\Vll- \Zl  \right] 2\Me}
           {\SLAMS \SLAME}
\nll
  &=& \frac{(S  + 2\Me) ( S \ym-\Vl)+2\Me \Zll}
           {\SLAMS \sqrt{ (S\ym -\Vl)^2+ 4\Me(S\ym-\Vl )}}.
\ea

and
\ba
  \sin^2 \theta^{^{\rm BREM}}_e
  &=&  \frac{ 4 \Me}{\LAMS \LAME}
       \Bigl\{ \Qe\Bigl[\LAMS-\Qe(S+\Me+\Mm)
              - \Zll(S+2\Me) \Bigr] - \Me\Zll\Bigr\}.
\ea

 In the URA in $ \Me $,
 it becomes
\ba
   \sin^2 \theta^{^{\rm BREM}}_e
   \approx
   \frac{4 \Me}{S^2 \Qe}
   \Bigl[ S^2 -\Qe \Mm -S(S\ym - \Vl + \Zll)\Bigr].
\label{sinBREMz}
\ea
%-------------------

   We may extract the correct boundaries for the invariant $z_1$,
 from three quantities:  $ \left(\theta_e^{^{\rm BORN}}\right)^2,
\left(\theta_e^{^{\rm BREM}}\right)^2$ and the difference
$ \bar \theta $.

 From
~(\ref{sinBREMz}) we get:
\ba
  \left(\theta_e^{^{\rm BREM}}\right)^2 =
     \frac{4 \Me}{S}  
     \Biggl[ \frac{S(1-\ym) +V_2 -\Zl}{S\ym-V_1} - \frac{\Mm}{S}  \Biggr].
\label{BREMtheta}
\ea

  In the Born approximation  $ V_1=V_2=z_1=0$ it reduces to
\ba
  \left(\theta_e^{^{\rm BORN}}\right)^2 =
     \frac{4 \Me}{S}  
     \Biggl[ \frac{1-\ym}{\ym}  - \frac{\Mm}{S}  \Biggr].
\label{BORNtheta}
\ea

  Due to the fact that
 the differences between $\left| \theta^{^{\rm BORN}}_e-\theta^{^{\rm BREM}}_e \right|$
 has to be smaller than some value,
 which is determined by experiment, we have:
\ba
 \left| \theta^{^{\rm BORN}}_e-
 \theta^{^{\rm BREM}}_e \right| \leq \bar \theta.
\ea
 One  gets two cases
\ba
 \theta_e^{^{\rm{BORN}}}    \geq
       \theta_e^{^{\rm{BREM}}}.
\label{first}
\ea
and
\ba
   \theta_e^{^{\rm{BREM}}}    \geq
 \theta_e^{^{\rm{BORN}}}   .
\label{two}
\ea
%-----------------------------------------------------------------------------
 With ~(\ref{BREMtheta}),
~(\ref{BORNtheta}) and
~(\ref{first}) we arrive at
\ba
   \left( z_1^c \right)^{\max}  &=&
            \Vll +\Vl \frac{(1-\ym)}{\ym}
     +\frac{S \Qe}{4 \Me} \; \ANCE \; (2 \theta^{^{\rm BORN}}_e -\ANCE),
\ea
 while from~(\ref{two}) we receive
\ba
   \left( z_1^c \right)^{\min} &=&
            \Vll +\Vl\frac{(1-\ym)}{\ym}
     -\frac{S \Qe}{4 \Me} \; \ANCE \; (\ANCE +2 \theta^{^{\rm BORN}}_e).
\ea

\item{Kinematical limitation}

 Also the kinematic boundaries
 $ \Zl^{\max}$ and
 $ \Zl^{\min}$ defined in~(\ref{ax}),
 should be taken into account.

\end{itemize}

  The $ \theta_e $ cut conditions is active, if at least one of 
\bq
\left( z_1^c \right)^{\min},\left( z_1^c \right)^{\max}\in [z_1^{\min},z_1^{\max}], 
\eq
and
\bq
  \left( z_1^c \right)^{\min} \leq \left( z_1^c \right)^{\max} .
\eq
% \begin{itemize}
% \item
%  {$
%  \left( \Zl^{\min}\right)^{\rm cut} \in
%   \Biggl[ \left(\Zl^{\min}\right)^{\rm kin},
%           \left(\Zl^{\max}\right)^{\rm kin}   \Biggr]
%      $}
% \item
%  {$
% \left( \Zl^{\max}\right)^{\rm cut}  \in
%   \Biggl[  \left(\Zl^{\min}\right)^{\rm kin},
%            \left(\Zl^{\max}\right)^{\rm kin}   \Biggr]
%      $}
% \item
%     {$
%  \left( \Zl^{\min}\right)^{\rm cut} \leq
%  \left( \Zl^{\max}\right)^{\rm cut} ,
%     $}
% \end{itemize}
% where $\left( \Zl^{\max,\min}\right)^{\rm kin}$ are defined in ~(\ref{ax}).
Then the absolute maximum of $z_{1}$
should be a minimun of the two possible maxima of $z_1$
\bq
\hat{z}^{\rm \max}_1  =
 \min\left[ z^{\rm \max}_{1}, \left(z^c_1\right)^{\rm \max} \; \right],
\label{zmaxcut}
\eq
and the absolute minimum of $ z_{1} $
should be a maximum of the two possible minima of $\Zl$
\bq
\hat{z}^{\rm \min}_1 =
   \max \left[ z^{\rm \min}_{1}, \left( z^c_1 \right)^{\rm \min} \; \right].
\label{zmincut}
\eq
%-----------------
%\input{m_lkrf    }
%-----------------
%-----------------------------------------------
\section{R-frame kinematics \label{Rframe} 
\label{rfrm}
 }
%-----------------------------------------------
In two applications in this paper, we will need another frame
 than the laboratory frame.
When  calculating the soft photon contribution and a table of two-fold
integrals within the completely analytic approach,
we will also make use of the so called R-frame, which is defined by 
\bq  \vec{p_2} +\vec{p} = 0,
\label{def_Rfrr}
\eq
  or
\bq
 \vec Q= \vec{p_1} +\vec{k_1} - \vec{k_2}=0.
\label{Rfr_00}
\eq

First, we introduce coordinates of 4-vectors $p_1, k_1, k_2, p $,
shown in the figure,
\ba
 p_1 &=& \left(0,0,|\vec{p_1}|,p_1^0\right)      
\nll
 k_i &=& \left(0,|\vec{k_i}| \sin\theta_i,|\vec{k_i}| \cos\theta_i,k_i^0\right)
\nll
 p   &=& p^0 \left(\sin\theta_R\sin\varphi_R, \sin\theta_R\cos\varphi_R,
                                                 \cos\theta_R ,1\right),      
\nll
 p_2 &=& \left(-p^0 \sin\theta_R\sin\varphi_R,
    - p^0 \sin\theta_R\cos\varphi_R, -p^0 \cos\theta_R ,p_2^0\right).
\label{rfr_v}
\ea

\begin{figure}[bhtp]
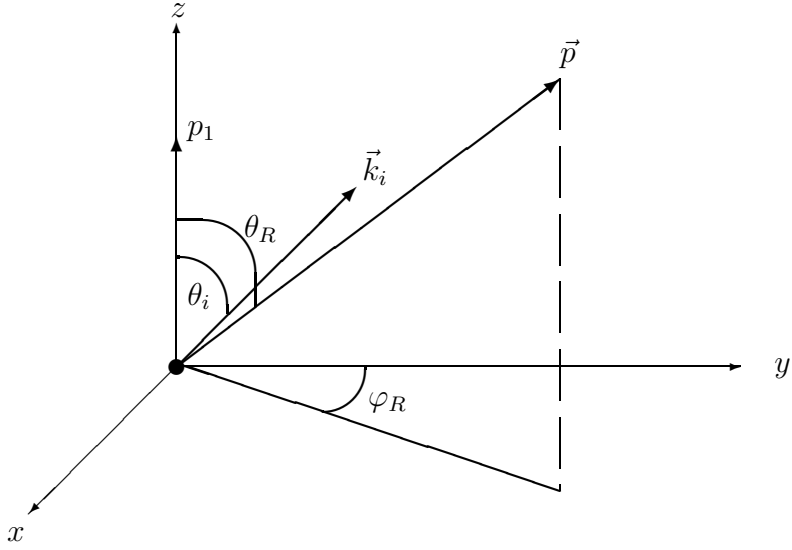

\begin{minipage}[bhtp]{14.5cm}{
%---                        
\vspace*{-2.5cm}
\begin{center}
%-----------------------------------
\begin{Feynman}{145,60}{-30,15}{1.5}
\thicklines
%---
%%%\put(45,7.5){\line(-3, 1){50}}              % \Lambda
\put(34,11.5){\line(-3,1){33.5}}              % \Lambda
%---
\put(0,22.5){\line(4, 3){34}}               % k_2
\put(34,49.5){${\vec p}  $}                   %
\put(34,48){\vector(4, 3){0}}               %
%-------------------------------------------
\put(16.5,39){${\vec k_i}$}
\put(00,22.5){\line(1,1){15.1} }
\put(16,38.5){\vector(1,1){0}}
\put(1,43)  {$  p_1 $     }
\put(-0,43  ){\vector(0,1){0}}
%---
\put(0 ,27.2){\oval(9 ,10)[tr]}               %
\put(1 ,28){${\theta_i }$}                    % \theta_i
%---                                          %
\put(0 ,27.8){\oval(14,15.5)[tr]}             %
\put(6 ,34){${\theta_R}$}                     % \theta
%---
\put(13 ,22.5   ){\oval(7.5 ,8   )[br]}              %
\put(17  ,19){${\varphi_R}$}
%---
\thinlines
%---
\put(0 ,22.5){\line(0,1){30} }        % axis X
\put(-0,53  ){\vector(0,1){0}}    %
\put(-.5,53.5){$ z $       }        % AXIS Z
\put( 00,22.5){\line  (-1,-1){13.1}}  % axis Y,Y_R
\put(-13, 9.5){\vector(-1,-1){0 }}    %
\put(-15, 7){$x                  $} % AXIS X
%----
\put(00,22.5){\line  (1,0){50}}       % AXIS Y
\put(50,22.5){\vector(1,0){0} }       %
\put(53,22  ){$y$     }               % AXIS Y
\put(00,22.5){\circle*{1.5}}
\put(34,  48 ){\line(0,-1){4.5}}
\put(34,  42 ){\line(0,-1){4.5}}
\put(34,  36  ){\line(0,-1){4.5}}
\put(34,  30  ){\line(0,-1){4.5}}
\put(34,  24  ){\line(0,-1){4.5}}
\put(34,  18  ){\line(0,-1){4.5}}
\put(34,  12 ){\line(0,-1){0.5}}
\thicklines
%----
\end{Feynman}
%---------------
\end{center}
}\end{minipage}
%---------------
\vspace*{2.cm}
\caption{\it The R-frame.}
\label{figA4}
\end{figure}
%-------------
%By definition  of the R-frame, shown in Fig.1,

 As may be seen from  figure 6, we have choosen the $z$-axis of the R-frame along the vector
$\vec{p}_1$ and matched the R-frame $(z,y)$-plane with the plane spanned by 
vector $\vec{p}_1$ and one of the vectors $\vec{k}_1$ or $\vec{k}_2$.
This trick is possible since we will use the R-frame only in analytical 
calculations of tables of bremsstrahlung hard and soft integrals,
in which we will always `decouple' $z_1$ and $z_2$ using an invariant 
partial fraction decomposition while calculating hard integrals and 
Feynman parametrization for the soft.
Finally, the photonic vector $\vec{p}$ is directed arbitrarily and
its angles are unlimited
\ba
&&0 \leq \varphi_R \leq 2\pi,
\nll
&&0 \leq \theta_R  \leq \pi.
\label{anglim}
\ea  

Actually we are using two R-frames, different for $z_1$ and $z_2$,
containing integrands with different pairs of $\theta_R,\;\varphi_R$,
both varying within the full solid angle $4\pi$.

The vector coordinates~(\ref{rfr_v}) contain, apart from
 angular variables,
the energies $p^0_1,k^0_{1,2}$ and vector moduli
$|\vec{p}_1|,|\vec{k}_{1,2}|$. 
They can be derived using an invariant language.
First we consider
\bq
\tau=-(p_2+p)^2=(p^0_2+p^0)^2.
\eq
Therefore,
\bq
p^0_2+p^0 = \sqrt{\tau}.
\label{renergy}
\eq
To derive an energetic coordinate of a 4-momentum, we consider the 4-scalar
product of this 4-momentum with
the four vector $-2(p_2+p)$ and write it twice: through invariants, and
in the R-frame, using~(\ref{renergy}). We give the simplest example:
\ba
-2(p_2+p).p&=&V_2 \quad-\quad \mbox{through~invariants,}
\nll
           &=&2(p^0_2+p^0)p^0\;=\;2\sqrt{\tau}p^0 \quad - \quad
\mbox{in~the~R-frame}.
\ea

Let us introduce the following $\lambda$-functions:
\ba
\lambda_{k_1} &\equiv& \lambda \left[-(p_2+p-k_1)^2, \tau, -k_1^2 \right]
               = \left[S(1-\ym )+V_2\right]^2 -4\mmu\tau,
\nll
\lambda_{k_2} &\equiv& \lambda \left[-(p_2+p-k_2)^2, \tau, -k_2^2 \right]
  = \left(S-V_2\right)^2         -4\mmu\tau,
\nll
\lambda_{p1 } &\equiv& \lambda \left[-(p_2+p-p_1)^2, \tau, -p_1^2 \right]
 = \left(S\ym +2 \me\right)^2   -4\me \tau=\lambda_\mu.
\label{lambda_add}
\ea

  In this way, the following table had been derived:
\bq
\displaystyle
%==========================================================
\begin{array}{rclcrcl}
%==========================================================
\vphantom{\int\limits_t^t}
|\vec{k_1}| &=&
             \frac{\ds \sqrt{\lambda_{k_1}}}{\ds 2\sqrt\tau},
&\hspace{1cm}&
 k^0_1 &=& \frac{\ds S(1-\ym)+V_2}{\ds 2\sqrt\tau} =
           \frac{\ds \EKl}{\ds 2\sqrt\tau},
%============
\nll
%============
\vphantom{\int\limits_t^t}
|\vec{k_2}| &=&
             \frac{\ds \sqrt{\lambda_{k_2}}}{\ds 2\sqrt\tau},
&\hspace{1cm}&
 k^0_2 &=& \frac{\ds S-V_2}{\ds 2\sqrt\tau}=
                            \frac{\ds \EKll}{\ds 2\sqrt\tau},
%============
\nll
%============
\vphantom{\int\limits_t^t}
|\vec{p_1}| &=&
              \frac{\ds \sqrt{\lambda_{\mu}}}{\ds 2\sqrt\tau},
&\hspace{1cm}&
 p^0_1 &=& \frac{\ds S\ym +2\me}{\ds 2\sqrt\tau} 
        \;=\; \frac{\ds\EPl}{\ds 2\sqrt\tau},
%============
\nll
%============
\vphantom{\int\limits_t^t}
|\vec{p_2}| &=& \frac{\ds V_2}{\ds 2\sqrt\tau},
&\hspace{1cm}&
 p^0_2 &=& \frac{\ds V_2+2\me }{\ds 2\sqrt\tau},
%============
\nll
%============
\vphantom{\int\limits_t^t}
 |\vec{p}|&=&|\vec{p}_{2}|,
&\hspace{1cm}&
 p^0 &=& \frac{\ds V_2}{\ds 2\sqrt\tau}.
%============
\end{array}
\label{Rfr_v0}
\eq

Now we write invariant $S$ in terms of R-frame variables
\bq
S   = -2 p_1 k_1=2(p_1^0 k_1^0-|\vec{p_1}| |\vec{k_1}| \cos\theta_1).
\label{Rfr_d}
\eq
By using~(\ref{Rfr_v0}), we get
\bq
 S = \frac{\EPl \EKl-\sqrt{\lambda_{\mu}\lambda_{k_1}}\cos\theta_1}{2 \tau},
\label{Rfr_S}
\eq
from which we derive an expression for $\cos\theta_1$
in the R-frame in terms of invariants:
\ba
\cos\theta_1 = \frac{\EPl \EKl -2S\tau}
              {\sqrt{\lambda_{\mu}} \sqrt{\lambda_{k_1}}}.
\label{Rfr_cos1}
\ea
 
Then we write the invariant $S_1$
\bq
S_1  = -2 p_1 k_2=2(p_1^0 k_2^0-|\vec{p_1}| |\vec{k_2}| \cos\theta_2).
\label{Rfr_s1}
\eq
By using~(\ref{Rfr_v0}), we get
\bq
 S_1
\label{Rfr_S11}
    = \frac{\EPl \EKll-\sqrt{\lambda_{\mu}\lambda_{k_2}}\cos\theta_2}{2 \tau},
\eq
from where we derive an expression for $\cos\theta_2$
in R-frame:
\ba
\cos\theta_2 = \frac{\EPl \EKll -2S_1\tau}
                    {\sqrt{\lambda_{\mu}} \sqrt{\lambda_{k_2}}}.
\label{Rfr_cos2}
\ea

 Now we can write down invariants $z_1$ and $z_2$ in their two own 
R-frames correspondingly
\ba             
z_1&=& -2k_1 p = 2p^0\left(k^0_1-|\vec{k}_1|\cos\theta_1\cos\theta_R
                      -|\vec{k}_1|\sin\theta_1\sin\theta_R\cos\varphi_R\right),                                       
\label{invz1}
\\
z_2&=& -2k_2 p = 2p^0\left(k^0_2-|\vec{k}_2|\cos\theta_2\cos\theta_R
                      -|\vec{k}_2|\sin\theta_2\sin\theta_R\cos\varphi_R\right).
\label{invz2}
\ea

The invariant $V_1$
\bq
V_1 = -2 p_1p =2 p^0 \left(p^0_1-|\vec{p}_1|\cos\theta_R\right)
\label{invV1}
\eq
looks formally the same in both R-frames.

Finally, relations ~(\ref{invV1}),~(\ref{invz2}) allow to perform
 ${\ds  \int dV_1 dz}$ in term of  ${\ds\int d\cos\theta_R}d\varphi_R $ 
\bq
 {\ds \int^{V_1^{\max}}_{V_1^{\min}} dV_1
      \int^{z_1^{\max}}_{z_1^{\min}} \frac{dz_1}{\pi\sqrt{R_z}}
 }
 \equiv \frac{V_2}{\tau}
\int_{-1}^{+1}d\cos\theta_R
\int_0^{2\pi}d\varphi_R.
\label{379}
\eq
This property was widely used when we discussed the treatment of the 
infrared divergent part.
%--------------------
%\input{m_testv1_ne}
%--------------------
\section{Analytic approach. Tables of integrals
\label{ddd}
}
\subsection{First and second analytic integrations. The R-integrals
\label{f_s_a}
}
\bq
\Biggl[{\cal{A}}\Biggr] _R =
         {\ds \int^{V_1^{\rm{max}}}_{V_1^{\rm{min}}} dV_1
         \int^{z_1^{\rm{max}}}_{z_1^{\rm{min}}}\frac{dz_1}{\pi\sqrt{R_z}}
         {\cal{A}}}                                            
\eq
%--
\ba
%==========================================================
\begin{array}{lclclclclclcl}
%==========================================================
       1)&&
        \Biggl[ 1\Biggr]_R
      %AINT( 1)
               &=& {\ds \frac{\Vll }{\TAU} \SLAMU}                 
    %-------
&& \hspace{1cm}
       2)&&
      \Biggl[\Vl \Biggr]_R
      %AINT( 2)
            &=&{\ds \frac{1}{ 2}\EPl \VllS \SLAMU} 
\nll
 \sskip
       3)&&
      {\ds \Biggl[\frac{1}{\Vll}\Biggr] _R}
      %AINT( 3)
            &=& {\ds \LPl}  
&& \hspace{1cm}
       4)&&
     %------
      {\ds \Biggl[ \frac{\AME}{\VlS} \Biggr] _R }
      %AINT( 4)
            &=& {\ds \frac{1}{\Vll} \SLAMU}                      
\nll
 \sskip
       5)&&
     %------
     {\ds \Biggl[ \frac{1}{\Qe} \Biggr]_R}
      %AINT( 5)
            &=& {\ds \LQE} 
&& \hspace{1cm}
     6)&&
     %------
     {\ds \Biggl[ \frac{\AME}{\Qes} \Biggr]_R}
      %AINT( 6)
            &=& {\ds \frac{\Vll}{\QMS} \SLAMU}   
\nll
 \sskip
       7)&&
     %------
     {\ds \Biggl[ \frac{1}{\Zl}\Biggr]_R}
      %AINT( 7)
            &=&{\ds \LKl}           
&& \hspace{1cm}
       8)&&
     %------
     {\ds \Biggl[ \frac{1}{\Zll}\Biggr]_R }
      %AINT( 8)
        &=& {\ds  \LKll}
\nll
 \smskip
       9)&&
     %-------
     \Biggl[ \frac{\AMU}{\ZlS}\Biggr]_R
      %AINT( 9)
               &=&{\ds \frac{\displaystyle 1}{\Vll}}    
&& \hspace{1cm}
       10)&&
     %-------
      {\ds \Biggl[ \frac{\AMU}{\ZllS}\Biggr]_R}
      %AINT(10)
               &=&{\ds \frac{1}{\Vll}}    
\end{array}
\ea

\ba
%==========================================================
\begin{array}{lclclclcl}
%==========================================================
     &&  11)&&
     %-------
       {\ds \Biggl[\frac{1}{\Zl \Zll}\Biggr]_R}
      %AINT(11)
%  &=&{\ds\frac{2}{\Vll \SMM} \LOG{\frac{\SMM+\QlM}{\SMM-\QlM}}
           & = &{\ds  \frac{2}{V_2}\LLM}
\nll
   \smskip
     &&  12)&&
     %-------
       {\ds \Biggl[ \frac{\AMU  \Vl}{\ZlS} \Biggr]_R}
      %AINT(12)
               &=&{\ds  \frac{1}{\PKlS}
                 \left[ S \EKl  - 2 \AMU \EPl
                  +\AMU (\EPl \EKl - 2 S \TAU )
                 \left[ \frac{1}{\Zl}\right]_R
                 \right]}
\nll
   \smskip
     &&  13)&&
     %-------
      {\ds \Biggl[ \frac{\Vl}{\Zl}\Biggr]_R}
      %AINT(13)
              &=&{\ds \frac{ \EPl \EKl-2 S \TAU}{\PKlS \PPl}
      \Biggl[ 1 \Biggr]_R
              + \left(S \EKl - 2 \AMU \EPl\right) 
       \frac{\Vll}{\PKlS}\Biggl[ \frac{1}{\Zl}\Biggr]_R}
 \nll
  \smskip
     &&  14)&&
     %-------
      {\ds \Biggl[ \frac{\Vl}{\Zll}\Biggr]_R}
     %AINT(14)
          &=&{\ds \frac{ \EPl \EKll-2 \Sl \TAU}{\PKllS \PPl}
         \Biggl[ 1 \Biggr]_R}
        + {\ds \left[ \Sl \EKll -2 \AMU \EPl \right]
         \frac{\Vll}{\PKllS}\Biggl[ \frac{1}{\Zll}\Biggr]_R }
     %-------
\end{array}
\ea

\ba
%==========================================================
\begin{array}{lclclclclclclcl}
%==========================================================
&& \hspace{-16mm}
       15)&&
     %-------
     {\ds \Biggl[ \frac{1}{\Vl \Zl}\Biggr]_R}
      %AINT(15)
           &=& {\ds \frac{1}{V_2}  \LSS}
&& \hspace{6mm}
       16)&&
     %-------
     {\ds \Biggl[ \frac{1}{\Vl \Zll}\Biggr]_R}
      %AINT(16)
           &=& {\ds \frac{1}{V_2} \LSSl }   
\nll
   \smskip
&& \hspace{-16mm}
       17)&&
     %-------
    {\ds \Biggl[ \frac{\ds 1}{\Qe\Zl}\Biggr]_R}
      %AINT(17)
           &=&  {\ds  \frac{1}{\QM} \LSSl}   
&& \hspace{6mm}
       18)&&
     %-------
    {\ds \Biggl[ \frac{ 1}{\Qe\Zll} \Biggr]_R}
      %AINT(18)
           &=&{\ds\frac{1}{\QM} \LSS}      
\end{array}
\ea

\ba
%==========================================================
\begin{array}{lclclcl}
%==========================================================
       19)&&
     %-------
{\ds \Biggl[\frac {\ds 1}{\Qes\Zl}\Biggr]_R}
      %AINT(19)
&=&{\ds \frac{\Vll N_{k_1} }{\AME  Q^6_\mu \ALX}   }
 %   \left[S^2 \YMU (1-\YMU)-2 \AME \QM\right]}   \nll   \smskip
 + {\ds \frac{\ds 1}{\QM\ALX}
                \left[{ \Sl \EKl-2 }\AMU
            \left(\frac{   S\YMU \Vll}{\QM}
                +2 \AME \right)\right]
      \Biggl[ \frac{\ds 1}{\Qe\Zl}\Biggr]_R}                  \nll
   \smskip
       20)&&
     %-------
    {\ds\Biggl[\frac{1}{\Qes \Zll} \Biggr]_R}
      %AINT(20)
    &=&
    { \ds \frac{\Vll  N_{k_2}}{ \AME Q^6_\mu  \ALS} }
%(S^2 \YMU + 2 \AME \QM)} \nll    \smskip
    +    {\ds
        \frac{1}{\QM\ALS}
        \left[S\EKll-2 \AMU
        \left(\frac{S\YMU\Vll}{\QM}+2\AME\right)\right]
        \Biggl[ \frac{1}{\Qe\Zll} \Biggr]_R}     \nll
   \smskip
       21)&&
    {\ds
    \Biggl[ \frac{1}{\Qe \ZlS}\Biggr]_R}
    % AINT(21)
    &=&{\ds\frac{N_{k_1}}{\QM\ALX}
%      \left[S^2 \YMU (1-\YMU)-2 \AME \QM \right]
       \Biggl[ \frac{1}{\Qe\Zl}\Biggr]_R}
    {\ds
+ \frac{1}{\QM\ALX} \left(\frac{\Sl \Vll+\ALX}{\AMU\Vll}
       -\frac{ 2 S \YMU}{\QM} \right)}                        \nll
   \smskip
       22)&&
     %-------
     {\ds
    \Biggl[ \frac{1}{\Qe \ZllS}\Biggr]_R }
      %AINT(22)
       &=&{\ds \frac{N_{k_2}}{\QM \ALS}
% \left( S^2 \YMU + 2 \AME \QM\right)
    \Biggl[ \frac{1}{\Qe\Zll} \Biggr]_R}
   +  {\ds
 \frac{1}{\QM \ALS}\left(\frac{-S \Vll+\ALS}{\AMU \Vll}
        - \frac{2 S \YMU}{\QM} \right)}                      \nll
%    \end{array}
%\ea
%
%\ba
%==========================================================
%   \begin{array}{lclclcl}
%==========================================================
%   \vph
   \smskip
       23)&&
     %-------
    {\ds
      \Biggl[ \frac{1}{\Qes \ZlS}\Biggr]_R }
      %AINT(23)
       &=& {\ds \frac{1}{\QM \ALX}\Biggl[ \frac{S \Vll }{\QM}
       - \Sl + 2 \AME }                                  \nll
   \smskip
          &&  &&   +
     {\ds
         \frac{3 N_{k_1}}
%{ S^2 \YMU (1-\YMU)-2\AME \QM}
{\QM\ALX}
       \left(\ALX + \Sl \Vll
       - \frac{ 2 \AMU S \YMU\Vll}{\QM}\right)\Biggr]
         \Biggl[ \frac{1}{\Qe\Zl} \Biggr]_R}                 \nll
   \smskip
          &&  &&   +
     {\ds
           \frac{1}{\AMU \QMS }
      \Biggl(  \frac{2\Sl + \Vll - 4 \AMU}{\ALX}+\frac{1}{\Vll} \Biggr)
       +\frac{\Vll \LAMU}{\AME Q^8_\mu \ALX}}              \nll
   \smskip
          &&  &&   +
     {\ds
         \frac{12 \Vll}{Q^8_\mu \ALXS}
       \Biggl( \AMU\LAMU+\AME\QMS-S \Sl \QM\Biggr)}  \nll
   \smskip
     24)&&
     %-------
    {\ds
     \Biggl[ \frac{1}{\Qes \ZllS} \Biggr]_R}
      %AINT(24)
       &=&{\ds  \frac{1}{\QM \ALS}
       \Biggl[\frac{ \Sl \Vll}{\QM}- S-2 \AME}
       \nll
   \smskip
      &&   &&    +
        {\ds
    \frac{ 3 N_{k_2}}
%{S^2\YMU+2\AME\QM}
{\QM\ALS}
     \left( \ALS- S \Vll - \frac{2\AMU S \YMU\Vll}{\QM}\right)
     \Biggr]
     \Biggl[ \frac{1}{\Qe\Zll} \Biggr]_R}               \nll
   \smskip
       &&  &&   +
      {\ds
         \frac{1}{ \AMU \QMS} \Biggl(\frac{ -2 S +\Vll-4\AMU}{\ALS}
        +\frac{1}{\Vll}\Biggr)
        +\frac{\Vll \LAMU}{\AME Q^8_\mu \ALS}}             \nll
   \smskip
        && && +
      {\ds
          \frac{12 \Vll}{Q_\mu^8 \ALSS}
          \Biggl( \AMU \LAMU+\AME \QMS-S \Sl \QM \Biggr) }
\end{array}
\ea
%-----------------
with
\ba
      \EKl   &=& \Sl + \Vll, \qquad     
      \EKll \;=\; S-\Vll,    \qquad       
      \EPl  \;=\; S \YMU+2 \AME,  
      \nll   
      \LKl   &=& {\ds\frac{1}{\PKl}\ln\frac{(\EKl+\PKl)^2}{\AMU\TAU}},
      \qquad
      \LKll \;=\;
           {\ds \frac{1}{\PKll}\ln \frac{(\EKll+\PKll)^2}{4\AMU\TAU}},
      \nll
      \LPl   &=& {\ds\ln\frac{(\SLP+2\AME)^2}{4\AME\TAU}},
      \qquad
      \LQE  \;=\;{\ds\ln\frac{ \Vll(\SLP)+2\AME\QlM}
                              { \Vll(\SLM)+2\AME\QlM}},   
      \nll
      \LLM   &=& {\ds\frac{1}{\SMM}
                  \ln{\frac{\SMM+\QlM}{\SMM-\QlM}} },      
      \nll
%----
\LSS  &=& {\ds\frac{1}{\SLS}}\LOG\frac{\ds  (S+\SLS)^2}{4\AME\AMU},
\qquad\qquad
\LSSl\;=\;{\ds\frac{1}{\SLX}}\LOG\frac{\ds(\Sl+\SLX)^2}{4\AME\AMU},
\nll
%----
     N_{k_1} &=&  S\Sl \YMU -2 \AME \QM,  \qquad\qquad\quad\;
     N_{k_2}\;=\; S^2 \YMU + 2 \AME \QM,                   
\nll   
    \Sl      &=& S(1-\YMU),               \qquad\qquad\qquad\qquad\quad 
    \TAU    \;=\; \Vll+\AME,                               
\nll 
\lambda_{k1} &=& \EKl -4 \AMU\TAU,        \qquad\qquad\qquad\quad\;    
\lambda_{k2}\;=\;\EKll-4 \AMU\TAU,                         
\nll
\lambda_{m}  &=& {\QMS+4 \AMU\QM}.
\label{Rintegrals}
\ea

In ~(\ref{Rintegrals}) we defined the objects
$\LKl,\;\LKll,\;\LPl,\;\LQE,\;\LLM,\;\LSS,\;\LSSl$, which will be used as 
integrands in the foregoing table.
On top of them, we will also use two differences
\ba
\DKl  &=& \LKl - \LSS, \qquad\qquad
\DKll\;=\;\LKll- \LSSl.
\label{dk12}
\ea
%-----------------
%\input{m_v2_1l_ne}
%-----------------
\subsection{The third analytic integration. The $V_2$-integrals
\label{third_a}
}
\ba
%==========================================================
\begin{array}{lclclcl}
\vph
 && {\ds \Biggl[ A \Biggr]_V } &=& {\ds \int_0^{V_2^{\max}}dV_2 A }\lnl
%==========================================================
 \vph
   1) && {\ds \Biggl[  1 \Biggr]_V }
               &=& {\ds S \YMU-\QMMAX }               \lnl
%-----------------
   2) && {\ds \Biggl[\frac{1}{\QM}\Biggr]_V}
              &=& {\ds
                     -\ln \frac{\QMMAX}{ S \YMU}}     \lnl
%-----------------
   3) && {\ds\Biggl[  \frac {1}{\QMS} \Biggr]_V }
               &=&  {\ds
                 \frac{1}{S} \left(\frac{S}{\QMMAX}
                -\frac{1}{\YMU}\right)}               \lnl
%-----------------
   4) &&{\ds \Biggl[  \frac{1}{Q^6_\mu }  \Biggr]_V }
               &=&{\ds \frac{1}{2S^2}\left( \frac{S^2}{\QMMAXS}
                -\frac{1}{\YMUS} \right)}             \lnl
%-----------------
   5) && \Biggl[ {\ds \frac{1}{\QMSS}}  \Biggr]_V
             &=& {\ds \frac{1}{ 3 S^3}
            \left(\frac{S^3}{\QllMMAXlll}-\frac{1}{\YMUlll}\right)} \lnl
%-----------------
   6) && {\ds \Biggl[\LLM \Biggr]_V}
               &=&{\ds \frac{1}{2}}
  \left({\ds \ln^2
  \frac{\TMAX+1}{\TMAX-1}
                     - \ln^2
  \frac{\TMIN+1}{\TMIN-1}}
  \right)  \lnl
%-----------------
    7) &&  {\ds \Biggl[ \frac{\LLM}{\QllM} \Biggr]_V }
               &=&{\ds\frac{1}{2  \AMU}

                 \Biggl(\TMIN \ln\frac{\TMIN+1}{\TMIN-1}
               - \TMAX  \ln \frac{\TMAX+1}{\TMAX-1}
               - \ln \frac{\QllMMAX}{ S \YMU} \Biggr) }       \lnl
\end{array}
\ea
%--
\ba
%==========================================================
\begin{array}{lclclcl}
%==========================================================
%-----------------
    8) &&  {\ds  \Biggl[ \frac{\LLM}{\QMS} \Biggr]_V }
               &=&{\ds -\frac{1}{ 12 \AMUS}  }
  {\ds \Biggl[
           \left(\frac{2\RMU}{\YMU}-1\right)\TMAX \;
                  \ln\frac{\TMAX+1}{\TMAX-1} }                \lnl
      && && - {\ds \left( \frac{2 S \RMU}{\QllMMAX}-1 \right) \TMIN \;
                  \ln\frac{\TMIN+1}{\TMIN-1}}
            - {\ds\ln\frac{\QllMMAX}{S\YMU}
               -\frac{2 S \RMU}{\QllMMAX}+ \frac{2 \RMU}{\YMU} \Biggr] } \lnl
%-----------------
   9.1)&&  {\ds \Biggl[ \frac{1}{\LAMl} \Biggr]_V}
               &=&{\ds \frac{1}{\SDETl}
       \left(\arctan \frac{\BLAMl}{\SDETl}
    -  \arctan \frac{\BLAMl - \VllMAX }{\SDETl}\right)} \lnl
%-----------------
   9.2)&& {\ds \Biggl[ \frac{1}{\LAMl} \Biggr]_V}
               &=&{\ds \frac{1}{2\SDETl} \left( \ln\frac{\BLAMl-\SDETl}
                                                        {\BLAMl+\SDETl}
              -\ln\frac{\BLAMl-\VllMAX-\SDETl}
                       {\BLAMl-\VllMAX+\SDETl} \right)} 
\lnl
  10)&& {\ds \Biggl[\frac{1}{\LAMll} \Biggr]_V}
           &=&{\ds \frac{1}{2\SDETll} \Biggl(
                \ln \frac{\BLAMll-\SDETll}
                          {\BLAMll+\SDETll}}
        {\ds - \ln \frac{\BLAMll-\VllMAX - \SDETll}
                         {\BLAMll-\VllMAX + \SDETll} \Biggr) }
 \lnl
%-----------------
  11)&& {\ds\Biggl[ \frac{ \Vll}{\LAMl}\Biggr]_V}
      &=& {\ds \frac{1}{2}
\ln \left[\frac{(\Sl+\VllMAX)^2-4 \AMU \VllMAX}{S_1^2}\right]
                                                 + \BLAMl }
                 {\ds \Biggl[ \frac{1}{\LAMl} \Biggr]_V }     \lnl
%-----------------
  12) & & {\ds \Biggl[  \frac{ \Vll}{\LAMll} \Biggr]_V}
      &=& {\ds \frac{1}{2}
 \ln \left[\frac{(S-\VllMAX)^2-4 \AMU \VllMAX}{S^2} \right]
                            +\BLAMll }
         {\ds \left[\frac{ 1}{\LAMll} \right]_V }         \lnl
%-----------------
  13) & & {\ds \Biggl[ \LKl \Biggr]_V  }
      &=&{\ds    -\ln \left|\frac{\TlMAX+1}{\TlMAX-1}\right|
              \ln \left|\frac{\TlMAX-1}{2 \RMUl} \right| }
       -{\ds \SPENCE \left[\frac{2 (1+ \RMUl)}{\TlMAX+1}\right]
            +\SPENCE\left( \frac{2 \RMUl }{\TlMAX-1}\right)} \lnl
%-----------------
  14) & & {\ds \Biggl[ \LKll \Biggr]_V  }
      &=& {\ds \ln \frac{\TllMAX+1}{\TllMAX-1}
                \ln \frac{\TllMAX-1  }{2 \RMU}   }
         + {\ds \SPENCE\left[ \frac{2(1+\RMU)}{\TllMAX+1}\right]
             -\SPENCE\left(\frac{2 \RMU }{\TllMAX-1}\right)} \lnl
%-----------------
 15)  & & {\ds \Biggl[ \frac{1}{\ALAMl} \LKl  \Biggr]_V  }
          &=& {\ds -\frac{2}{S^2_1}
 \Biggl[
       \frac{-\TlMAX}{\ALlMAX}
    \ln\left|\frac{2 \RMUl (\TlMAX+1)}{(\TlMAX-1)(\TlMAX-\TOl)}\right|}
 \lnl   &&  &&
       + {\ds \frac{1}{2 (\TOl+1)}
    \ln\left|\frac{(\TlMAX-\TOl)^2 (\TlMAX-1)}{(\TlMAX+1) \ALlMAX}\right|}
 \lnl   &&  &&
       {\ds  + \frac{1}{ \TOlS-1}
    \ln\left|\frac{(\TlMAX-\TOl) (\TlMAX-1)}{\ALlMAX}\right| }
 \lnl   &&  &&
 {\ds  -\left| \frac{1}{\XTl-\XTll}
    \ln\left(\frac{\TlMAX-\XTl}{\TlMAX-\XTll}\right)  
            +\frac{1}{\XTl-\XTll}
    \ln\left(\frac{\TOl-\XTl}{\TOl-\XTll}\right)\right|} \Biggr]
%-----------------
\end{array}
\ea
%-----------------
\ba
%==========================================================
\begin{array}{lclclcl}
%==========================================================

%-----------------
  16)&& {\ds \Biggl[ \frac{ 1}{\ALAMll} \LKll    \Biggr]_V }
              &=&{\ds \frac{2}{S^2} \Biggr[
            -\frac{\TllMAX}{\ALllMAX}
  \ln \left| \frac{2 \RMU (\TllMAX+1)}
                  {(\TllMAX-1) (\TllMAX-\TOll)}\right|}  \lnl
 &&  && + {\ds \frac{1}{ 2 (\TOll+1)}
  \ln\left|\frac{(\TllMAX-\TOll)^2 (\TllMAX-1)}
                  {(\TllMAX+1) \ALllMAX}\right|}        \lnl
 &&  && + {\ds  \frac{1}{\TOllS-1}
  \ln\left|\frac{(\TllMAX-\TOll)(\TllMAX-1)}
                  {\ALllMAX}\right|}
% \lnl  &&  &&
 - {\ds \frac{1}{\RTl-\RTll}
  \ln\left|\frac{\TllMAX-\RTl}{\TllMAX-\RTll}\right|  } \Biggl]
\lnl
  17)&& {\ds \Biggl[ \frac{\Vll}{\ALAMl} \LKl       \Biggr]_V}
      &=& {\ds \frac{2}{\Sl}  \Biggl[
    -\frac{1}{\ALlMAX}
 \ln\left| \frac{2 \RMUl  (\TlMAX+1)}
            {(\TlMAX-1)(\TlMAX-\TOl)}\right|}
\lnl   && &&  +{\ds \frac{1}{\TOlS-1}
 \ln\left| \frac{(\TlMAX-\TOl)(\TlMAX-1)}{\ALlMAX}\right|}
    +{\ds \frac{1}{2 (\TOl+1)}
 \ln \left|\frac{\TlMAXS-1}{\ALlMAX}\right|}                  \lnl
  && &&   +{\ds  \left| \frac{1}{\XTl-\XTll}
 \ln\left(\frac{\TlMAX-\XTl}{\TlMAX-\XTll}\right)
     +\frac{1}{\XTl-\XTll}
 \ln\left(\frac{\TOl -\XTl}{\TOl-\XTll}\right)\right|} \Biggr]\lnl
%-----------------
  18)& & {\ds \Biggl[  \frac{\Vll}{\ALAMll} \LKll \Biggr]_V}
     &=& {\ds \frac{2}{S} \Biggl[
         -\frac{1}{\ALllMAX}
 \ln\left|\frac{2\RMU(\TllMAX+1)        }
               {(\TllMAX-1)(\TllMAX-\TOll)}\right|}
\lnl && &&{\ds         +\frac{1}{\TOllS -1}
 \ln\left|\frac{(\TllMAX-\TOll) (\TllMAX-1)}{\ALllMAX}\right|}
\lnl && &&
       +{\ds \frac{1}{2(\TOll+1)}
 \ln\left|\frac{ (t_2^{\max})^2-1}{\ALllMAX}\right|
          +\frac{1}{\RTl-\RTll}
 \ln\left|
\frac{\TllMAX-\RTl}{\TllMAX-\RTll}
     \right| \Biggr] }
\lnl
%-----------------
  19)&&  { \ds \Biggl[ \frac{1}{\QllM}\LKl\Biggr]_V }
     &=& { \ds -\frac{1}{\XllDl }
  \Bigl[ F_{L_k}(\XlTl,\XlTll,\TOl,\TlMAX)
       - F_{L_k}(\XlTl,\XlTll,\TOl,\TOl  )     \Bigr] }
\lnl
%-----------------
  20)&&  {\ds \Biggl[ \frac{1}{\QllM}\LKll \Biggr]_V }
     &=& {\ds  \frac{1}{\XllDll}
 \Bigl[ F_{L_k}(\XllTl,\XllTll,\TOll,\TllMAX)
     -  F_{L_k}(\XllTl,\XllTll,\TOll,\TOll  )  \Bigr] }
\lnl
  21)&&  {\ds \Biggl[ \frac{1}{\tau^2} \Biggr]_V  }
           &=&{\ds \frac{1}{\AME} }
\lnl
%-----------------
  22)&&  {\ds \Biggl[\frac{1}{\TAU} \Biggr]_V  }
%       &=& {\ds\ALME+\ALVM  }
        &=& {\ds {\it L}_{+} }
\lnl
%------------------
   23)&&  {\ds \Biggl[ \LPl \Biggr]_V}
%          &=& \left( \ALME - \ALVM + 1 \right)
           &=& \left( {\it L}_{-}   + 1 \right)
               {\ds \Biggl[  1   \Biggr]_V    }
\lnl
%-----------------
  24)&& {\ds \Biggl[\frac{1}{\QllM}\LPl \Biggr]_V}
%  &=& {\ds \left( \ALME -\ln \frac{\VllMAX}{S\YMU} \right)}
   &=& {\ds  {\it L}_{-}                                   }
       {\ds \Biggl[\frac{1}{\QM}\Biggr]_V}
       {\ds   + \SPENCE \left(\frac{\VllMAX}{S\YMU}\right)}
\end{array}
\ea
%-----------------
\ba
%==========================================================
\begin{array}{lclclcl}
%==========================================================
  25)&& {\ds \Biggl[ \frac{1}{\QMSS}\LPl \Biggr]_V}
%   &=& {\ds \left( \ALME-\ALVM \right) }
    &=& {\ds {\it L}_{-}  }
   {\ds\Biggl[  \frac {1}{\QMS} \Biggr]_V }
        +{\ds \frac{1}{S\YMU}  }
     {\ds \Biggl[\frac{1}{\QM}\Biggr]_V}
 \lnl
%---------------------
  26)&& {\ds \Biggl[  \LQE \Biggr]_V}
%       &=& {\ds \left( \ALME+\ALVM+1 \right) }
        &=& {\ds \left( {\it L}_{+}+1 \right) }
    {\ds \Biggl[  1 \Biggr]_V }
        +{\ds 2 \left(\VllMAX- S\YMU\right) }
     {\ds \Biggl[\frac{1}{\QM}\Biggr]_V}
\lnl  27)&&  {\ds \left[ \frac{1}{\Vll} \LQE \right]_V}
      &=&{\ds \frac{1}{2}
%            \left(\ALME+\ALVM \right)^2
                  {\it L}^2_{+}
+\Fl+2 \SPENCE\left(\frac{\VllMAX}{S\YMU}\right) }
\lnl
  28)&&  {\ds \Biggl[\frac{1}{\Vll}  \Biggr]_V}
      &=&{\ds \ln \frac{\VllMAX}{\VllMIN}}                \lnl
  29)&& {\ds \Biggl[\frac{1}{\Vll} \LPl \Biggr]_V}
              &=&{\ds 2 \ALME      }
        {\ds \Biggl[\frac{1}{\Vll}  \Biggr]_V}
%      -{\ds \frac{1}{2}\left(\ALME+\ALVM\right)^2-\Fl}   \lnl
       -{\ds \frac{1}{2}  {\it L}^2_{+} - \Fl}            \lnl
  30)&&  {\ds \Biggl[\frac{1}{\Vll} \DKl  \Biggr]_V}
              &=&{\ds \frac{1}{\Sl}
                \Biggl[
 \frac{1}{2}\ln\left|\frac{\Sl^2}{\AME\AMU}\right|
 \ln\left|\frac{\Sl^2 \AME}{\VllMAXS \AMU}\right| - \Fl }
         {\ds -\frac{1}{2}
 \ln^2\left|\frac{-\Sl(\TlMAX-\TOl)}{2  (-\Sl+\AMU)} \right|}
\lnl && &&{\ds   +\SPENCE\left(\frac{\TlMAX-\TOl}{1-\TOl}\right)
             -\SPENCE\left(\frac{\TlMAX-\TOl}
{-1-\TOl}\right)\Biggr]}
\lnl
  31)&&   {\ds \Biggl[   \frac{1}{\Vll}\DKll \Biggr]_V}
               &=& {\ds \frac{1}{S} \Biggl[ \frac{1}{2}
 \ln\frac{S^2}{\AME \AMU}
    \ln\frac {S^2 \AME}{\VllMAXS\AMU}
    -\frac{1}{2} \ln^2\frac{S (\TllMAX-\TOll)}{2 (S+\AMU)}-\Fl
         \Biggr]}                            \lnl
    && &&{\ds +\SPENCE\left(\frac{\TllMAX-\TOll}{1-\TOll}\right)
              -\SPENCE\left(\frac{\TllMAX-\TOll}{-1-\TOll }\right) }
  \lnl
%-----------------------------------
  32)&& {\ds \Biggl[ \frac{1}{\Vll}\ALM  \Biggr]_V}
             &=&{\ds \frac{1}{\SLAMM}
          \Biggl\{ \ln\frac{\SLAMM+S\YMU}{\SLAMM- S\YMU}
     \left[ \ln\frac{(\TAMAX-\TAO)\SLAMMlll}{\VllMAXS {m_\mu^4 }}
           +\ln\frac{\VllMAX}{\VllMIN} \right]  }  \lnl
     && && {\ds+ \SPENCE\left(\frac{\TAMAX-\TAO}{1-\TAO}\right)
               - \SPENCE\left(\frac{\TAMAX-\TAO}{-1-\TAO}\right)}\Biggr\}
%-----------------------------------
\end{array}
\ea
with
\ba
     % \QllMMAX &=& S \YMU \left(1-\frac{\epsilon}{1000}\right),  \nll
       \QllMMIN &=& \frac{\AMU \YMUS}{1-\YMU},          \nll
       \VllMAX  &=&  S \YMU-\frac{\AMU \YMUS}{1-\YMU},  \qquad\qquad\quad
       \VllMIN \;=\; 2 m_e \bar\omega,                  \nll
%\ea
%ultrarelativistic approximation: $ m_e = 0 $
%\ba
       \RMU   &=&  \frac{\AMU}{S},  \qquad\qquad\qquad\qquad\qquad
       \RMUl \;=\;-\frac{\AMU}{\Sl},                           \nll
%       {y_\mu}_1&=&  1-\YMU,                                  \nll
%       \Sl    &=& S(1- \YMU),                                 \nll
       \QllMMAX&=& S \YMU-\VllMAX,                             \nll
       t^{\max}&=& \sqrt{1+\frac{4 \AMU}{S\YMU}},  \qquad\qquad\qquad\;
       \TMIN  \;=\;\sqrt{1+\frac{4 \AMU}{\QllMMAX}},           \nll
      %----------------
       \DETl   &=& 4 \AMU [ S (1-\YMU)-\AMU],       \qquad
       \DETll \;=\;4 \AMU (S+\AMU),                           \nll
       B_{k_1} &=& - S_1 + 2 \AMU,                  \qquad\qquad\;
       B_{k_2}\;=\;   S   + 2 \AMU,                            \nll
       {\it L}_{+} &=& {\ds  \ALME+\ALVM  },        \qquad
       {\it L}_{-}\;=\; {\ds  \ALME-\ALVM },
\ea
and the additional notation
%-------15'
       \ba
      \TOl   &=&  1+2 \RMUl,                                    \nll
      \XTl   &=&  1+2\left[\RMUl+ \sqrt{\XMUl (1+\XMUl)}\right],\qquad
      \XTll \;=\; 1+2\left[\RMUl- \sqrt{\XMUl (1+\XMUl)}\right],\nll
      \ALlMAX&=&  (\TlMAX-1)^2-4  \RMUl \TlMAX,                 \nll
%---
      \TOll   &=& 1+2\RMU,                                      \nll
      \RTl    &=& 1+2\RMU+2\sqrt{\RMU (1+\RMU)},   \qquad\qquad
      \RTll  \;=\;1+2\RMU-2\sqrt{\RMU (1+\RMU)},                \nll
      \ALllMAX&=& (\TllMAX-1)^2-4\RMU \TllMAX,                  \nll
%-------16''
      \XlTl  &=& \frac{-\Sl+\XSDl}{S\YMU},   \qquad\qquad\qquad\qquad\;\;
      \XlTll\;=\;\frac{-\Sl-\XSDl}{S\YMU},                      \nll
%-----17:
      \TlMAX &=&
      \frac{\sqrt{(\Sl+\VllMAX)^2-4\AMU\VllMAX}-\Sl}{\VllMAX}, 
\nll
      \TllMAX&=&\frac{S-\sqrt{(S-\VllMAX)^2-4\AMU\VllMAX}}{\VllMAX},
\nll
%----30:
      \TlMIN  &=& \frac{2 \Sl+\VllMIN-4 \AMU}
                   {\sqrt{(\Sl+\VllMIN)^2-4\AMU\VllMIN}+\Sl},   \nll
%----31:
      \TllMIN &=& \frac{2 S-\VllMIN+4 \AMU}
                       {S+\sqrt{(S-\VllMIN)^2-4 \AMU \VllMIN}}, \nll
%----32:
      \SLAMMS &=& { S^2 y^2_\mu + 4 \AMU S\YMU},                \nll
      \TAO    &=& \frac{ S \YMU + 2 \AMU}{\SLAMM},              \nll
      \TAMIN  &=&\frac{ 2 (S \YMU+2 \AMU)-\VllMIN},
     {\SLAMM+\sqrt{\SLAMMS-2 \VllMIN (S \YMU+2\AMU)+\VllMINS}}, \nll
      \TAMAX &=&\frac{\SLAMM -
                \sqrt{\SLAMMS-2\VllMAX (S\YMU+2 \AMU)+\VllMAXS}} {\VllMAX},
                                                                \nll
F_{L_k} & = &
          \ln \frac{(\TO-1)(\XTl+1)   }
                    {(\XTl-1)(\XTl-\TO)}
           \ln\left(\TA-\XTl\right)
      -\ln\frac{(\TO-1)(\XTll+1)}{(\XTll-1)(\XTll-\TO)}
                         \ln(\TA-\XTll)
\nll &&       - \SPENCE\left(\frac{\TA-\XTl }{-1-\XTl  }\right)
              + \SPENCE\left(\frac{\TA-\XTll}{-1-\XTll }\right)
              + \SPENCE\left(\frac{\TA-\XTl }{ 1-\XTl  }\right)
\nll &&
              - \SPENCE\left(\frac{\TA-\XTll}{1-\XTll  }\right)
              + \SPENCE\left(\frac{\TA-\XTl }{\TO-\XTl }\right)
              - \SPENCE\left(\frac{\TA-\XTll}{\TO-\XTll}\right).
\ea
%--------------------
%\input{m_soft}
%--------------
\section{Soft contribution to $d\sigma^{\mr{IR}}$}
%----------------------------------------------------
In~(\ref{br_ird}) the expression for $d\sigma^{\mr{IR,soft}}$ reads:
\ba
d\sigma^{\rm{IR,soft}} &=&\frac{2^7\pi^3\alpha^3}{\lambda_S} {\cal{B}}
\left(\Qmul F^{\mr{IR}}_{\mu\mu}
+ \Qmue F^{\mr{IR}}_{\mu e}
+ \Qel F^{\mr{IR}}_{e e}
\right)
\theta(\varepsilon-p^0)
 d\Gamma_3.
\label{ir_soft}
\ea

 We substitute now the phase space $d\Gamma_3$,~(\ref{phvbr}), which
in the soft photon limit can be factorized
\ba
  d\Gamma_{3} = (2\pi)^4 \frac{d^3 \vec {k_2}}{(2\pi)^3 2k_2^0}
                          \frac{d^3 \vec {p_2}}{(2\pi)^3 2p_2^0}
                          \frac{d^3 \vec {p}}{(2\pi)^3 2p^0}
                          \delta(k_1+p_1-k_2-p_2)
= d\Gamma_{2} \frac{d^3 \vec {p}}{(2\pi)^3 2p^0}. 
\label{phsoft}
\ea
Using definition~(\ref{born1}) and relation~(\ref{bornfac}), 
we straightforwardly derive from~(\ref{ir_soft})
\ba
  d\sigma^{\rm{IR,soft}} \Rightarrow
  d\born\;\frac{\alpha}{\pi}\;
\delta^{\mr{IR,soft}},
\label{br_s}
\ea
where( see also~\cite{mimi})$\;$:
\ba
\delta^{\rm{IR,soft}}
     = % \int^1_0 d\alpha\;
       \frac{2}{\pi} \int\;\frac{d^3 \vec {p }}{2p^0}\;
       F^{\mr IR}  \; \theta(\varepsilon-p^0).
\label{FIR}
\ea

Then, the function $ F^{\rm{IR}} $ takes the unique form
\ba
   F^{\rm{IR}}  &=&
   \Qmul \biggl(\frac{k_1}{2k_1 p }- \frac{k_2}{2k_2 p}  \biggr)^2
   +
   2 \Qmue
\biggl(\frac{k_1}{2k_1 p }- \frac{k_2}{2k_2 p}  \biggr)
\biggl(\frac{p_1}{2p_1 p }- \frac{p_2}{2p_2 p}  \biggr)   
\nll
&& + \Qel  \biggl(\frac{p_1}{2p_1 p }-\frac{p_2}{2p_2 p}  \biggr)^2
\label{FIR_l}
\\
&& = \Qmul F^{\rm{IR}}_{\mu\mu}
   + \Qmue F^{\rm{IR}}_{\mu e}
   + \Qel  F^{\rm{IR}}_{ee}.
\label{FIR_ll}
\ea
%----------------------

The symbol $\Rightarrow$ in~(\ref{br_s}) means that instead of $z$-integrated
IR-factors~(\ref{brem20_frm})-(\ref{brem22_frm}), we use them in a
completely differential form~(\ref{FIR_l}); they depend on two photonic
 angles in the R-frame, $\vartheta_R,\varphi_R$,
and on the photonic energy, $p^0$.
%The fourth variable $\alpha$ stands
%for a Feynman parameter which will be introduced, when necessary, below. 

Since the energy of
emitted soft photons is limited within a narrow interval
\ba
0 \leq p^0  \leq  \varepsilon
\label{s27} 
\ea
it is always possible to choose $\varepsilon$ small enough to ensure the 
phase space of soft photons be a sphere non-limited by experimental cuts.
In other words, the phase space of soft photons is {\em isotropic} which
allows to choose the $z$-axis along a convenient direction, differently for 
every term in~(\ref{FIR_l}).
 while performing angular integrations in~(\ref{FIR}).
  Therefore, the invariant variables $V_1, V_2, z_1$ and $z_2$ 
might be expressed in terms 
of only {\em the polar angle} $\cos\vartheta_R\equiv \xi$ and made
independent of the {\em azimuthal angle} $\varphi_R$  
%--------------
\ba
V_2 &=& -2 p_2 p  = 2 p_2^0 p^0 = 2 m_e p^0 ,
\label{s29}  \\
V_1 &=& -2 p_1 p  = 2 p^0 (p_1^0 - |\vec{p_1}| \xi ),
\label{s30}  \\
z_1 &=& -2 k_1 p  = 2 p^0 (k_1^0 - |\vec{k_1}| \xi ),
\label{s31}  \\
z_2 &=& -2 k_2 p  = 2 p^0 (k_2^0 - |\vec{k_2}| \xi ).
\label{s32}
\ea
%--------------

% We see that $F^{\rm IR}$ may be parametrization such that they depend on
%only one $\cos\vartheta_R=\xi$ and on variables independent of photon (see
%~(\ref{stst}).

In the R-frame
\ba \vec{p_2}+\vec{p} = 0
\ea
and in soft limit $ p\rightarrow 0 $ R-frame degenerates to the rest 
frame of $\vec{p}_2$
\ba  
p_2^0= m_e.
\ea
This is why $V_2$ in~(\ref{s29}) is angular independent.

 We will use the dimensional regularization for the infrared divergences
and rewrite the photonic phase space as follows (we took also $2p^0$ out of
every invariant variable in~(\ref{s29})-(\ref{s32}))
\ba
\frac{2}{\pi}
       \int \frac{ d^3\vec{p} }{(2p^0)^3}\;\Rightarrow
\frac{4\pi}{(2\sqrt\pi)^n\Gamma(n/2-1)}
\int^{\varepsilon}_0\frac{ (p^0)^{n-5}dp^0 }{\mu^{n-4}}
\int^{2\pi}_{0}d\varphi_R
\int^{\pi}_{0}(\sin\vartheta_R)^{n-3}d\vartheta_R.
\label{ndim}
\ea
Here $\gamma$
is the Euler constant and $\mu$  is an arbitrary parameter
with a dimension of mass.

 Now integration in~(\ref{FIR}) over $p^0$ is performed
straightforwardly:

%----------------------
\ba
\delta^{\mr{IR,soft}} =
               \frac{1}{2}
               \int\limits_{-1}^{+1} d\xi
               \left[ P^{\mr{IR}} + \ln \frac{\varepsilon}{\mu}
               + \frac{1}{2} \ln(1-\xi^2) \right] {\cal{F}^{\mr{IR}}}.
\label{s82}
\ea
%----------------------
In~(\ref{s82}) 
\ba
{\cal{F}}^{\mr {IR}} =  (2p^0)^2 F^{\mr {IR}}
\label{twofirs}
\ea
and the typical pole term
%----------------------
\ba
 P^{\mr IR} = \frac{1}{n-4} + \frac{1  }{2} \gamma
                            + \ln\frac{1}{2\sqrt\pi}
\label{s81}
\ea
%----------------------
represents the infrared divergences at $n=4$. 

 From~(\ref{s29})--(\ref{s32}) entering~(\ref{FIR_ll})
and from~(\ref{s82}) we see that only those
integrals over $\xi$ may occur, which are presented in
Appendix D.2 of~\cite{mimi}).
%----------------------
 Writing $\cal{F}^{\rm{IR}}$ in a form similar to~(\ref{FIR_ll}),
we will calculate the
three contributions $\delta_{\mu\mu}^{\rm{IR,soft}}, 
\delta_{\mu e}^{\rm{IR,soft}}$
and $\delta_{e e}^{\rm{IR,soft}}$ separately.
Before this, however, we present a collection of formulae in the R-frame
in the soft photon limit.

\subsection{R-frame kinematics for the calculation of the 
{\em soft} photon contribution}
%---------------------------------------------------------------------------
The R-frame is defined by
\ba  \vec{p_2} +\vec p = 0
\label{def_Rfr}
\ea
or
\ba
 \vec Q= \vec{p_1} +\vec{k_1} - \vec{k_2}=0.
\label{Rfr_0}
\ea
Since the R-frame is isotropic, there is no need to fix its $z$-axis 
along a given direction, 
say along $\vec{p}_1$\footnote{In this subsection all 4-momentum
coordinates are understood in the R-frame.}
\ba
 p_1 &=& \left(0,0,|\vec{p_1}|,p_1^0\right).
\ea
It might be equally chosen along $\vec{k}_1$,
\ba
 k_1 &=& \left(0,0,|\vec{k_1}|,k_1^0\right),
\ea
or along $\vec{k}_2$, then
\ba
 k_2 &=& \left(0,0,|\vec{k_2}|,k_2^0\right),
\ea
or along any linear combination of any vectors, 
say $\vec{k}_\alpha=\alpha \vec{k}_1+(1-\alpha) \vec{k}_2$, then
\ba
k_\alpha=\left(0,0,|\alpha \vec{k}_1+(1-\alpha) \vec{k}_2|,
 \alpha k^0_1+(1-\alpha) k^0_2\right).
\label{kalpha}
\ea
 So, we have indeed many R-frames, which differ one from 
another by a spatial rotation and when
we write an arbitrarily oriented photonic 4-momentum as
\ba
 p &=& p^0 (\sin\vartheta_R \sin\varphi_R,\sin\vartheta_R\cos\varphi_R,
                                                 \cos\vartheta_R,1),
\label{Rfr_1}
\ea
one should understand that
in every R-frame one has its own angles $\vartheta_R,\varphi_R$ which
 vary within {\em the same limits}
-- covering the full solid angle.
In this way, we arrive at equations~(\ref{s29})-(\ref{s32}) for 
invariants $V_2,V_1,z_1,z_2$
with formally one parameter $\xi$.

 In the expression~(\ref{s82}), which has to be integrated over
$\xi$ with~(\ref{twofirs}) and~(\ref{FIR_l}),
enter the energies 
$p^0_1,k^0_{1,2}$ and moduli
$|\vec{p}_1|,|\vec{k}_{1,2}|$
( see ~(\ref{s29})-(\ref{s32})).

 All the energies and momenta moduli depend only on three
invariants: $S, y$ and $V_2$. We make no distinction between $y$ and $\YMU$
in the soft photon kinematics.
In the soft photon problem, we neglect
the small invariant $V_2$ as compared to the others. 
In this limit, the table~(\ref{Rfr_v0}) reduces to
\bq
\displaystyle{
%==========================================================
\begin{array}{rclcrcl}
%==========================================================
\vphantom{\int\limits_t^t}
 |\vec{k_1}| &=&
              \ds{\frac{\ds \sqrt{\lambda_{ l}}}{\ds 2m_e}},
&\hspace{.5cm}&
 k^0_1 &=& \ds{\frac{\ds S_1}{\ds 2m_e}},
 \nll
%============
\nll
%============
%\vphantom{\int\limits_t^t}
|\vec{k_2}| &=&
           \ds{\frac{\ds  \sqrt{\ds \lambda_{_S}}}{\ds 2m_e}},
&\hspace{.5cm}&
 k^0_2 &=& \ds{\frac{\ds S}{\ds 2m_e}},
 \nll
%============
\nll
%============
%\vphantom{\int\limits_t^t}
|\vec{p_1}| &=&
           \ds{\frac{\ds \sqrt{\lambda^0_{e}}}{\ds 2m_e}},
 &\hspace{.5cm}&
 p^0_1 &=& \ds{\frac{\ds S y +2\me}{\ds 2m_e}},
 \nll
%============
\nll
%============
%\vphantom{\int\limits_t^t}
  |\vec p_2|   &=& \ds{\frac{\ds V_2}{\ds 2\sqrt{\ds\tau}}},
 &\hspace{.5cm}&
 p^0_2 &=& m_e,
\nll
%============
\nll
%============
%\vphantom{\int\limits_t^t}
 |\vec{p}| &=& |\vec{p}_2|,
 &\hspace{.5cm}&
  p^0   &=&\ds{ \frac{\ds V_2}{\ds 2m_e}}.
\label{stst}
\end{array}}
\eq

\subsection{Muonic current}
%-----------------------------
In $\it muonic$ current we are dealing with
\ba
  F^{\rm{IR}}_{\mu\mu}&=&
    \left(\frac{k_1    }{2k_1p         }\right)^2
  + \left(\frac{k_2    }{2k_2p         }\right)^2
  -      \frac{2k_1k_2}{(2k_1p)(2k_2p)},
 \ea
see~(\ref{FIR_l}). 
By applying the Feynman parameterization for the last term, we have
\ba
  F^{\rm{IR}}_{\mu\mu}&=& 
  - \frac{\Me}{(2k_1p)^2}
  - \frac{\Me}{(2k_2p)^2}
  - \int^1_0 d\alpha  \frac{2k_1k_2}{(2k_\alpha p)^2},
\ea
\noindent
where a new $4$-vector was introduced as discussed in ~(\ref{kalpha}):
%---
\ba
 k_\alpha  =  \alpha k_1 + (1-\alpha) k_2.
\label{k_alpha}
\ea

Therefore, using all soft machinery described above 
we can write $\delta^{\rm{IR,soft}}$  as follows
%-----
\ba
\delta_{\mu\mu}^{\rm{IR,soft}}&=&
\label{03}
  {\tt Q}_\mu^2  \Biggl\{
   \left( P^{\mr{IR}} + \ln\frac{\varepsilon}{\mu} \right)
   \frac{1}{2}
   \int\limits_{ 0}^{1} d\alpha
   \int\limits_{-1}^{1} d\xi \; {\cal{F}}^{\rm{IR}}_{\mu\mu}
  +\frac{1}{4}
   \int\limits_{ 0}^{1} d\alpha
   \int\limits_{-1}^{1} d\xi
   \ln (1-\xi^2)\;  {\cal{F}}^{\rm{IR}}_{\mu\mu}  \Biggr\},
\nll
\ea
%---
see~(\ref{s82}), with
\ba
{\cal{F}}^{\rm{IR}}_{\mu\mu} =
\label{148}
-\frac{\mmu}{{k_1^0}^2} \frac{1}{(1-\beta_1\xi)^2}
-\frac{\mmu}{{k_2^0}^2} \frac{1}{(1-\beta_2\xi)^2}
+\frac{S y+2\mmu}{{k_\alpha^0}^2} \frac{1}{( 1 - \beta_\alpha \xi)^2}.
\ea
On passing, we used
\ba
 -2k_\alpha p &=&  2p^0 (k^0_\alpha - |\vec k_\alpha| \xi)
             \;=\; 2p^0  k^0_\alpha(1 - \beta_\alpha  \xi),   \\
          z_1\;=\;
 -2k_1      p &=&  2p^0 (k^0_1      - |\vec k_1     | \xi)
             \;=\; 2p^0  k^0_1     (1 - \beta_1       \xi),   \\
          z_2\;=\;
 -2k_2      p &=&  2p^0 (k^0_2      - |\vec k_2     | \xi)
             \;=\; 2p^0  k^0_2     (1 - \beta_2       \xi),
\ea
with three velocities
\ba
 \beta_1  &=& \frac{|\vec k_1|}{k^0_1}\;=\;\frac{\sqrt{\lambda_l}}{S_1},  
\label{betaeel}   \\
 \beta_2  &=& \frac{|\vec k_2|}{k^0_2 }\;=\;\frac{\sqrt{\lambda_{_S}}}{S},
\label{betaeell}  \\
 \beta_\alpha &=& \frac{|\vec k_\alpha|}{k^0_\alpha }.
\label{betaeea}
\ea

 We also have from~(\ref{k_alpha}) the following relations:
\ba
k^0_\alpha   &=&\frac{S\alpha+S_1(1-\alpha)}{2m_e},
\nll
 -k_\alpha^2 &=&\mmu + \alpha(1-\alpha) S y.                 
\ea
%---
Therefore
\ba
\beta_\alpha &=&\frac{\sqrt{\left[S\alpha+S_1(1-\alpha)\right]^2
             - 4\me\left[\mmu + \alpha(1-\alpha) S y \right]}}
                        {S\alpha+S_1(1-\alpha)}.
\label{betak12al}
\ea

  Using the table of integrals from Appendix D.2 of~\cite{mimi} we obtain
%---
\ba
\delta_{\mu\mu}^{\rm{IR,soft}}&=&
\label{s149}
  {\tt Q}_\mu^2  \int\limits_{0}^{1}d\alpha \Biggl\{
   \left( P^{\mr{IR}} + \ln\frac{2\varepsilon }{\mu} \right)
\left[ - \frac{\mmu}{{k_1^0}^2} \frac{1}{(1-\beta_1^2)}
       - \frac{\mmu}{{k_2^0}^2} \frac{1}{(1-\beta_2^2)}
+\frac{S y+2\mmu}{{k_\alpha^0}^2} \frac{1}{(1-\beta^2_\alpha)}
                                                        \right] \nll
&&+ \frac{1}{2\beta_1}\frac{\mmu}{{k_1^0}^2} \frac{1}{(1-\beta_1^2)}
       \ln\frac{1+\beta_1}{1-\beta_1}
+  \frac{1}{2\beta_2}\frac{\mmu}{{k_2^0}^2} \frac{1}{(1-\beta_2^2)}
       \ln\frac{1+\beta_2}{1-\beta_2}                               \nll
&&-  \frac{1}{2\beta_\alpha}\frac{S y+2\mmu}{{k_\alpha^0}^2}
   \frac{1}{(1-\beta_\alpha^2)}
       \ln\frac{1+\beta_\alpha}{1-\beta_\alpha}\Biggr\}.
\ea
%---

We have from~(\ref{betaeel})--(\ref{betaeea})
%---
\ba
\label{s150}
(k_1^0)^2 (1-\beta_1^2) &=& -k_1^2 \;=\; \mmu, \\
\label{s151}
(k_2^0)^2 (1-\beta_2^2) &=& -k_2^2 \;=\; \mmu, \\
\label{s152}
(k_\alpha^0)^2 (1-\beta_\alpha^2) &=& -k_\alpha^2.
\ea
%---
And the expression~(\ref{03}) becomes
%---
\ba
\delta_{\mu\mu}^{\rm{IR,soft}}&=&
\label{dmumu_2}
 {\tt Q}_\mu^2  \Biggl\{
   \left( P^{\mr{IR}} + \ln\frac{2\varepsilon}{\mu} \right)
\left[ -2 + \left(S y+2\mmu\right)
    \int\limits_{0}^{1}
    \frac{d\alpha}{\mmu + \alpha(1-\alpha)S y } \right] \nll
&&+ \frac{1}{2\beta_1} \ln\frac{1+\beta_1}{1-\beta_1}
 +  \frac{1}{2\beta_2} \ln\frac{1+\beta_2}{1-\beta_2}  \nll
&&+ \frac{S y+2\mmu}{2}
    \int\limits_{0}^{1}       
    \frac{d\alpha}{\beta_\alpha [\mmu + \alpha(1-\alpha)S  y]}
 \ln\frac{ 1 - \beta_\alpha }{ 1 + \beta_\alpha }  \Biggr\}.
\ea
%---

Now we use the URA in the electron mass
\ba
 \frac{1}{2\beta_1}\ln\frac{1+\beta_1}{1-\beta_1}
 &\approx&
   =\ln\frac{S_1}{m_e m_\mu},                     \\
 \frac{1}{2\beta_2}\ln\frac{1+\beta_2}{1-\beta_2}
 &\approx&
   =\ln\frac{S}{m_e m_\mu}.
\ea

The first integral we calculated precisely in~\cite{mimi}
%---
\ba
\label{s154}
\left(S y + 2\mmu\right)
\int\limits_{0}^{1}\frac{d\alpha}{ [\mmu + \alpha(1-\alpha) S y] } =
    \frac{1+\beta^2}{\beta} L_\beta 
\ea
with
\ba
 \beta   &=& \sqrt{1+\frac{4\mmu}{S y  }}, \nll
 L_\beta &=& \ln \frac{\beta+1}{\beta-1}.
\ea

With the second integral the situation is more complicated.
 Defining it as in~\cite{mimi},
\ba
S_{\Phi}\equiv \frac{1}{2}
\left(S y+2\mmu\right)
\int\limits_{0}^{1}\frac{d\alpha}{\beta_\alpha [\mmu + \alpha(1-\alpha)S y]}
 \ln\frac{ 1 - \beta_\alpha }{ 1 + \beta_\alpha },
\ea
we cannot take over the analogous result from~\cite{mimi} since there that
integral was calculated in the URA in the leptonic
mass $m$.
Here we need result exact in $m_\mu$ (see also~\cite{BarShum}).
 Making use of URA in $m_e$, with the aid of~(\ref{betak12al}) we get
\ba
S_{\Phi} \approx 
\left(S y+2\mmu\right)\int\limits_{0}^{1}
\frac{d\alpha}{\mmu + \alpha(1-\alpha)S y}
 \ln\frac{\me\left[\mmu + \alpha(1-\alpha)S y\right]}
         {\left[S\alpha+S_1(1-\alpha)\right]^2}.
\ea
The expression for $S_{\Phi}$ simplifies drastically and can be calculated
straightforwardly
\ba
S_{\Phi}&=&
 \frac{S y + 2\mmu}{\sqrt{\lambda^0_m}}
 \Biggl\{
 \ln \frac{\mmu(S y + 4\mmu)}{S^2(1-y)(1-y\alpha_1)(1-y\alpha_2)}
 \ln\frac{\alpha_2}{(-\alpha_1)}                                        \nll
&&-\frac{1}{2} \ln^2\left[\frac{\alpha_2}{(-\alpha_1)(1-y)}\right]
  +\ln(1-y)    \ln        \frac{(1 - y)(1 - y \alpha_1)}{(1-y\alpha_2)}
  -\Litwo(1)                                                            \nll
&& +\Litwo \Biggl[\frac{     (-\alpha_1)  }{(1-y) \alpha  _2} \Biggr]
+\Litwo \Biggl[\frac{(1-y)(-\alpha_1)  }{      \alpha  _2} \Biggr]
-\Litwo \Biggl[\left(\frac{-\alpha_1}{\alpha_2}\right)^2 \Biggr]
 \Biggr\}.   
\ea
with
\ba
\alpha_{1,2}=\frac{1\mp\beta}{2}.
\ea

Collecting all terms together, we finally have
\ba
\delta^{\rm{IR,soft}}_{\mu\mu}
 &=&
  {\tt Q}^2_\mu\Biggl\{
  2\Biggl( P^{\rm{IR}} + \ln \frac{2 \varepsilon }{\mu}\Biggr)
   \Biggl( \frac{1+\beta^2}{2\beta} L_\beta -1         \Biggr)
 + \ln\frac{S^2(1- y)}{\me\mmu} + S_{\Phi}            \Biggr\}.
\ea

 From the virtual photon correction to the muon vertex we have
 the contribution
%---
\ba
\delta^{\rm{vert}}_{\mu \mu} &=&
 {\tt Q}^2_\mu\Biggl\{-2
 \Biggl( P^{\rm{IR}} + \ln \frac{m_\mu}{\mu}      \Biggr)
 \Biggl(\frac{1+\beta^2}{2\beta}L_\beta -  1 \Biggr)  
+\frac{3}{2}\beta L_{\beta} -2
                                                      \nll
 &&-  \frac{1+\beta^2}{2\beta}
            \Biggl[     
   L_{\beta} \ln\frac{4 \beta^2}{\beta^2-1}
 + \Litwo \left(\frac{1+\beta  }{1-\beta  } \right)
 - \Litwo \left(\frac{1-\beta  }{1+\beta  } \right)
            \Biggr]\Biggr\}.
 \ea

%---
The infrared divergence and the scale parameter $\mu$ cancel exactly 
in the sum of these two contributions. The sum reads
\ba
\delta^{^{\rm{VR}}}_{\mu\mu}&=&
\delta^{\rm{IR,soft}}_{\mu\mu} +
\delta^{\rm{vert}}_{\mu\mu} 
\nll
&=&
 {\tt Q}^2_\mu
\Biggl\{
 \Biggl[\frac{1+\beta^2}{2\beta}L_\beta -  1 \Biggr]
\left(2 \ln\frac{V_2^{\min}}{S}- \ln(1-y)\right)
+\frac{3}{2}\beta L_{\beta} -2
                                                      \nll
&& +  \frac{1+\beta^2}{2\beta}
            \Biggl[     
   L_{\beta}  \ln\frac{1-y}{(1-y\alpha_1 )(1-y\alpha_2 )}
- \frac{1}{2}\ln^2(1-y)  \nll
&& + \ln(1-y)  
\ln\frac{(1-y)(1-y \alpha_1 )}{(1-y \alpha_2 )}
 +\Litwo \left[\frac{(-\alpha_1)}{(1-y)\alpha_2} \right] \nll
&& +\Litwo \left[\frac{(1-y)(-\alpha_1)}{\alpha_2} \right]
 - 2 \Litwo \left(\frac{(-\alpha_1)}{\alpha_2} \right)
            \Biggr]  \Biggr\}  .
\label{dvrmm}
\ea
%---------------------------------
\subsection{Electronic current}
%---------------------------------
The correction $\delta_{ee}^{\rm{IR,soft}}$ is very easy to calculate.
%----------------------
\ba
F^{\rm{IR}}_{e e} =
   \left(\frac{p_1}{2p_1 p}\right)^2
 + \left(\frac{p_2}{2p_2 p}\right)^2
 - \frac{2 p_1 p_2}{(2 p_1 p)(2 p_2 p)}.
\label{fii}
\ea
Since $2 p_2 p$ is independent of $ \xi = \cos\vartheta_R $, 
the $\alpha$-parameterization
is not needed here and $\delta_{ee}^{\rm{IR,soft}}$ reads
%----------------------
\ba
\delta_{ee}^{\rm{IR,soft}}&=&
\label{s91ee}
  {\tt Q}_e^2 \Biggl[ \left( P^{\rm IR} + \ln\frac{\varepsilon}{\mu} \right)
    \frac{1}{2}\int\limits_{-1}^{+1} d\xi
    {\cal{F}}^{\rm{IR}}_{ee}
   + \frac{1}{4}\int\limits_{-1}^{+1}
    d\xi \ln(1-\xi^2) {\cal{F}}^{\rm{IR}}_{ee}  \Biggr].
\ea
%----------------------

 Introducing
\ba
 \beta=\frac{|\vec{p_1}|}{p^0_1}
  =
 \frac{\sqrt {S^2 y^2 + 4\me S y}}{ S y + 2\me},
\label{beta}
\ea
we receive
%----------------------
\ba
{\cal{F}}^{\rm{IR}}_{ee} &=&
 - \frac{\me}{{p^0_1}^2}   \frac{1}{(1-\beta\xi)^2}
 - 1 + \frac{2m_e p^0_1}{m_e p^0_1}
       \frac{1}{(1-\beta\xi)  }
   =
       - 1 + \frac{2}{1-\beta\xi}
\label{fee_2}
       - \frac{\me}{{p_1^0}^2} \frac{1}{(1-\beta\xi)^2}.
\nll
\label{fee_1}
\ea

Using the table of
integrals from Appendix D.2 of~\cite{mimi} we derive
%----------------------
\ba
\delta_{ee}^{\mr{IR,soft}}&=&
  {\tt Q}_e^2 \Biggl\{
 \left( P^{\mr IR} + \ln\frac{2  \varepsilon}{\mu} \right)
 \left[ - 1 + \frac{1}{\beta} \ln{\frac{1+\beta}{1-\beta}}
            - \frac{\me}{{p_1^0}^2} \frac{1}{(1-\beta^2)}\right]
\\   &&
          + 1 + \frac{1}{2\beta}
              \left[\litwo\left(\frac{2\beta}{\beta-1}\right)
                   -\litwo\left(\frac{2\beta}{\beta+1}\right)\right]
          + \frac{\me}{{p_1^0}^2}\frac{1}{2\beta}\frac{1}{(1-\beta^2)}
              \ln{\frac{1+\beta}{1-\beta}}  \Biggr\}. \nonumber
\label{s93}
\ea

 In the URA in  $\me$ we obtain
%----------------------
\ba
    \frac{1}{\beta} \ln \frac{1+\beta}{1-\beta}
 \approx 2\ln\frac{S y}{m_e^2}
\label{J_ee}
\ea
%----------------------
and
%----------------------
\ba
     \litwo\left(\frac{2\beta}{\beta-1}\right)
   - \litwo\left(\frac{2\beta}{\beta+1}\right)
\label{s98}
%&=&
%   2 \litwo\left( \frac{1-\beta}{1+\beta}\right)
% - 2 \litwo(1) - \frac{1}{2}\ln^2 \frac{1+\beta}{1-\beta}    \\
%&&+ 2 \ln\frac{1-\beta}{1+\beta} \ln(1-\frac{1-\beta}{1+\beta})
 \approx - 2 \litwo(1) - 2 \ln^2 \frac{S y}{\me}.  
\ea

 So, we can write
%----------------------
\ba
\delta_{ee}^{\rm{IR,soft}}&=&
\label{s99}
2  Q_e^2 \Biggl\{
 \left( P^{\mr IR} + \ln\frac{2\varepsilon }{\mu} \right)
 \left( - 1 +  \ln\frac{S y}{\me}\right)
          + 1 + \ln \frac{S y}{\me} - \ln^2 \frac{S y}{\me}
             - \litwo(1) \Biggr\}. \nll
\ea
%----------------------
%where \ba
%\label{s100}
%\litwo(1) = \frac{\pi^2}{6}. \ea

  The corresponding virtual photon correction to  the electron vertex
has the following form:
%----------------------
\ba
\delta_{ee}^{\mr{vert}}&=&
\label{ee_v}
   Q_e^2 \Biggl\{2
 \left( P^{\mr IR} + \ln\frac{m_e}{\mu} \right)
 \left( 1-\ln\frac{S y}{\me}\right)          
-2 + \frac{3}{2} \ln\frac{S y}{\me}
           - \frac{1}{2} \ln^2\frac{S y}{\me}
           + \litwo(1) \Biggr\}.     \nll
\ea
%----------------------

The complete answer is:
 \ba
\delta^{^{\rm{VR}}}_{ee}&=&
\delta_{ee}^{\rm{IR,soft}} + \delta_{ee}^{\rm{vert}}
\nll
  &=&
\label{dvree}
   Q_e^2 \Biggl[
  \left(\ln\frac{S y}{\me}-1\right)
  \Biggl(  \ln  \frac{S  y        }{\me}
        + 2\ln  \frac{\ds V^{\min}_2}{\me} \Biggr)      
  -1 +    \frac{3}{2} \ln  \frac{S y}{\me}
          - \frac{1}{2} \ln^2\frac{S y}{\me}
                       \Biggr].
\ea
%-----------------------------------
\subsection{$\mu e$ interference}
%-----------------------------------
In the $\mu e$ interference the expression for $F^{\rm{IR}}_{\mu e}$ reads
%---
\ba
F^{\rm{IR}}_{\mu e} =  
   \frac{2k_1p_1}{(2k_1p)(2 p_1 p)}
 - \frac{2k_1p_2}{(2k_1p)(2 p_2 p)}
 - \frac{2k_2p_1}{(2k_2p)(2 p_1 p)}
 + \frac{2k_2p_2}{(2k_2p)(2 p_2 p)}.
\label{f_mue}
\ea
%---
We introduce the $\alpha$ parameterization with the aid of
two new 4-vectors 
\ba
k_{1\alpha} &=& k_1 \alpha+p_1(1-\alpha),         \\
k_{2\alpha} &=& k_2 \alpha+p_2(1-\alpha),        
\ea
resulting in
\ba
{F}^{\rm{IR}}_{\mu e}=
 \frac{S_1}{2k^0_1p^0(1-\beta_1\xi)}
-\frac{S  }{2k^0_2p^0(1-\beta_2\xi)}+
\int\limits_{0}^{1}d\alpha
\left[
        \frac{S_1}{2(k_{2\alpha}p)^2}
-       \frac{S  }{2(k_{1\alpha}p)^2}
\right],
\ea
and $\delta_{\mu e}^{\rm{IR,soft}}$ becomes
\ba
\delta_{\mu e}^{\rm{IR,soft}}&=&
  \Qmue \Biggl[ \left( P^{\rm{IR}}
 +\ln\frac{\varepsilon}{\mu} \right)
  \frac{1}{2}\int\limits_{0}^{1} d\alpha \int\limits_{-1}^{+1} d\xi
  {\cal{F}}^{\rm{IR}}_{\mu e}                            
   +\frac{1}{4}\int\limits_{0}^{1} d\alpha\int\limits_{-1}^{+1}
  d\xi \ln(1-\xi^2) {\cal{F}}^{\rm{IR}}_{\mu e}  \Biggr]
\label{s91}
\ea
with
%---
\ba
{\cal{F}}^{\rm{IR}}_{\mu e}
                   &=&
   \frac{2}{1-\beta_1 \xi} - \frac{2}{1-\beta_2 \xi}
+  \frac{S_1}{(k^0_{2\alpha})^2 (1-\beta_{2 \alpha} \xi)^2}
  -\frac{S  }{(k^0_{1\alpha})^2 (1-\beta_{1 \alpha} \xi)^2}.
\ea
%---
Here
\ba
-k_{1\alpha}^2 & = & \Mm\alpha^2 + \me(1-\alpha)^2 -2k_1p_1\alpha(1-\alpha)  
\nll
               & = & \Mm\alpha^2 + \me(1-\alpha)^2 +S      \alpha(1-\alpha), 
\nll
 k^0_{1\alpha} & = & \frac{S_1\alpha+(S y+2\me)(1-\alpha)}{2m_e}          , \\
-k_{2\alpha}^2 & = & \Mm\alpha^2 + \me(1-\alpha)^2 -2k_2p_1\alpha(1-\alpha)  
\nll
               & = & \Mm\alpha^2 + \me(1-\alpha)^2 +S_1 \alpha(1-\alpha),    
\nll
 k^0_{2\alpha} & = & \frac{S  \alpha+(S y+2\me)(1-\alpha)}{2m_e}. 
\ea
%----
\ba
\beta_{1\alpha}    &=&\frac {\left| k_{1\alpha}\right|}{ k^0_{1\alpha}}
   \\
     &=& \frac{\sqrt{\ds  \Bigl[S_1\alpha+(S y+2\me)(1-\alpha)\Bigr]^2
  -4\me \left[\Mm\alpha^2 + \me(1-\alpha)^2 +S \alpha(1-\alpha)\right]}}
                       {S_1\alpha+(S y+2\me)(1-\alpha)},
\nll
\beta_{2\alpha}
     &=&\frac {\left| \vec k_{1\alpha} \right| }{ k^0_{1\alpha}}
   \\
    &=&\frac{\sqrt{ \ds  \Bigl[S  \alpha+(S y+2\me)(1-\alpha)\Bigr]^2   
  -4\me \left[\Mm\alpha^2 + \me(1-\alpha)^2 +S_1 \alpha(1-\alpha)\right]}}
                       {S  \alpha+(S y+2\me)(1-\alpha)}.
\nonumber
\ea

  Using the table of 
integrals from Appendix D.2 of~\cite{mimi} we derive
\ba
\delta_{\mu e}^{\rm{IR,soft}}&=&\Qmue
\Biggl\{
\left( P^{\mr{IR}} + \ln\frac{2\varepsilon }{\mu} \right)
\left[  
        \frac{1}{2\beta_1} \frac{1+\beta_1}{1-\beta_1}
      - \frac{1}{2\beta_2} \frac{1+\beta_2}{1-\beta_2}
      - S  
\int\limits^1_0 
 \frac{d\alpha}{-k_{1\alpha}^2}
      + S_1 
\int\limits^1_0 
 \frac{d\alpha}{-k_{2\alpha}^2}        
\right]
\nll &&
-\frac{2}{\beta_2}\left[\Litwo\left(\frac{2\beta_2}{\beta_2-1}\right)
                         -\Litwo\left(\frac{2\beta_2}{\beta_2+1}\right)\right]
+\frac{2}{\beta_1}\left[\Litwo\left(\frac{2\beta_1}{\beta_1-1}\right)
                         -\Litwo\left(\frac{2\beta_1}{\beta_1+1}\right)\right]
\nll &&
-\frac{S }{2}
\int\limits^1_0 
\frac{d\alpha } {-k_{1\alpha}^2 \beta_{1\alpha}}
\ln\frac{1-\beta_{1\alpha}}{1+\beta_{1\alpha}}
+\frac{S_1}{2}
\int\limits^1_0 
\frac{d\alpha } {-k_{2\alpha}^2\beta_{2\alpha}}
\ln\frac{1-\beta_{2\alpha}}{1+\beta_{2\alpha}}
\Biggr\}.
\label{softme1}
\ea

 In the URA in $ \me $
\ba
S \int_0^1 \frac{d\alpha}{-k_{1\alpha}^2}
&\approx&
 \ln  \frac{S^2  }{\Mm \me},
\\
S_1\int_0^1 \frac{d\alpha}{-k_{2\alpha}^2}
&\approx&
  \ln \frac{S_1^2}{\Mm \me},
\\
\frac{S}{2}
\int \frac{d\alpha}
          {-k_{1\alpha}^2\beta_{1\alpha}}
\ln\frac{1-\beta_{1\alpha}}{1+\beta_{1\alpha}}
&\approx&
S_{\Phi}(S,S_1),
\\
\frac{S}{2}
\int \frac{d\alpha}
          {-k_{2\alpha}^2\beta_{2\alpha}}
\ln\frac{1-\beta_{2\alpha}}{1+\beta_{2\alpha}}
&\approx&
S_{\Phi}(S_1,S).
\ea 
 Where we introduced the generic function $S_\Phi(I,{\hat{I}})$ 
\ba
S_{\Phi}(I,{\hat{I}})&=&\frac{I}{2}\int\limits^1_0 
\frac{d\alpha}{ \Mm \alpha^2 + \me (1-\alpha)^2 + {I} \alpha (1-\alpha) }
\ln\frac{\me\left[\Mm\alpha^2 + \me(1-\alpha)^2 + {I} \alpha (1-\alpha) \right] }
        {\left[{\hat{I}}\alpha + (S y+2\me)(1-\alpha)\right]^2}
\nll
& = & \frac{1}{2}
\ln\frac{\me I}{S^2 y^2}
\ln\frac{ I^2 }{\me \Mm}
-\frac{1}{4}\ln^2\frac{\Mm }{I} 
-\frac{1}{4}\ln^2\frac{\me }{I}
-\litwo\left( 1-\frac{Sy\Mm}{I \hat I } \right)
\nll
&& +\ln \frac{\hat I }{S y } \ln \frac{\Mm }{I}
-\frac{1}{2}\ln^2 \frac{\hat I }{S y }. 
\ea
The expression~(\ref{softme1}) reduces in the URA in $m_e$ to
\ba
\delta_{\mu e}^{\rm{IR,soft}}
&=&  
\left( P^{\mr{IR}} + \ln\frac{2\varepsilon }{\mu} \right)  4\ln(1-y)
+2\ln(1- y)\left(\ln\frac{\me}{S}-\ln y \right).
\label{d_mue_soft}
\ea

In the next Appendix on the box contribution we derive
%---
\ba
\delta_{\mu e}^{\rm box}&=&{\tt Q}_e {\tt Q}_\mu
  \Biggl\{ \Biggl[
  \left(-P^{\mr{IR}} -\frac{1}{2} \ln\frac{Q^2}{\mu^2} \right)
    4\ln (1-y)
                  \Biggr]  + B^{\rm{fin}}_{\mu e}  \Biggr\},
\label{d_mue_box}
\ea
and  the following short final expression is obtained
\ba
\delta^{^{\rm{VR}}}_{\mu e}&=&
\delta_{\mu e}^{\rm IR,soft} + \delta_{\mu e}^{\rm box} = {\tt Q}_e {\tt Q}_\mu
  \Biggl\{ 
    4\ln (1-y) \ln \frac{V_2^{\min}}{S y}
                   + B^{\rm{fin}}_{\mu e}  \Biggr\}.
\label{dvrme}   
\ea
 The finite contribution from box diagram, $B^{\rm{fin}}_{\mu e}$,
is presented in Appendix F.
%-------------------
%\input{m_boxes}
%----------------
\section{Two photon exchange contribution}
%-----------------------------------------
%\newcommand{\DET  }{\mbox{${\Delta_4 }  $}}
%\newcommand{\DETQ2}{\mbox{${\frac{\Delta_4 }{Q^2}$}}
\newcommand{\Ki  }{\mbox{${    k}_{i } $}}
\newcommand{\Pj  }{\mbox{${    p}_{j } $}}
\newcommand{\VK  }{\mbox{${  V^{ij}_k }  $}}
\newcommand{\Vk  }{\mbox{${  V^{ij}_k }  $}}
\newcommand{\VQ  }{\mbox{${  V^{ij}_q }  $}}
\newcommand{\Vq  }{\mbox{${  V^{ij}_q }  $}}
\newcommand{\VP  }{\mbox{${  V^{ij}_p }  $}}
\newcommand{\Vp  }{\mbox{${  V^{ij}_p }  $}}
\newcommand{\SKi }{\mbox{${  S_{k_i}}$}}
\newcommand{\SPj }{\mbox{${  S_{p_j}}$}}
%---
\newcommand{\Tkk }{\mbox{${T^{ij}_{kk}}$}}
\newcommand{\Tqk }{\mbox{${T^{ij}_{qk}}$}}
\newcommand{\Tkp }{\mbox{${T^{ij}_{kp}}$}}
%---
\newcommand{\Tpp }{\mbox{${T^{ij}_{pp}}$}}
\newcommand{\Tqp }{\mbox{${T^{ij}_{qp}}$}}
%---
\newcommand{\Tqq }{\mbox{${T^{ij}_{qq}}$}}
%-----
%--------------------------------------------
The two-photon exchange contribution is described by the two box diagrams:
{\it direct box} and {\it crossed box}. For the sake of symmetry, it is
convenient to duplicate the number of diagrams and to deal with four
diagrams shown in the figure below: \\

\vspace{.5cm}

\unitlength=1.00mm
\linethickness{0.6pt}
\begin{picture}(125.00,132.00)
\bezier{24}(25.00,100.00)(23.00,102.00)(25.00,104.00)
\bezier{24}(25.00,104.00)(27.00,106.00)(25.00,108.00)
\bezier{24}(25.00,108.00)(23.00,110.00)(25.00,112.00)
\bezier{24}(25.00,112.00)(27.00,114.00)(25.00,116.00)
\bezier{24}(25.00,116.00)(23.00,118.00)(25.00,120.00)
\bezier{24}(25.00,120.00)(27.00,122.00)(25.00,124.00)
\bezier{24}(45.00,100.00)(43.00,102.00)(45.00,104.00)
\bezier{24}(45.00,104.00)(47.00,106.00)(45.00,108.00)
\bezier{24}(45.00,108.00)(43.00,110.00)(45.00,112.00)
\bezier{24}(45.00,112.00)(47.00,114.00)(45.00,116.00)
\bezier{24}(45.00,116.00)(43.00,118.00)(45.00,120.00)
\bezier{24}(45.00,120.00)(47.00,122.00)(45.00,124.00)
\bezier{24}(90.00,100.00)(88.00,102.00)(90.00,104.00)
\bezier{24}(90.00,104.00)(92.00,106.00)(90.00,108.00)
\bezier{24}(90.00,108.00)(88.00,110.00)(90.00,112.00)
\bezier{24}(90.00,112.00)(92.00,114.00)(90.00,116.00)
\bezier{24}(90.00,116.00)(88.00,118.00)(90.00,120.00)
\bezier{24}(90.00,120.00)(92.00,122.00)(90.00,124.00)
\bezier{24}(110.00,100.00)(108.00,102.00)(110.00,104.00)
\bezier{24}(110.00,104.00)(112.00,106.00)(110.00,108.00)
\bezier{24}(110.00,108.00)(108.00,110.00)(110.00,112.00)
\bezier{24}(110.00,112.00)(112.00,114.00)(110.00,116.00)
\bezier{24}(110.00,116.00)(108.00,118.00)(110.00,120.00)
\bezier{24}(110.00,120.00)(112.00,122.00)(110.00,124.00)
\put(10.00,76.00){\vector(1,0){5.00}}
\put(15.00,76.00){\line(1,0){16.00}}
\put(31.00,76.00){\vector(1,0){5.00}}
\put(36.00,76.00){\line(1,0){19.00}}
\put(55.00,76.00){\vector(1,0){5.00}}
\put(10.00,46.00){\vector(1,0){5.00}}
\put(15.00,46.00){\line(1,0){16.00}}
\put(31.00,46.00){\vector(1,0){5.00}}
\put(36.00,46.00){\line(1,0){19.00}}
\put(55.00,46.00){\vector(1,0){5.00}}
\put(75.00,76.00){\vector(1,0){5.00}}
\put(80.00,76.00){\line(1,0){16.00}}
\put(96.00,76.00){\vector(1,0){5.00}}
\put(101.00,76.00){\line(1,0){19.00}}
\put(120.00,76.00){\vector(1,0){5.00}}
\put(75.00,46.00){\vector(1,0){5.00}}
\put(80.00,46.00){\line(1,0){16.00}}
\put(96.00,46.00){\vector(1,0){5.00}}
\put(101.00,46.00){\line(1,0){19.00}}
\put(120.00,46.00){\vector(1,0){5.00}}
\put(75.00,100.00){\vector(1,0){5.00}}
\put(80.00,100.00){\line(1,0){16.00}}
\put(96.00,100.00){\vector(1,0){5.00}}
\put(101.00,100.00){\line(1,0){19.00}}
\put(120.00,100.00){\vector(1,0){5.00}}
\put(75.00,128.00){\vector(1,0){5.00}}
\put(80.00,128.00){\line(1,0){16.00}}
\put(96.00,128.00){\vector(1,0){5.00}}
\put(101.00,128.00){\line(1,0){19.00}}
\put(120.00,128.00){\vector(1,0){5.00}}
\put(10.00,128.00){\vector(1,0){5.00}}
\put(15.00,128.00){\line(1,0){16.00}}
\put(31.00,128.00){\vector(1,0){5.00}}
\put(36.00,128.00){\line(1,0){19.00}}
\put(55.00,128.00){\vector(1,0){5.00}}
\put(10.00,100.00){\vector(1,0){5.00}}
\put(15.00,100.00){\line(1,0){16.00}}
\put(31.00,100.00){\vector(1,0){5.00}}
\put(36.00,100.00){\line(1,0){19.00}}
\put(55.00,100.00){\vector(1,0){5.00}}
\bezier{24}(44.00,52.00)(44.00,55.00)(41.00,55.00)
\bezier{24}(41.00,55.00)(38.00,55.00)(38.00,58.00)
\bezier{24}(38.00,58.00)(38.00,61.00)(35.00,61.00)
\bezier{24}(35.00,61.00)(32.00,61.00)(32.00,64.00)
\bezier{24}(32.00,64.00)(32.00,67.00)(29.00,67.00)
\bezier{24}(29.00,67.00)(26.00,67.00)(26.00,70.00)
\bezier{24}(26.00,52.00)(26.00,55.00)(29.00,55.00)
\bezier{24}(29.00,55.00)(32.00,55.00)(32.00,58.00)
\bezier{24}(32.00,58.00)(32.00,61.00)(35.00,61.00)
\bezier{24}(35.00,61.00)(38.00,61.00)(38.00,64.00)
\bezier{24}(38.00,64.00)(38.00,67.00)(41.00,67.00)
\bezier{24}(41.00,67.00)(44.00,67.00)(44.00,70.00)
\bezier{24}(44.00,70.00)(44.00,73.00)(47.00,73.00)
\bezier{24}(47.00,73.00)(50.00,73.00)(50.00,76.00)
\bezier{24}(26.00,70.00)(26.00,73.00)(23.00,73.00)
\bezier{24}(23.00,73.00)(20.00,73.00)(20.00,76.00)
\bezier{24}(44.00,52.00)(44.00,49.00)(47.00,49.00)
\bezier{24}(47.00,49.00)(50.00,49.00)(50.00,46.00)
\bezier{24}(26.00,52.00)(26.00,49.00)(23.00,49.00)
\bezier{24}(23.00,49.00)(20.00,49.00)(20.00,46.00)
\bezier{24}(109.00,52.00)(109.00,55.00)(106.00,55.00)
\bezier{24}(106.00,55.00)(103.00,55.00)(103.00,58.00)
\bezier{24}(103.00,58.00)(103.00,61.00)(100.00,61.00)
\bezier{24}(100.00,61.00)(97.00,61.00)(97.00,64.00)
\bezier{24}(97.00,64.00)(97.00,67.00)(94.00,67.00)
\bezier{24}(94.00,67.00)(91.00,67.00)(91.00,70.00)
\bezier{24}(91.00,52.00)(91.00,55.00)(94.00,55.00)
\bezier{24}(94.00,55.00)(97.00,55.00)(97.00,58.00)
\bezier{24}(97.00,58.00)(97.00,61.00)(100.00,61.00)
\bezier{24}(100.00,61.00)(103.00,61.00)(103.00,64.00)
\bezier{24}(103.00,64.00)(103.00,67.00)(106.00,67.00)
\bezier{24}(106.00,67.00)(109.00,67.00)(109.00,70.00)
\bezier{24}(109.00,70.00)(109.00,73.00)(112.00,73.00)
\bezier{24}(112.00,73.00)(115.00,73.00)(115.00,76.00)
\bezier{24}(91.00,70.00)(91.00,73.00)(88.00,73.00)
\bezier{24}(88.00,73.00)(85.00,73.00)(85.00,76.00)
\bezier{24}(109.00,52.00)(109.00,49.00)(112.00,49.00)
\bezier{24}(112.00,49.00)(115.00,49.00)(115.00,46.00)
\bezier{24}(91.00,52.00)(91.00,49.00)(88.00,49.00)
\bezier{24}(88.00,49.00)(85.00,49.00)(85.00,46.00)
\bezier{24}(25.00,124.00)(23.00,126.00)(25.00,128.00)
\bezier{24}(45.00,124.00)(43.00,126.00)(45.00,128.00)
\bezier{24}(90.00,124.00)(88.00,126.00)(90.00,128.00)
\bezier{24}(110.00,124.00)(108.00,126.00)(110.00,128.00)
\put(12.00,132.00){\makebox(0,0)[cc]{$k_1$}}
\put(12.00,94.00){\makebox(0,0)[cc]{$p_1$}}
\put(58.00,132.00){\makebox(0,0)[cc]{$k_2$}}
\put(58.00,94.00){\makebox(0,0)[cc]{$p_2$}}
\put(35.00,94.00){\makebox(0,0)[cc]{$p_1-p$}}
\put(35.00,132.00){\makebox(0,0)[cc]{$k_1+p$}}
\put(25.00,131.00){\makebox(0,0)[cc]{$\alpha$}}
\put(25.00,96.00){\makebox(0,0)[cc]{$\alpha^\prime$}}
\put(45.00,96.00){\makebox(0,0)[cc]{$\beta^\prime$}}
\put(45.00,131.00){\makebox(0,0)[cc]{$\beta$}}
\put(20.00,114.00){\makebox(0,0)[cc]{$p$}}
\put(53.00,114.00){\makebox(0,0)[cc]{$q+p$}}
\put(82.00,114.00){\makebox(0,0)[cc]{$q+p$}}
\put(114.00,114.00){\makebox(0,0)[cc]{$p$}}
\put(22.00,67.00){\makebox(0,0)[cc]{$p$}}
\put(51.00,67.00){\makebox(0,0)[cc]{$q+p$}}
\put(84.00,67.00){\makebox(0,0)[cc]{$q+p$}}
\put(113.00,67.00){\makebox(0,0)[cc]{$p$}}
\put(77.00,132.00){\makebox(0,0)[cc]{$k_1$}}
\put(123.00,132.00){\makebox(0,0)[cc]{$k_2$}}
\put(100.00,132.00){\makebox(0,0)[cc]{$k_2-p$}}
\put(90.00,131.00){\makebox(0,0)[cc]{$\beta$}}
\put(110.00,131.00){\makebox(0,0)[cc]{$\alpha$}}
\put(77.00,94.00){\makebox(0,0)[cc]{$p_1$}}
\put(123.00,94.00){\makebox(0,0)[cc]{$p_2$}}
\put(100.00,94.00){\makebox(0,0)[cc]{$p_2+p$}}
\put(90.00,96.00){\makebox(0,0)[cc]{$\beta^\prime$}}
\put(110.00,96.00){\makebox(0,0)[cc]{$\alpha^\prime$}}
\put(12.00,80.00){\makebox(0,0)[cc]{$k_1$}}
\put(58.00,80.00){\makebox(0,0)[cc]{$k_2$}}
\put(35.00,80.00){\makebox(0,0)[cc]{$k_1+p$}}
\put(20.00,79.00){\makebox(0,0)[cc]{$\alpha$}}
\put(50.00,79.00){\makebox(0,0)[cc]{$\beta$}}
\put(12.00,40.00){\makebox(0,0)[cc]{$p_1$}}
\put(58.00,40.00){\makebox(0,0)[cc]{$p_2$}}
\put(35.00,40.00){\makebox(0,0)[cc]{$p_2+p$}}
\put(20.00,42.00){\makebox(0,0)[cc]{$\beta^\prime$}}
\put(50.00,42.00){\makebox(0,0)[cc]{$\alpha^\prime$}}
\put(77.00,80.00){\makebox(0,0)[cc]{$k_1$}}
\put(123.00,80.00){\makebox(0,0)[cc]{$k_2$}}
\put(100.00,80.00){\makebox(0,0)[cc]{$k_2-p$}}
\put(85.00,79.00){\makebox(0,0)[cc]{$\beta$}}
\put(115.00,79.00){\makebox(0,0)[cc]{$\alpha$}}
\put(77.00,40.00){\makebox(0,0)[cc]{$p_1$}}
\put(123.00,40.00){\makebox(0,0)[cc]{$p_2$}}
\put(100.00,40.00){\makebox(0,0)[cc]{$p_1-p$}}
\put(85.00,42.00){\makebox(0,0)[cc]{$\alpha^\prime$}}
\put(115.00,42.00){\makebox(0,0)[cc]{$\beta^\prime$}}
\put(30.00,67.00){\vector(-1,0){2.00}}
\put(42.00,67.00){\vector(-1,0){2.00}}
\put(93.00,67.00){\vector(1,0){2.00}}
\put(105.00,67.00){\vector(1,0){2.00}}
\put(44.00,117.00){\vector(1,-1){2.00}}
\put(89.00,117.00){\vector(1,-1){2.00}}
\put(109.00,111.00){\vector(1,1){2.00}}
\put(24.00,111.00){\vector(1,1){2.00}}
\end{picture}

\vspace{-3cm}

This is why the two-photon exchange matrix element is given by an expression,
which contains an extra factor $1/2$:
%\ba
%M &=&\bar{u}(k_2)
%\Biggl[
%         \gamma_\beta
%   \frac{\hat k_1 + \hat p + i m_\mu}{p^2+2 p k_1}
%         \gamma_\alpha
%      +  \gamma_\alpha
%   \frac{\hat k_2 - \hat p +i m_\mu}{p^2-2 p k_2}
%         \gamma_\beta
%  \Biggr]
% u(k_1)
%\nll
%&& \cdot \frac{1}{2 p^2(p+q)^2}
%\nll
%&& \cdot
%\bar{u}(p_2)
%\Biggl[    \gamma_\alpha
%   \frac{  \hat p_2 + \hat p + i m_e }{ p^2 + 2 p p_2 }
%           \gamma_\beta
%          +\gamma_\beta
%   \frac{  \hat p_1 - \hat p + i m_e }{ p^2 - 2 p p_1 }
%           \gamma_\alpha
%\Biggr] u(p_1)
% \ea
%---------------------------
\ba
M^{^{\rm{BOX}}} =&& \frac{1}{2} \int \frac{d^4 p}{p^2(p+q)^2}
\nll
&&\times \bar{u}(k_2)
   \Biggl[
   \frac{\gamma_\beta(2k_{1 \alpha}+\hat p \gamma_\alpha)}
        {p^2 + 2pk_1}
 + \frac{(2k_{2\alpha}-\gamma_\alpha \hat p) \gamma_\beta}
        {p^2 - 2pk_2}
   \Biggr] u(k_1)
\nll
&&\times \bar{u}(p_2)
   \Biggl[
   \frac{(2p_{2\alpha} + \gamma_\alpha \hat p)\gamma_\beta}
        {p^2 + 2pp_2}
 + \frac{\gamma_\beta(2p_{1\alpha} - \hat p\gamma_\alpha)}
        {p^2 - 2pp_1}
   \Biggr] u(p_1).
\label{mbox}
\ea
%-------------------
For later use we introduce the short hand notation for propagators
\ba
\prod (k_1) &=& p^2 + 2p k_1,  \nll
\prod (k_2) &=& p^2 - 2p k_2,  \nll
\prod (p_1) &=& p^2 - 2p p_1,  \nll
\prod (p_2) &=& p^2 + 2p p_2.
\ea
%-------------------

 Now we separate the infrared divergent (IRD) part of the two-photon exchange
contribution. There are two IRD-poles: 1) one is located at $p \to 0$ and
2) another one -- at $(p+q) \to 0$.

\noindent At $p \to 0$, one has
\ba
\frac{1}{2}\frac{1}{p^2q^2}
   \Biggl[
   \frac{4k_1 p_2} {\ds \prod(k_1) \prod(p_2) }
 + \frac{4k_1 p_1} {\ds \prod(k_1) \prod(p_1) }
 + \frac{4k_2 p_2} {\ds \prod(k_2) \prod(p_2) }
 + \frac{4k_2 p_1} {\ds \prod(k_2) \prod(p_1) }
   \Biggr] \gamma_\beta \otimes \gamma_\beta ,
\label{pole_1}
\ea

\noindent While at $p \to q$, one has
\ba
\frac{1}{2}\frac{1}{q^2(p+q)^2}
   \Biggl[
   \frac{4k_1 p_2} {\ds \prod(k_1) \prod(p_2) }
 + \frac{4k_1 p_1} {\ds \prod(k_1) \prod(p_1) }
 + \frac{4k_2 p_2} {\ds \prod(k_2) \prod(p_2) }
 + \frac{4k_2 p_1} {\ds \prod(k_2) \prod(p_1) }
   \Biggr] \gamma_\beta \otimes \gamma_\beta .\nll
\label{pole_2}
\ea
For the latter term, we perform the substitution $p=p^{'}-q$ and
observe that at this substitution the propagators transform as follows
\ba
\prod (k_1) &\longrightarrow & \prod(-k_2),  \nll
\prod (k_2) &\longrightarrow & \prod(-k_1),  \nll
\prod (p_1) &\longrightarrow & \prod(-p_2),  \nll
\prod (p_2) &\longrightarrow & \prod(-p_1).
\ea
With one more substitution $p^{'}\to -p$, and using equalities
$k_2.p_2=k_1.p_1,\;k_2.p_1=k_1.p_2$,
we see that the second pole
gives exactly the same contribution as the first one.
%-------------------
Therefore, the full IRD-part of the two-photon exchange reads:
%----------------
\ba
\frac{1}{q^2}\frac{1}{p^2}
   \Biggl[
   \frac{4k_1 p_2} {\ds \prod(k_1) \prod(p_2) }
 + \frac{4k_1 p_1} {\ds \prod(k_1) \prod(p_1) }
 + \frac{4k_2 p_2} {\ds \prod(k_2) \prod(p_2) }
 + \frac{4k_2 p_1} {\ds \prod(k_2) \prod(p_1) }
   \Biggr] \gamma_\beta \otimes \gamma_\beta.
\label{pole_12}
\ea

Now we add and subtract the IRD-part of the two-photon exchange and
when adding it we use~(\ref{pole_12}), while when subtracting it we use
the sum of~(\ref{pole_1}) and~(\ref{pole_2}).
Using the identity
\ba
\frac{1}{p^2(p+q)^2}-\frac{1}{p^2q^2}-\frac{1}{q^2(p+q)^2}
\equiv \frac{-2p.(p+q)}{q^2p^2(p+q)^2},
\ea
we arrive at the final expression for the two-photon exchange
contribution before integration over $d^{4}p$.

\ba
M^{^{\rm{BOX}}}&\sim&
   \frac{4} { q^2 p^2 }
   \Biggl[
   \frac{ - S } {\ds \prod(k_1) \prod(p_1) }
 + \frac{ - S_1}{\ds \prod(k_1) \prod(p_2) }
   \Biggr] \gamma_\beta \otimes \gamma_\beta \nll
&&-\frac{4 p (p+q)} { p^2 (p+q)^2 q^2  }
   \Biggl[
   \frac{ - S  }{\ds \prod(k_1) \prod(p_1) }
  +\frac{ - S_1}{\ds \prod(k_1) \prod(p_2) }
   \Biggr] \gamma_\beta \otimes \gamma_\beta \nll
&&+\frac{1} {  p^2 (p+q)^2 }
   \Biggl[
   \frac{ k_{1 \alpha}}{\ds \prod(k_1) }
  +\frac{ k_{2 \alpha}}{\ds \prod(k_2) }
   \Biggr] \gamma_\beta \otimes
   \Biggl[
   \frac{ \gamma_\alpha \hat p \gamma_\beta } { \ds \prod(p_2) }
  -\frac{ \gamma_\beta  \hat p \gamma_\alpha} { \ds \prod(p_1) }
   \Biggr]                                   \nll
&&+\frac{1} { p^2 (p+q)^2 }
   \Biggl[
   \frac{ \gamma_\beta \hat p \gamma_\alpha } { \ds \prod(k_1) }
  -\frac{ \gamma_\alpha \hat p \gamma_\beta } { \ds \prod(k_2) }\Biggr]
   \otimes \gamma_\beta
   \Biggl[
   \frac{ p_{2 \alpha} } { \ds \prod(p_2) }
  +\frac{ p_{1 \alpha} } { \ds \prod(p_1) }
   \Biggr]                               \nll
&&+\frac{1} {2 p^2 (p+q)^2 }
   \Biggl[
   \frac{ \gamma_\beta  \hat p \gamma_\alpha } { \ds \prod(k_1) }
  -\frac{ \gamma_\alpha \hat p \gamma_\beta  } { \ds \prod(k_2) }\Biggr]
   \otimes
   \Biggl[
   \frac{ \gamma_\alpha \hat p \gamma_\beta } { \ds \prod(p_2) }
  -\frac{ \gamma_\beta  \hat p \gamma_\alpha} { \ds \prod(p_1) }\Biggr].
   \label{bf_int}
\ea

The first raw of this formula represents the IRD-contribution,
the second raw stands for the IR-free scalar, the next two represent
the IR-free vector, and finally the last one is an IR-free tensor.

 For the ${\rm{IR}}$-divergent part we introduce the well-known presentation
\ba
 J_3^{^{\rm{IR}}} \left( k_1,p_1 \right) &=&
 \frac{16\pi^2}{i} 
 \int\frac{d^4 p}{(2\pi)^4 p^2 \prod(k_1)\prod(p_1) }
  \nll &=&
  P_{\rm IR}(\mu) \int_0^1 \frac{dy}{(-K_y^2)}
               +\frac{1}{2}\int_0^1 \frac{dy}{(-K_y^2)}\ln\frac{(-K_y^2)}{q^2}
               + {\cal O}(n-4).
\ea
where
\ba
-K^2_y &=& \AMU y^2 + \AME(1-y)^2-Sy(1-y).
\ea
We will use the short hand notations
\ba
P^{\rm IR}_{S} &\equiv& P^{\rm IR}(S) \;=\;-P_{\rm{IR}}(\mu)\int_0^1\frac{dy}{(-K_y^2)},
\nll
K_S &\equiv& K(S) \;=\; \int\frac{dy}{(-K^2_y)} \ln \frac{(-K^2_y)}{q^2},
\ea
Then all needed IRD-integrals are
\ba
   J_3^{^{\rm IR}}(k_1,p_1) &=&
   J_3^{^{\rm IR}}(k_2,p_2) = -P^{\rm IR}_{S  }+\frac{K_S      }{2},\\
   J_3^{^{\rm IR}}(k_1,p_2) &=&
   J_3^{^{\rm IR}}(k_2,p_1) =  P^{\rm IR}_{S_1}+\frac{K_{S_{1}}}{2}.
\ea
with
%---
\ba
P^{\rm IR}(S_1)&=&-P^{\rm IR}(-S_1),
\nll
K_{S_1} &=& K(-S_1).
\ea

For the IR-finite scalars
\ba
 J_4^{^{\rm IR}} \left( k_i p_j \right)
   =\frac{16 \pi^2}{i}
 \int\frac{2p.(p+q) d^4 p}{(2\pi)^4 p^2(p+q)^2\prod(k_i)\prod(p_j)},
\ea
we derived
\ba
&& J_4(k_1,p_1) = J_4(k_2,p_2) = K_{S}, \\
&& J_4(k_1,p_2) = J_4(k_2,p_1) = K_{S_{1}},
\ea

\subsection{Vectorial integrals}
%-------------------------------
We begin  with the reduction of vector integrals. First we define the coefficients
of the vector reduction
%------------------
\ba
 V^{ij}_{\mu} &=&
   \frac{16 \pi^2}{i} \int
   \frac{d^4 p\;p_\mu }{(2\pi)^4 p^2 (p+q)^2 \prod(k_i)\prod(p_j) }
  \nll &&
   = (V_q)^{ij} q_\mu+(V_k)^{ij} (k_i)_\mu + (V_p)^{ij} (p_j)_\mu,
\label{bector}
\ea
%-------------------------
The system of linear equations for these coefficients in
terms of scalar integrals, derived by contraction of~(\ref{bector}) 
with $q_{\mu},\;(k_i)_{\mu},\;(p_j)_{\mu}$, reads
\ba
 && 2 q^2  \Vq +2 \SKi q\cdot \Ki\Vk  +  2 \SPj q\cdot\Pj \Vp
           - J_4(k_i,p_j) + 2 J_3^{^{\rm IR}}(k_i,p_j)=0
\nll &&
 2 \SKi q\cdot k_i \Vq+ 2 \SKi^2 k_i^2  \Vk
        + 2 \SKi \SPj k_i\cdot p_j \Vp
        - J_3(p_j) + J_3^{^{\rm IR}}(k_i,p_j)=0,
\nll &&
 2 \SPj  q \cdot p_j \Vq + 2 \SKi \SPj k_i\cdot p_j \Vk
        + 2 \SPj^2 p_j^2  \Vp
        - J_3(k_i)  + J_3^{^{\rm IR}}(k_i,p_j)=0.
\label{vystem}
\ea
Here
\ba
S_{k_1}&=&-1,\qquad S_{k_2}\;=\;+1,
\nll
S_{p_1}&=&+1,\qquad S_{p_2}\;=\;-1.
\ea
%-------------------------
Two additional types of scalar integrals
%-----------------------------
\ba
J_3\left( k_i  \right)
   =\frac{16\pi^2}{i}  \int
   \frac{d^4 p}{(2\pi)^4 p^2 (p+q)^2 \prod(k_i) },
\label{j3intk}
\ea
and
\ba
J_3\left( p_j  \right)
   =    \frac{16\pi^2}{i} \int
   \frac{d^4 p}{(2\pi)^4 p^2 (p+q)^2\prod(p_j) },
\label{j3intp}
\ea
are introduced. They safisfy the equalities
%-----------------------------
\ba
 J_3\left( k_1  \right) = J_3\left( k_2  \right)
\ea
and
\ba
 J_3\left( p_1  \right) = J_3\left( p_2  \right),
\ea
and may be described by one generic formula:
%-----------------------------
\ba
J_3\left( q_i  \right)
   =\int_0^1
    \frac{dy}{-q^2y -q_i^2(1-y)+(q-q_i)^2 y(1-y)}
    \ln\frac{(q-q_i)^2 y(1-y)-q_i^2(1-y) }{q^2 y}. \nll
\ea
with $q_i=(k_1,\;-k_2,\;p_2,\;-p_1)$.

The solution of the system~(\ref{vystem}) is presented in subsection 
F.4 of this Appendix.

%----------- TENSOR !!!
\subsection{Tensorial integrals}
%-------------------------------
Define the coefficients of tensor reduction
%-------------------------
\ba
 T^{ij}_{\mu\nu} &=&
      \frac{16 \pi^2}{i}  \int
  \frac{d^4 p \; p_\mu p_\nu}{(2\pi)^4 p^2 (p+q)^2 \prod(k_i)\prod(p_j)}
  \nll && =   T_0 \delta_{\mu\nu}
   + T_{qq}^{ij} q_\mu q_\nu
   + T_{kk}^{ij} (k_i)_\mu (k_j)_\nu
   + T_{qp}^{ij} (p_i)_\mu (p_j)_\nu
  \nll &&
   + T_{qk}^{ij} \left[q_\mu(k_i)_\nu + q_\mu(k_i)_\mu \right]
   + T_{qp}^{ij} \left[q_\mu(p_j)_\nu + q_\mu(p_j)_\mu \right]
  \nll &&
   + T_{kp}^{ij} \left[(k_i)_\mu (p_j)_\nu + (k_i)_\nu (p_j)_\mu \right].
\label{bensor}
\ea
%------------------------

 The system of equations for the coefficients $T^{ij}$ was derived in
the following way:
%-------------------------
\begin{itemize}
\item
{ the first equation was received by multiplying~(\ref{bensor})
  with $\delta_{\mu\nu}$
\ba
 J_3^{^{\rm IR}}\left( k_i p_j \right)
   = 4 T_0^{ij}
   +   q^2 T_{qq}^{ij}
   +  (k_i)^2 T_{kk}^{ij}
   +  (p_j)^2 T_{pp}^{ij}
   +   q^2 T_{qk}^{ij}
   +   q^2 T_{qp}^{ij}
   +   I_{ij}  T_{kp}^{ij},
\label{fromdelta}
\ea}
%---
\item
 {the next three equations were received by multiplying~(\ref{bensor})
  with  $q_{\nu}$
 \ba
  && \int  \frac{d^4 p  \left[ 2pq+q^2+p^2-q^2-p^2 \right] p_\mu }
                {(2\pi)^4 p^2 (p+q)^2 \prod(k_i)\prod(p_j)       }
 \nll
%---
 && =\int\frac{d^4p}{(2\pi^4)}\frac{p^\mu}{p^2 \prod(k_i) \prod(p_j)}
-    \int\frac{d^4p}{(2\pi^4)}\frac{p^\mu}{(p+q)^2 \prod(k_i) \prod(p_j)}
  \nll &&
-q^2 \int\frac{d^4p}{(2\pi^4)}\frac{p^\mu}{p^2 (p+q)^2 \prod(k_i) \prod(p_j)}
\nll &&
   = 2 q_\mu T_0^{ij}
   + 2 q^2 q_\mu T_{qq}^{ij}
   +   q^2 T_{kk}^{ij} (k_i)_\mu
   +   q^2 T_{pp}^{ij} (p_j)_\mu
\nll &&
   + T_{qk}^{ij} \left[q_\mu q^2 + 2 q^2 (k_i)_\mu \right]
   + T_{qp}^{ij} \left[q_\mu q^2 + 2 q^2 (p_j)_\mu \right]
\nll &&
   + q^2 T_{kp}^{ij} \left[(k_i)_\mu + (p_j)_\mu   \right],
\ea}
%---
\item{another six equations were received by multiplying~(\ref{bensor})
 with  $(k_i)_{\nu}$ and $(p_j)_{\nu}$.}
\end{itemize}

The latter nine equations are 
\ba
% 369
% procedure test(Ki,SKi,Pj,SPj,T1,T2,T3,DET,H1,H2,H3,Tkk,Tqk,Tkp)
 && q^2 \Tkk + 2 q^2 \Tqk + q^2 \Tkp + q^2 V_k(k_i, p_j) = 0
 \nll  &&
 2 k_i^2 \Tkk + q^2 \Tqk + 2 \SKi  \SPj k_i \cdot p_j \Tkp
 %\nll  &&
            + V_{1k}(k_i,p_j) + q^2 V_q(k_i,p_j) + J_3^{^{\rm IR}}(k_i,p_j)=0
 \nll  &&
 2 \SKi  \SPj   k_i \cdot p_j \Tkk  + q^2 \Tqk + 2 p_j^2  \Tkp
 %\nll  &&
            -    V_{2k}(k_i, p_j)+ V_{1k}(k_i, p_j)=0
 \nll
%-----
% 258
% procedure test(Ki,SKi,Pj,SPj,T1,T2,T3,DET,H1,H2,H3,Tpp,Tqp,Tkp)
 &&   q^2 \Tpp + 2 q^2 \Tqp  + q^2 \Tkp  + q^2 V_p(\Ki,\Pj )=0
 \nll
 &&  2 \SKi \SPj  \Ki\cdot \Pj  \Tpp + q^2 \Tqp +2 k_i^2  \Tkp
 %\nll  &&
           - V_{3p}(\Ki,\Pj ) + V_{1p}(\Ki,\Pj) =0
 \nll
 &&    2 \Pj \cdot\Pj  \Tpp + q^2 \Tqp + 2 \SKi \SPj \Ki \cdot \Pj \Tkp
 %\nll  &&
    + V_{1p}(\Ki,\Pj) + J_{3}{^{IR}}(\Ki,\Pj ) + q^2 V_q(\Ki,\Pj)=0
\nll
%----- &&
% 4710
% procedure test(Ki,SKi,Pj,SPj,T1,T2,T3,DET,H1,H2,H3,Tqq,Tqk,Tqp)
 && 2 q^2 \Tqq  + q^2 \Tqk  + q^2 \Tqp -  V_{1k}(\Ki,\Pj )
 %\nll  &&
 -   V_{1p}(\Ki,\Pj )+2 q^2 V_q(\Ki,\Pj)=0
 \nll  &&
    q^2 \Tqq  + 2 k_i^2  \Tqk  + 2 \SKi  \SPj  \Ki \cdot \Pj  \Tqp
 \nll  &&
  - V_{3q}(\Ki,\Pj )-V_{1k}(\Ki,\Pj )-V_{1p}(\Ki,\Pj )-J_3{^{\rm IR}}(\Ki,\Pj)=0
 \nll  &&
    q^2 \Tqq  + 2 \SKi \SPj \Ki \cdot \Pj \Tqk  + 2 \Pj^2 \Tqp
 \nll  &&
    - V_{2q}(\Ki,\Pj )- V_{1k}(\Ki,\Pj ) - V_{1p}(\Ki,\Pj)-J_3^{^{\rm IR}}(\Ki,\Pj)
    =0
\label{tystem}
\ea
Together with the tenth equation~(\ref{fromdelta}), they
can be solved and the answer 
may be expressed in terms $V^{ij}$ and additional 
vectors arising from various pinches of the box diagrams.

%All coefficients $V^{ij}$ and
%$T^{ij}$ are presented in~(\ref{vij})--(\ref{tij}).
%By definition for any $a$ it may be write
%$V_{a}^{ij}=V_{a}(k_i,p_j)$ and the same for tensor
%$T_{a}^{ij}=T_{a}(k_i,p_j)$.

\subsection{The box contribution to the cross-section}
%-----------------------------------------------------
   Substituting all these solutions to the box amplitude and computing its 
interference with Born amplitude, we derive the contributions
to the unpolarized and polarized cross-section parths from  the
two-photon exchange diagrams to
the differential cross-section of elastic $\mu e$ scattering
%----------------------------------------------------------
%file support: analytic approach, mueana:subroutine: MUEBOX
  \newcommand{\KSOL }{\mbox{$K_{_{SOL}} $}}
  \newcommand{\YllMU}{\mbox{$y_{\mu_2} $}}
%----------------------------------------------------------
%\ba
%\frac{d\sigma^{^{\rm{BOX}}}}{dy} &=& \frac{\alpha^3}{S}{\Qmue}
% \Biggl[ 2 (S \PIRSS-\Sl \PIRSl)\frac{d\sigma^{^{\rm{BORN}}}}{dy\quad}
%         +{\cal B}_{unp} + \Pe \Pm {\cal B}_{pol} \Biggr]
%\ea
\ba
 \frac{d\sigma^{^{\rm{BOX}}}_{_{\rm{SIURA}}}}{dy\qquad}
 \Biggl[ -4
     \PIRMU   \ln(1-y)
\frac{d\sigma^{^{\rm{BORN}}}}{\ds dy\quad}
         +{\cal B}_{unp} + \Pe \Pm {\cal B}_{pol} \Biggr],
\ea
where unpolarized and polarized non-factorized contributions
${\cal B}_{unp}$ and ${\cal B}_{pol}$ are:
\ba
{\cal B}_{unp}
	  &=&  \left(1-\frac{2}{y}+ \frac{2\Rm}{y}\right)
		      (S\KSO+\Sl\KSl)
	      + 4 \left[-1+\frac{2}{y}-2 \Rm
                  \left(1+\frac{1}{\yml} \right) \right]
	     \nll
	   && + 2 \left( \frac{\yml}{y} - 2\Rm\right)\LSO
	      + 2 \left( \frac{1}{y} - \frac{ \Rm}{\yml}\right) \LSl
	      \nll
	   && + 4 \Rm \left[- 1 - \frac{1}{\yml}
		   + \frac{2(2-y)}{y(y+4\Rm)} \right]
			      \ln\frac{Q^2}{\Mm}
	    + 2 \left(- 1 + \frac{2}{y}\right) S(\KSO - y \JPl)
	     \nll
           && + 2 \left[- 2 + y - 4\Rm
		 \left(\frac{\yml}{y}
		  +\frac{1+2\Rm}{y+4\Rm} \right) \right] S \JKl,
             \nll
{\cal B}_{pol} &=&  \left( -1+ \frac{2}{y}-2 \Rm\right) (S\KSO+\Sl\KSl)
	     \nll
	   && + 4\left[-1 + 2 \Rm
	     \left(-1+\frac{2}{y}+\frac{1-\Rm}{\yml}-\Rm\right) \right]
	     \nll
	   && + 2 \left[\frac{\yml}{y} (1+2\Rm)-2\Rms\right] \LSO
	    + 2 \left[ \frac{2 \Rm (1-\Rm)}{\yml}
			-\frac{1-2\Rm}{y}\right] \LSl
	     \nll
	   && + 4 \Rm \left( \frac{2}{y}
	      +\frac{1-\Rm}{\yml}
	      -\frac{2(1+2 \Rm)}{y+4 \Rm} -\Rm \right)
		\ln\frac{Q^2}{\Mm}
	     \nll
	   && + 4 \frac{\Rm y (1+2\Rm)}{y+4\Rm} S \JKl,
\label{exbox}
\ea
with
% all definitions from mueana.f:
\ba
 R_\mu & = & \frac{\AMU}{S},  \nll
 Y   & = & \frac{1}{1+R_\mu}.
\ea
%Objects listed below, arose from integration over $d^{4}p$.
%They typically enter in couples.
%\ba
%  S\PIRSS - \Sl\PIRSl &=&
% \PIRMU
%\left[ \ln\frac{S^2  }{\me\mmu}
%      -\ln\frac{\Sl^2}{\me\mmu}\right]
%\;=\; - \PIRMU 2 \ln(1-y).
%\ea
There are two useful combinations:
\ba
 \Sl \KSl+S\KSO &=&
  -\ln  \frac{Q^4}{S\Sl} \ln \frac{\Sl}{S}
           -2 \Litwo\left(1-\frac{\AMU}{\Sl}\right)
           +2 \Litwo\left(1+\frac{\AMU}{  S}\right)+\pi^2,    \\
%---
S(\KSO - y \JPl) &=&
          -   \ln^2\frac{Q^2}{S}+\frac{1}{2}\ln^2\frac{Q^2}{\AMU}
          + 2 \Litwo\left(1+\frac{\AMU}{S}\right)+\frac{\pi^2}{3}.
\ea
Here
\ba
 \JKl &=& J(\AMU)\;=\;\frac{1}{\SLAMM}
          \Biggl[ \ln \frac{\AMU}{Q^2} \ln {\YllMU}
                 -2 \Litwo(1-\YllMU)
                  -\frac{1}{2} \ln^2 {\YllMU}+\pi^2 \Biggr],\\
 J_P &=& J(\AME).
\ea
%---
and
\ba
y_{\mu_1}  &=& \frac{ Q^2+2\AMU +\SLAMM }{2 \AMU},   \\
y_{\mu_2}  &=& \frac{1}{y_{\mu_1}}.
\ea

The two-photon exchange contribution
simplifies drastically in the DOURA.
We list the answer because of its elegance.
\ba
\frac{d\sigma^{^{\rm{BOX}}}_{_{\rm DOURA}} }{dy\qquad} &=& \frac{\alpha^3}{S}{\Qmue}
      \Biggl\{-\PIRMU\left[\frac{2}{y^2}+
	        ( 1 + \Pe \Pm) \left(1-\frac{2}{y} \right)
		    \right] 8 \ln(1-y)
\nll
         && +(1-\Pe\Pm)\Biggl[ \left( 1 - \frac{2}{y}    \right)
             \left(\ln^2(1-y) - 2\ln (1-y) \ln y + \pi^2 \right) \nll
         && +\frac{2}{y}   \ln(1-y) - 2 \ln y  \Biggr]
\nll
         && +2 \left(1-\frac{2}{y}   \right)    \ln^2(1-y)
            -4 \frac{1-y}{y} \ln(1-y)\Biggr\}.
\ea
%-
%The corresponding Gram determinant will be $\Delta_4 = p_i \cdot p_j$
%---
%-------------------
%\input{m_fig      }
%-------------------
\section{Description of www-Figures}
%-----------------------------------
In March 1996, when the {\tt FORTRAN} code $\mu${\bf{e}}{\it{la}} was 
completed, a lot of figures for the radiative corrections to the
polarized cross-sections and the polarization asymmetry were produced
and put to a home-page of the theory group of DESY-Zeuthen,
{\tt http://www.ifh.de/}\~ {\tt bardin}.\\

\noindent
In this Appendix, we give a short guide to these figures. \\

\noindent All the cross-sections, radiative corrections
and asymmetries are shown as function of $y_{\mu}$.
The units are: cross-sections in microbarns, radiative corrections in
percent, asymmetries in absolute, dimensionless units varying from 0 to 1.

\noindent
Some legends:
\begin{itemize}
\item[-]
parallel, antiparallel means mutual longitudinal polarization orientations;
\item[-]
$P_e$ and $P_{\mu}$ are modules of polarization degrees, always written
in figures;
\item[-]
{\bf All corrections} mean that all 12 contributions (\ref{cont12}) are taken into account;
\item[-]
{\bf AN} means Analytic calculations;
\item[-]
when {\bf AN} is not written, the numerical calculations are meant.
\end{itemize}

The $\delta_{y_{\mu}}$ is defined by
\bq
\delta_{y_{\mu}} = 
\frac{\displaystyle \frac{d\sigma^{^{\rm{QED }}} }{dy_{\mu}\quad}}
     {\displaystyle \frac{d\sigma^{^{\rm{BORN}}} }{dy_{\mu}\qquad}} - 1, \; (\%).
\nonumber
\eq

\noindent
SIURA-series of 43 figures contain following plots:
\begin{itemize}
\item
1-3 Born cross-section and Born asymmetry;
\item
4-5 Results of completely integrated analytic calculations without cuts;
\item
6-7 Results of completely numerical calculations without cuts;
\item
8-17 Illustrate effects of some cuts separately, muon angular cut
being treated both analytically and numerically;
\item
18-19 are our main result: All corrections, All cuts (four indeed);
\item
20-27 illustrate four cases: $\mu+e$ corrections, $\mu$ corrections,
$e$ corrections and $\mu e$ interference corrections,
correspondingly, without cuts;
\item
28-35 the same, but with all cuts;
\item
36-43 is a  series of figures with `realistic' $P_e$ and $P_{\mu}$.
It is quite senceless for the asymmetry, where $P_e$ and $P_{\mu}$
cancel. We only `gain' an instability due to cancellation of nearly
equal numbers.
\end{itemize}
%--------------
%\input{m_bibl}
%--------------

%--------------------
\end{document}